\documentclass[final,5p,times,twocolumn,authoryear]{elsarticle}

\usepackage{aas_macros} 
\usepackage{amssymb}
\usepackage{amsmath}
\usepackage[dvipsnames]{xcolor} 
\usepackage{lipsum}

\usepackage{float}
\usepackage{hyperref}

\journal{Astronomy $\&$ Computing}

\begin{document}

\begin{frontmatter}



\title{Analysis of Galactic cirrus filaments in HSC-SSP high-resolution deep images using artificial neural networks}


\author[CAO_RAS]{Denis~M.~Poliakov\corref{cor1}}\ead{polyakovdmi93@gmail.com}
\author[CAO_RAS]{Anton~A.~Smirnov}
\author[CAO_RAS,SPbSU]{Sergey~S.~Savchenko}
\author[CAO_RAS,SPbSU]{Alexander A. Marchuk}
\author[BYU]{Aleksandr~V.~Mosenkov}
\author[CAO_RAS,SPbSU]{Vladimir~B.~Il'in}
\author[CAO_RAS]{George~A.~Gontcharov}
\author[CAO_RAS,SPbSU]{Daria~G.~Turichina}
\author[CAO_RAS,SPbSU]{Andrey~D.~Panasyuk}

\cortext[cor1]{Corresponding author}

\affiliation[CAO_RAS]{organization={Central (Pulkovo) Astronomical Observatory, Russian Academy of Sciences},
            addressline={Pulkovskoye chaussee 65/1}, 
            city={St. Petersburg},
            postcode={196140}, 
            country={Russia}
}

\affiliation[SPbSU]{organization={Saint Petersburg State University},
            addressline={Universitetskij pr. 28}, 
            city={St. Petersburg},
            postcode={198504}, 
            country={Russia}
}


\affiliation[BYU]{organization={Astrophysical Research Consortium, c/o Department of Astronomy, University of Washington, Box 351580},
            city={Seattle},
            postcode={WA 98195}, 
            country={USA}
}


\begin{abstract}
The existence of Galactic optical cirrus poses a challenge for observing faint objects within our Galaxy and dim extragalactic structures. To investigate individual cirrus filaments in the Hyper Suprime-Cam Subaru Strategic Program public data release 3 (HSC-SSP DR3)
we use a technique based on convolutional neural networks and ensemble learning. This approach allows us to distinguish cirrus filaments from foreground and background objects across the entire HSC-SSP, using optical images in the $g$, $r$, and $i$ wavebands.
A comparison with previous work using deep Sloan Digital Sky Survey Stripe~82 (SDSS Stripe~82) data reveals that the cirrus clouds identified in this study are highly consistent in location within the overlapping survey region. However, in the deeper HSC-SSP dataset, we were able to detect $4.5$ times more cirrus clouds.
Our study indicates that the sky background in HSC-SSP coadd images is over-subtracted, as evidenced by the surface brightness distribution in cirrus filaments and surrounding regions. Objects with surface brightness of $m = 29~\mbox{mag~arcsec}^{-2}$ near large filaments can be dimmed by over-subtraction of $0.5$ magnitude in the $r$ band.
This suggests that cirrus clouds should be taken into account in algorithms for estimating the sky background.
For practical use, we provide a catalog of filaments and a framework that allows one to train neural network models for segmenting cirri in HSC-SSP coadd images.
\end{abstract}


\begin{highlights}
\item An ensemble of deep neural networks was used to identify cirrus in HSC-SSP data
\item A catalog of cirrus filaments in deep optical HSC-SSP images has been created
\item Cirrus clouds in HSC-SSP images agree well with cirrus in SDSS Stripe 82 images
\item Large cirrus filaments may affect the quality of sky background estimation
\end{highlights}


\end{frontmatter}




\section{Introduction}
\label{sec:introduction}
Cirrus clouds are diffuse filamentary structures resembling Earth's atmospheric cirrus clouds.
Galactic cirri are observed predominantly at high Galactic latitudes ($b \gtrsim 20^{\circ}$) and have been identified in the infrared (IR) \citep{Low_etal1984, Kiss_etal2001, Kiss_etal2003, Martin_etal2010, Veneziani_etal_2010, Planck_etal2011, Penin_etal2012, Schisano_etal2020}, optical \citep{Vacouleurs_1955, Vacouleurs_1960, Vaucouleurs_Freeman_1972, Sandage_1976, Mattila_1979, Vries_1985, Laureijs_etal_1987, Witt_etal_2008, Ienaka_etal2013, Miville-Deschenes_etal2016, Stripe82_2020A&A...644A..42R, Gontcharov_etal_2022_S82_Wolf, Cirrus_2023MNRAS.519.4735S}, and ultraviolet \citep{Haikala_etal_1995, Witt_etal_1997, Gillmon_Shull2006, Boissier_2015, Akshaya_etal2019, Gontcharov_etal_2022_S82_Wolf}.
Various studies have also found that cirrus fluxes in the optical and IR spatially correlate with emission in lines of CO and H$_2$\citep{Weiland_etal1986, deVries_etal1987, Gillmon_Shull2006, Ienaka_etal2013, Stripe82_2020A&A...644A..42R}.
At optical wavelengths, emission associated with Galactic cirrus directly contributes to diffuse Galactic light (DGL). This links the observed filamentary structures with the formation of sky background at high Galactic latitudes.


The DGL is a part of the sky background (light from diffuse and unresolved sources) that consists of starlight scattered by and re-emitted as thermal radiation from interstellar dust in the diffuse interstellar medium (ISM).
The DGL was studied by \citet{Brandt_Draine_2012, Chellew_etal_2022}, and it was found to be consistent with the light spectrum produced by a dust model of \citet{Zubko_2004}.
A detailed study of the high Galactic latitude translucent cloud MBM32 by \citet{Ienaka_etal2013} found a correlation between emission at 100 $\mu m$ and intensity of the DGL. However, the slope of this correlation is twice that predicted by dust models \citep{Weingartner2001DustGD, Zubko_2004}.
MBM32 was also studied in \citet{Onishi_etal_2018} using near-IR observations at  Science and Technology Satellite-3 (STSAT-3).
The authors calculated the DGL spectrum from a recent model of \citet{WLJ_2015ApJ...811...38W} and compared it with their observations. The comparison revealed that the observed color of near-IR DGL is closer to the model spectra without very large grains.
Other cirrus clouds, called the Spider complex, were used by \citet{Bowes_Martin_2023ApJ...959...40B, Zhang_etal_2023ApJ...948....4Z} to test new approaches to constraining the interstellar dust and anisotropic interstellar radiation field models required to interpret different cirrus observations.
A number of studies \citep{1994A&A...291L...5B_Bernard, 1998_Szomoru, Veneziani_etal_2010, 2014A&A...566A..55P_Planck_Collaboration, 2017A&A...597A.130B_Bianchi} have investigated the relationship between the cirrus spectrum (emission and reflected spectra) and the properties of dust in filaments. These studies mainly used IR, microwave, and submillimeter data, except for \citet{1998_Szomoru}, which used spectrographic data in the wavelength range $4090~\mbox{\AA} - 9590~\mbox{\AA}$. These investigations focus on the properties of dust and ISM, such as temperature, spectral index, emissivity, and opacity.
In particular, consistent estimates of the dust temperature ($T \sim 20~\mbox{K}$) in high-latitude Galactic cirri were obtained in \citet{Veneziani_etal_2010, 2014A&A...566A..55P_Planck_Collaboration, 2017A&A...597A.130B_Bianchi}.


The complex form of cirrus clouds has been the subject of extensive research. They are typically observed as groups of numerous co-directed filaments, rather than distinctly shaped clouds. The fractal nature of the cirrus has been identified and confirmed in various studies \citep{Bazell_Desert1988, Falgarone_etal1991, Hetem_Lepine1993, Vogelaar_Wakker1994, Elmegreen_Falgarone1996, Sanchez_etal2005, Juvela_etal2018, Marchuk_etal2021}. This morphological complexity makes the identification and segmentation of cirrus filaments
a non-trivial task, particularly in deep wide-field imaging surveys. Such an appearance of these structures is believed to be a result of the diverse physical processes such as turbulence \citep{Padoan_etal_2001, Kowal_Lazarian_2007, Federrath_etal2009, Konstandin_etal2016, Beattie_etal2019a, Beattie_etal2019b}, shock waves \citep{Koyama_Inutsuka_2000}, cooling flows \citep{Vazquez_Semadeni_2007}, the instability of a self-gravitating sheet \citep{Nagai_1998}, and various instabilities in non-self-gravitating clumps, which arise because of the presence of magnetic fields \citep{Hennebelle_2013}.

The investigation of these structures is important for studying various extragalactic sources \citep{Cortese_etal_2010, Sollima_etal_2010, Rudick_etal_2010, Davies_etal_2010, Duc_etal_2018, Barrena_etal_2018}, since cirri are found in both high and low Galactic latitudes \citep{Barrena_etal_2018, Schisano_etal2020, Stripe82_2020A&A...644A..42R, 2024AJ....168...88Z}. The presence of cirrus in deep images hinders the identification of background objects, such as faint galaxies and tidal features in interacting galaxies.

Catalogues of cirrus filaments, such as those presented by \citet{Schisano_etal2020} and \citet{Cirrus_2023MNRAS.519.4735S}, are essential for accounting for the impact of cirrus on the observed properties of extragalactic objects. Additionally, studying the optical properties of cirrus clouds can help distinguish them from extragalactic sources.
Such investigation have been conducted in \citet{Stripe82_2020A&A...644A..42R, Mattila2023LightAC} and in our recent work \citep{Cirrus_2023MNRAS.519.4735S}. \citet{Stripe82_2020A&A...644A..42R} and \citet{Cirrus_2023MNRAS.519.4735S} focus in cirrus clouds in the Sloan Digital Sky Survey Stripe~82 (SDSS Stripe~82) region \citep{SDSS_DR7_2009ApJS..182..543A} and are based on deep images from SDSS Stripe~82 \citep{IAC_Stripe_82_2016MNRAS.456.1359F}.
In \citet{Cirrus_2023MNRAS.519.4735S}, we demonstrated that the colors of most detected filaments cluster around specific values and fall within the following ranges: $0.55 \leqslant g - r \leqslant 0.73$ and $0.01 \leqslant r - i \leqslant 0.33$. These results are largely consistent with those reported in \citet{Stripe82_2020A&A...644A..42R}, which were expressed as the inequality: $(r - i) < 0.43 \times (g - r) - 0.06$. Catalogues of cirrus filaments can also be utilized to determine interstellar extinction in cirrus. For instance, such estimations were conducted in \citet{Szomoru1999ExtinctionCD,Gontcharov_etal_2022_S82_Wolf} using the star count method with Wolf diagrams \citep{Wolf_1923}.

The creation of such catalogs on the scale of modern surveys requires robust and automated identification methods.
Over the past decade, various methods have been employed to detect such structures.
One approach was presented in \citet{Haigh_etal_2021A&A...645A.107H}, where the \texttt{NoiseChisel} \citep{NoiseChisel_2015ApJS..220....1A} and the \texttt{MTObjects} \citep{MTO_teeninga2013bi, MTO_teeninga2016bi} tools, with Bayesian optimized parameters, were successfully used for cirrus detection in the Stripe 82 region.
A different strategy was adopted by \citet{Schisano_etal2020}, where filaments in the Herschel infrared Galactic Plane Survey images \citep{Molinari_etal_2010} were identified using a Hessian matrix.
A similar method was used in \citet{Planck_colab_2016} and \citet{Soler_etal_2022} to investigate the relative orientation between dust structures and the magnetic field, as well as between the HI filamentary structures and the Galactic disc, respectively. In \citet{Menshchikov_2013}, a multi-scale, multi-wavelength filament extraction method was proposed. In \citet{Salji_etal_2015}, the authors developed and applied a Hessian-based ridge detection technique to extract filaments forming a large ``integral shaped filament'' in Orion A North. To study cirrus in far-IR imaging data of dust emission released by the Herschel Gould Belt Survey team \citep{Andre_etal_2010}, a complex multistep method was employed in \citet{Koch_Rosolowsky_2015}. This approach includes an arctan transformation of the image, Gaussian smoothing, adaptive thresholding, mathematical morphology \citep{Serra1983ImageAA}, and the Rolling Hough Transform \citep{Clark_etal_2014}.
The Rolling Hough Transform has also been used to remove cirrus from images obtained by the Dragonfly Telephoto Array \citep{Danieli_etal_2020} in recent studies \citep{2022ApJ...925..219L, 2025ApJ...979..175L}.

In recent years, neural networks have emerged as an effective tool for cirrus detection \citep{Alina_etal_2022, Cirrus_2023MNRAS.519.4735S, Zavagno_etal_2023, Richards_etal_gabor_attention_2024}.
\citet{Alina_etal_2022} showed the effectiveness of Mask R-CNN \citep{MaskR-CNN_2017} and U-Net \citep{U-Net_2015arXiv150504597R} architectures in identifying cirri and determining their orientation using IR data from Planck and Herschel space observatories. \citet{Zavagno_etal_2023} based their research on the skeletons of filaments obtained in \citet{Schisano_etal2020} and trained several neural networks using the U-Net and UNet++ \citep{Zhou2018UNetAN} architectures.
\citet{Richards_etal_gabor_attention_2024} proposed a gridded multi-scale architecture that uses three different attention modules, including a novel Gabor attention model. The architecture proposed in \citet{Richards_etal_gabor_attention_2024} combines computational efficiency, high segmentation quality, and the ability to work with large images.
In our recent work \citep{Cirrus_2023MNRAS.519.4735S} we identified cirrus filaments using U-Net based neural network architectures in deep images from SDSS Stripe~82.
Building upon this earlier work, the present study investigates cirrus filaments in optical images from the Hyper Suprime-Cam Subaru Strategic Program public data release 3 \citep[HSC-SSP DR3,][]{HSC-SSP_DR3_2022PASJ...74..247A} using the neural network method developed in \citet{Cirrus_2023MNRAS.519.4735S}.
We aim to evaluate the performance of our neural network models on new data, compare the results with previous studies, and compile a catalog of cirrus filaments. As HSC-SSP images are deeper than SDSS Stripe 82 images and have a higher resolution, we hope to detect more cirrus clouds than in \citet{Cirrus_2023MNRAS.519.4735S}.

The structure of this paper is as follows. In Section~\ref{sec:data}, we describe the data used and the preliminary steps required for identifying cirrus filaments and conducting their analysis. In Section~\ref{sec:auto_segment}, we give a description of the training dataset for the neural networks, their architecture, and the experimental results of their training. In Section~\ref{sec:results}, we present our main results.
In Section~\ref{sec:discussion}, we describe the properties of the identified cirrus filaments and compare them with those of cirrus clouds reported in \citet{Cirrus_2023MNRAS.519.4735S}, and then discuss the relationship between cirrus and background over-subtraction in the HSC-SSP data. We summarize our results in Section~\ref{sec:summary}.

\section{Data}
\label{sec:data}
In this work, we utilize deep images from the third data release of the HSC-SSP survey \citep{HSC-SSP_DR3_2022PASJ...74..247A}. The largest component of this survey (the Wide layer) consists of three separate regions of the sky: two equatorial areas near the equinox points (``Fall equatorial field'' and ``Spring equatorial field'') and an area near the HectoMAP region \citep{North_field_2011AJ....142..133G} (``North field''). Hereafter, for convenience, we will refer to these regions as the Fall region, the Spring region, and the North region, respectively.
A major advantage of the Fall region is its overlap with the SDSS Stripe~82 survey, which has been used to study cirrus in previous research \citep{Stripe82_2020A&A...644A..42R, Cirrus_2023MNRAS.519.4735S}. This enables us to utilize previously obtained data for the overlapping region as a reference for investigating cirrus in the rest of the HSC-SSP survey, while also allowing for a comparison of cirrus parameters derived from the two independent datasets.

\begin{figure*}[ht]
\begin{center}
\includegraphics[width=0.66\linewidth]{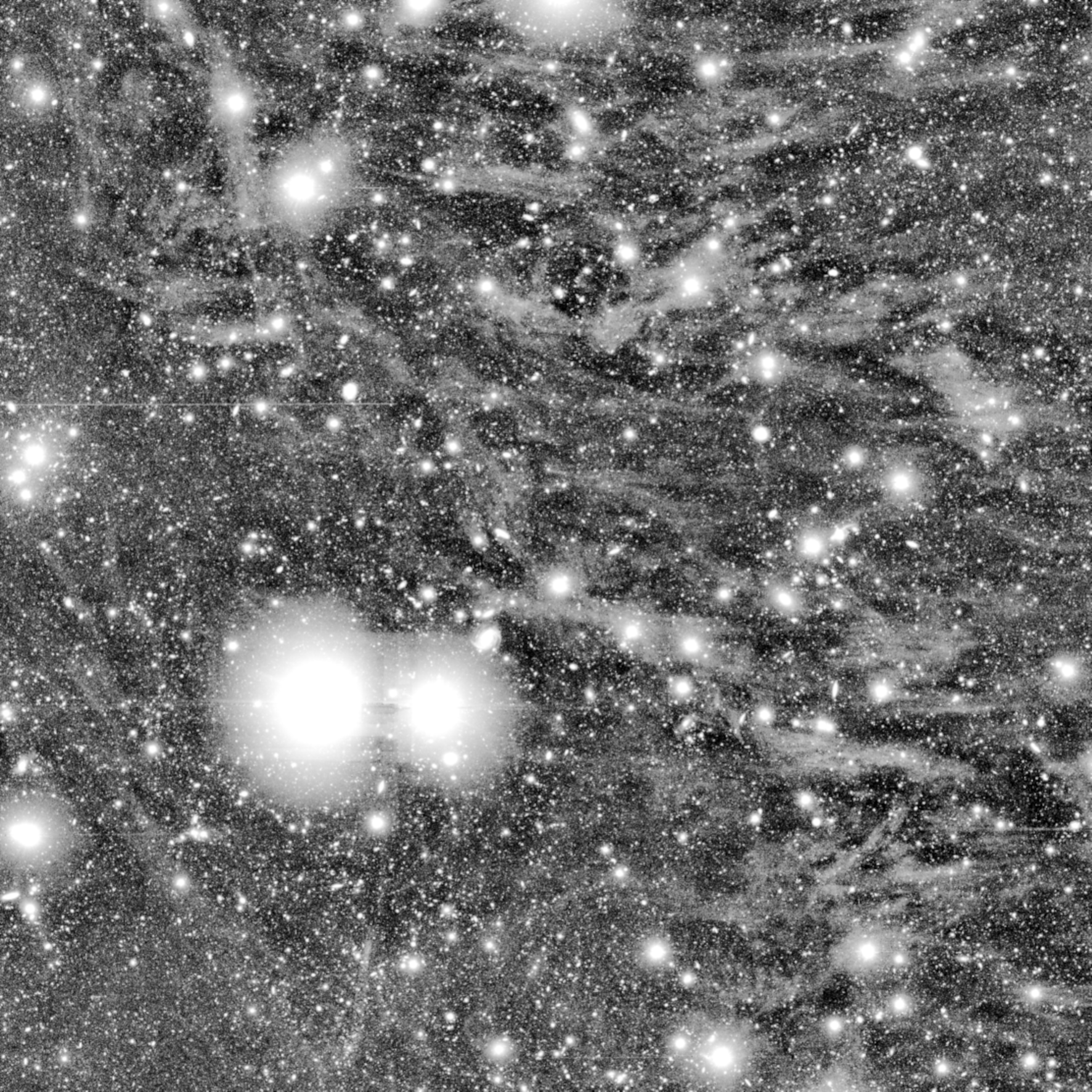}
    \caption{``global-sky'' combined image of the Field f1053 in the $r$ band. Contamination by sky over-subtraction is clearly seen in areas around cirrus filaments (darker areas).}
    \label{fig:f1053_example}
\end{center}
\end{figure*}

To achieve our goal of automatically detecting cirrus clouds, we utilize coadd optical images (the $g, r, i$ bands) and masks in the $r$ band.
Since cirrus filaments are faint, diffuse objects, their detection and analysis are significantly impacted by aggressive sky subtraction \citep{2019A&A...621A.133B}. To address this issue, we use ``global-sky'' coadd images provided by HSC-SSP.
The global sky subtraction technique in the HSC-SSP survey is designed to preserve the faintest extended features while minimizing over-subtraction around sources in the ``global-sky'' combined images. This technique was introduced in \citet{HSC-SSP_DR2_2019PASJ...71..114A} and improved upon in \citet{HSC-SSP_DR3_2022PASJ...74..247A}.
Unfortunately, this approach did not avoid sky over-subtraction around large cirrus filaments. Such contamination can clearly be seen in Fig.~\ref{fig:f1053_example}. The issue of sky over-subtraction is discussed in Section~\ref{sub_sec:over-subtraction}.

\begin{table*}[ht]
	\caption{List of regions in the HSC-SSP survey. The first column contains the names of the regions. The range of coordinates that defines the approximate boundaries of each HSC-SSP region is contained in the second column, the area of each region is given in the round brackets. The third column gives the number of fields in each region, for which there are images in $g$, $r$, $i$ bands.}
	\centering
    \begin{tabular}{l c c}
        \hline \hline
        Region name             & (RA, Dec)                                 & Number of fields \\
        \hline
        Intersection region & $-30^{\circ} \leq$ RA $\leq 40^{\circ}$, $-1.25^{\circ} \leq$ Dec. $\leq 1.25^{\circ}$ $\left(\simeq 160~\mbox{deg}^{2}\right)$   & 613 \\
        Fall plus region         &  $-30^{\circ} \leq$ RA $\leq 40^{\circ}$, $1.25^{\circ} \leq$ Dec. $\leq 7^{\circ}$ $\left(\simeq 400~\mbox{deg}^{2}\right)$ & 613 \\
        Fall minus region        & $27.5^{\circ} \leq$ RA $\leq 40^{\circ}$, $-7^{\circ} \leq$ Dec. $\leq -1.25^{\circ}$ $\left(\simeq 70~\mbox{deg}^{2}\right)$ & 253 \\
        \hline
        Spring region & $127.5^{\circ} \leq$ RA $\leq 232.5^{\circ}$, $-2^{\circ} \leq$ Dec. $\leq 5^{\circ}$ $\left(\simeq 730~\mbox{deg}^{2}\right)$ & 2442 \\
        North region             & $200^{\circ} \leq$ RA $\leq 250^{\circ}$, $42^{\circ} \leq$ Dec. $\leq 44.5^{\circ}$ $\left(\simeq 90~\mbox{deg}^{2}\right)$ & 410 \\

    \end{tabular}
    \label{tab:regions}
\end{table*}

\subsection{HSC-SSP DR3 images properties}
\label{sub_sec:hsc_ssp_images}
The ``global-sky'' coadd images have a common photometric zero-point of $27.0$ in broad bands \citep[$g, r, i, z, y$,][]{HSC-SSP_DR1_2018PASJ...70S...8A} and a pixel scale of $0.168$~arcsec. The average $3 \sigma$ surface brightness limits of the dataset $\mu_{\mathrm{lim}} (3 \sigma; 10'' \times 10'') = 30,~ 29.6,~ 29.4$ mag\,arcsec$^{-2}$ for the $g$, $r$, and $i$ bands, respectively.
We describe how these limits are calculated in \ref{app_sec:sb_limit}.
According to \citet{Stripe82_2020A&A...644A..42R}, the corresponding brightness limits of SDSS Stripe~82 are: $29.1$, $28.6$, and $28.2$ mag\,arcsec$^{-2}$. As a result, the HSC-SSP data are approximately 1 magnitude deeper than the SDSS Stripe~82 data.


To maintain compatibility with our previous work on the Stripe~82 survey, we chose to use the same tessellation scheme introduced in Section 3.1 of \citet{IAC_Stripe_82_2016MNRAS.456.1359F}, which is identical to the SDSS Stripe~82 scheme in the overlapping region.
In this scheme, the celestial sphere is divided into separate fields with a step of $0.5$ degrees in both declination and right ascension. Near the celestial equator, this results in nearly square fields with an area of 0.25 square degrees, while closer to the celestial poles, the fields become more elongated with a smaller area.



\subsection{Preparing field images}
\label{sub_sec:image_preparation}
The HSC-SSP survey provides many convenient Data Access Tools for HSC Data\footnote{\url{https://hsc-release.mtk.nao.ac.jp/doc/index.php/data-access\_\_pdr3/}}.
Before being stacked into coadd images, the individual images were processed through the HSC pipeline. The description of this data processing pipeline can be found in Section 3 of \citet{HSC-SSP_DR1_2018PASJ...70S...8A}. To download these ``global-sky'' coadd images and masks in FITS format, we used a command-line tool\footnote{\url{https://hsc-gitlab.mtk.nao.ac.jp/ssp-software/data-access-tools/-/tree/master/pdr3/downloadCutout/}} as a Python module.
This tool has an internal limit on the size of the image that can be requested, preventing the download of an entire field. To overcome this limitation, we requested overlapping image cutouts of $10.1^{'} \times 10.1^{'}$, with a stride of $10^{'}$ between the centers of adjacent cutouts. If all nine image cutouts covering a field were available, we merged them into a single image of the field using the \texttt{Swarp} utility \citep{SWarp_2002ASPC..281..228B}. Since our approach to the automatic segmentation of cirrus clouds relies on images in the $g$, $r$, and $i$ bands, we only considered fields with corresponding images available. The use of only three optical bands out of five available in the HSC-SSP DR3 ($g$, $r$, $i$, $z$, $y$ bands) is due not only to the fact that these three bands were used in previous work \citep{Cirrus_2023MNRAS.519.4735S}, but also due to two other reasons. First, the usage of three bands allows us to cover about $1050$~deg$^{2}$. According Table~2 in \citet{HSC-SSP_DR3_2022PASJ...74..247A}, the full-color area ($g$, $r$, $i$, $z$, $y$ bands) of the survey is only $670$~deg$^{2}$. Second, images in the $g$, $r$, $i$ bands are considerably deeper than in $z$ and $y$ bands. According to Table 1 in \citet{HSC-SSP_DR3_2022PASJ...74..247A}, $5$-sigma depths in $g$, $r$, $i$, $z$, $y$ bands are $26.5$, $26.5$, $26.2$, $25.2$ and $24.4$ mag\,arcsec$^{-2}$, respectively. Therefore, the faint cirri in the images in the $z$, $y$ bands will be less visible than those in the $g$, $r$, $i$ bands.

Additionally, we used source masks in the $r$ band. To prepare source masks, we used HSC-SSP masks in the $r$ band.
The HSC-SSP mask encodes the presence of certain objects or artifacts identified by the HSC-SSP pipeline using bits. A complete definition of all HSC-SSP mask bits can be found in the FITS file header.
We specifically chose the fifth, eighth, and ninth bits corresponding to weak sources, ``no-data'' areas, and bright sources, respectively, to construct source masks. Thus, the source mask is a subset of the HSC-SSP mask.

\begin{figure}
\begin{center}
\includegraphics[width=0.95\linewidth]{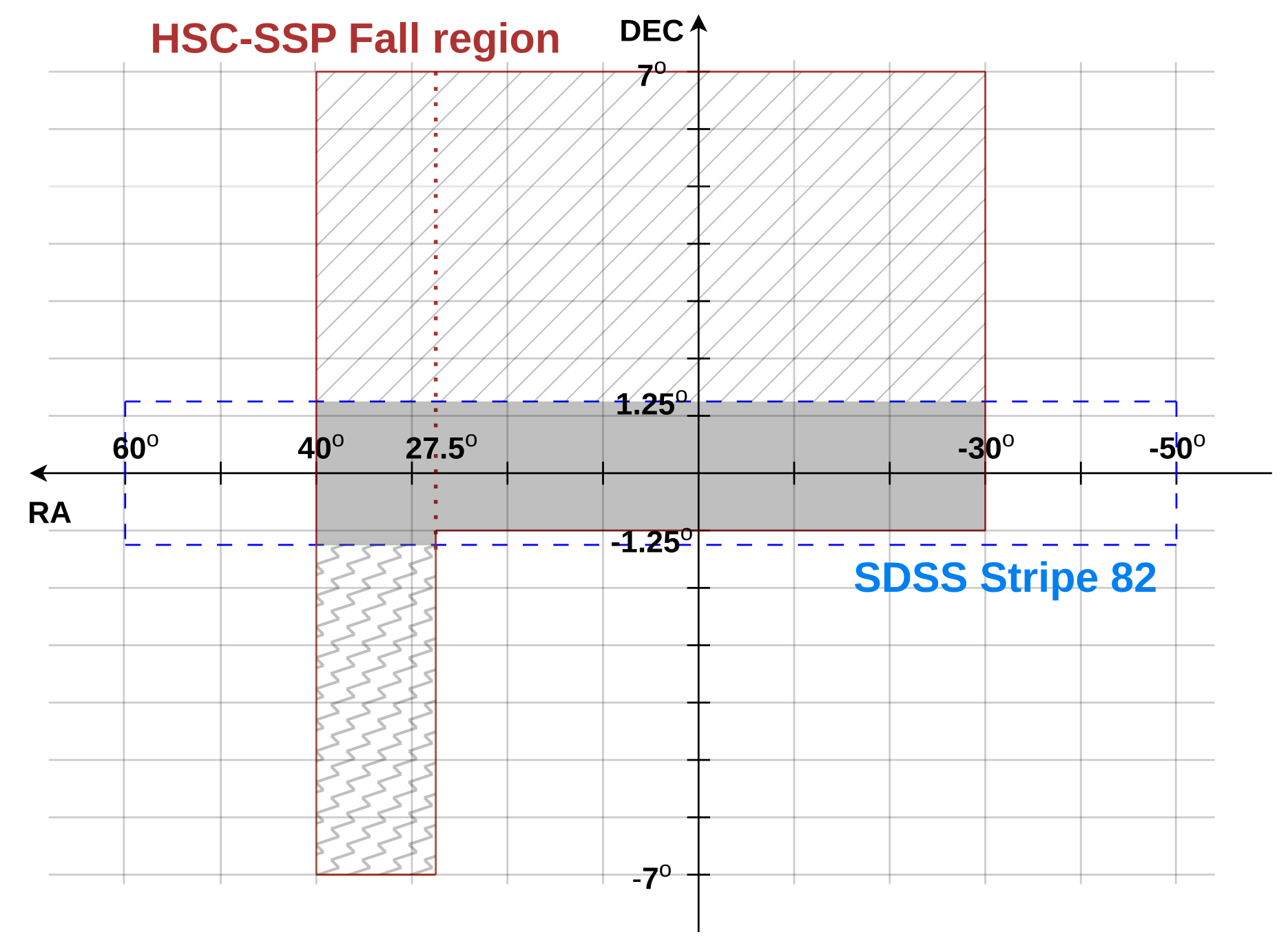}
    \caption{Schematic map of the HSC-SSP Fall region.
    The step of the coordinate grid along the DEC is $1^{\circ}$, and on the RA it is $10^{\circ}$.
    The SDSS Stripe 82 region is bounded by the dashed blue line.
    The HSC-SSP Fall region is bounded by the red solid line and divided into 3 regions. A hatched region is called the Fall plus region, a solid gray region is called the Intersection region, and a region filled with zigzag lines is called the Fall minus region.
    }
    \label{fig:fall_field_region}
\end{center}
\end{figure}

A total of 4,331 fields in the HSC-SSP dataset are suitable for our analysis. For convenience, we divided the HSC-SSP data into five regions.
Two of these correspond to the original HSC-SSP regions, namely the Spring and North regions (the frames of these regions are shown in Fig.~1 of \citet{HSC-SSP_DR3_2022PASJ...74..247A}), while the remaining three are subdivisions of the HSC-SSP Fall region.
The division scheme for the Fall region is illustrated in Fig.\ref{fig:fall_field_region}. The location and number of fields in each region are provided in Table\ref{tab:regions}.

\subsection{Intersection dataset preparation}
\label{sub_sec:hsc-ssp_s82_preparation}
\begin{figure}
\begin{center}
\includegraphics[width=0.95\linewidth]{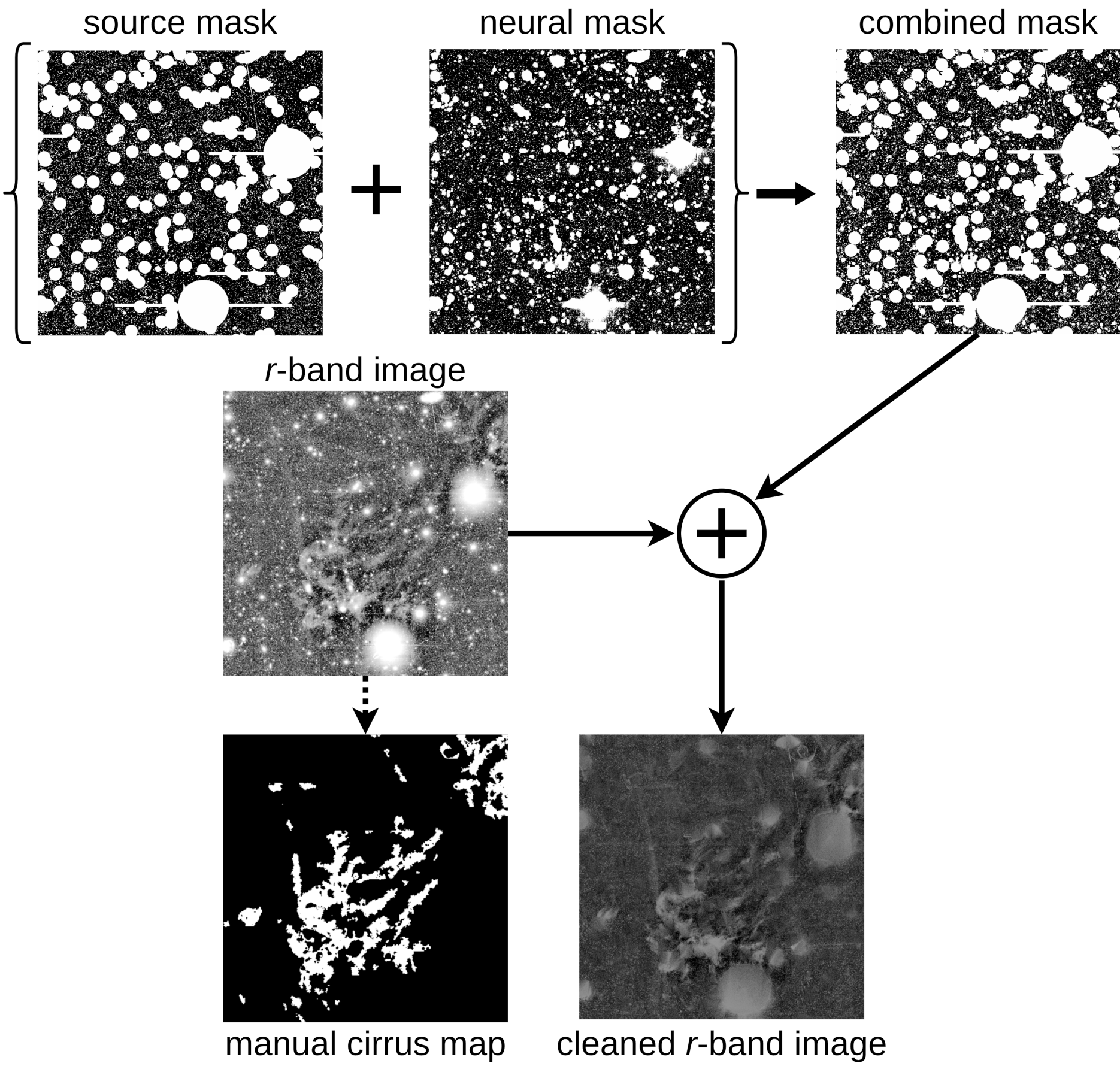}
    \caption{Diagram of the Intersection dataset preparation process. The top row of the diagram demonstrates the creation of a combined mask by the union of a neural mask and a source mask. The direct sum sign denotes the creation of a cleaned image in the $r$ band from an image in the same band with a combined mask. The dotted arrow demonstrates the creation of a cirrus map using manually selected and corrected contours in \texttt{DS9}.}
    \label{fig:diagram_Intersection_dataset}
\end{center}
\end{figure}

The first dataset we prepared consists of data from the intersection of the SDSS Stripe~82 region with HSC-SSP DR3 (hereafter referred to as the Intersection dataset). This dataset is used to train a neural network model for automatic cirrus segmentation and to compare the results with those from \citet{Cirrus_2023MNRAS.519.4735S}. Stripe~82 is a narrow region of the sky, spanning 110 degrees in width ($-50^{\circ} < \alpha < 60^{\circ}$) and 2.5 degrees in height ($-1.25^{\circ} < \delta < 1.25^{\circ}$). The Intersection dataset contains 613 fields covering $153.25$~deg$^{2}$. Despite these fields being located close to the equator, their sizes differ slightly (the dimensions of the images along the $x$ and $y$ axes are in the range of 10715 -- 10717 pixels). To enable a direct comparison of segmentation results with previous SDSS Stripe~82 data segmentation \citep{Cirrus_2023MNRAS.519.4735S}, the image sizes were reduced to $10715 \times 10715$ pixels by cropping pixels at the image border.
The data preparation process outlined below in this section is shown in Fig.~\ref{fig:diagram_Intersection_dataset}.

In previous work \citep{Cirrus_2023MNRAS.519.4735S}, we used cleaned images in the $r$ band to prepare a dataset for neural network training. Bright sources were masked in these images (cirrus filaments were left unmasked), and the background was interpolated inside the masked areas. Since one of the objectives of this study is to assess the impact of using cleaned images directly in a dataset for neural network training, we have also prepared cleaned images in the $r$ band.

The first step in preparing the cleaned images was to mask out all objects in the images except cirrus.
The prepared source masks cover all objects as well as some image artifacts.
Although these masks cover the vast majority of objects, some, particularly extended ones, have faint outer wings that extend beyond the masked regions.
To quantify the impact of these unmasked wings we have measured the surface brightness in a pixel-wide layers along the mask edges and found out that in r-band the median value for fifty random fields is $28.32$ mag\,arcsec$^{-2}$, and top (brightest) quartile is $28.16$ mag\,arcsec$^{-2}$, which is not negligible taken the survey depths.
To solve this problem, we utilized mask images previously prepared in \citet{Cirrus_2023MNRAS.519.4735S} to train a conditional Generative Adversarial Network (GAN) \citep{2014arXiv1406.2661G} for generating new masks. The GAN architecture consists of two networks: the generator and the discriminator. These networks are trained to compete with each other. In our case, the generator takes a field image as input and creates a mask image, while the discriminator distinguishes between real and generated masks. For our generator, we use the U-Net architecture  \citep{U-Net_2015arXiv150504597R}, while for the discriminator, we use a convolutional neural network with four convolutional blocks (two-dimensional convolution, batch normalization, and leaky ReLU activation) with 64, 128, 256 and 512 kernels. For the training purposes, we used the overlapping regions of SDSS Stripe~82 and HSC-SSP DR3 (Intersection region) to create 4500 training and 2500 testing samples. After completing the training, we generated new masks (``neural'' masks) for all HSC-SSP DR3 images.
As these masks are based on those from Smirnov et al., 2023, they cover all sources in the HSC-SSP images, regardless of their type. Cirrus filaments that overlap with sources are masked out, but those that are not overlapped with other sources are not.
As the final masks {of the background objects}, we used a union of the source masks prepared with HSC-SSP masks and our neural masks.
Hereafter, we refer to these masks as the ``combined'' masks. When we measured the surface brightness along the {pixel-wide} edges of our combined masks we found values which are lower by about 0.5 of a magnitude and with considerably lower scatter between fields: the median value is $28.86$ mag\,arcsec$^{-2}$ and the top (brightest) quartile is $28.75$ mag\,arcsec$^{-2}$, {which means that these new masks cover the background sources considerably better than the original ones.}

To obtain cleaned images, i.e. images where all the background objects except cirri are cleaned out, we use the following procedure. For each masked region of our combined mask we zero-out all the pixels from the corresponding area of an $r$-band image. Then we fill this gap by interpolating the fluxes of a boundary layer of this masked region back to its center. To suppress the random noise during this interpolation we use four pixel wide boundary layer and apply a median filter to it.

Training a neural network segmentation model requires field images along with corresponding segmentation maps (pixel-wise labels). The creation of accurate cirrus segmentation maps, hereafter referred to as \textit{cirrus maps}, is a complex and time-consuming task. The process of preparing cirrus maps is described below.

In our previous work \citep[see Section 2.2 in][]{Cirrus_2023MNRAS.519.4735S}, we developed a semi-automatic method for preparing cirrus maps. Using this method, we have prepared a training dataset, the same size as in the previous work \citep[see Section 3.1 in][]{Cirrus_2023MNRAS.519.4735S}.
Unfortunately, the models trained on this dataset reached a low intersection over union metric $\leqslant 0.2$ \citep{IoU1901}. Intersection over union metric (IoU) quantifies the degree of overlap between the predicted and actual cirrus map.
For comparison, the best model yielded an IoU of $0.576$ in \citet{Cirrus_2023MNRAS.519.4735S}. We assume that our IoU is lower due to both the properties of the data and the quality of the cirrus maps. Indeed, careful visual inspection showed that the cirrus maps do not accurately reproduce the filament boundaries in the HSC-SSP DR3 images.
Training a high-quality machine learning segmentation model using such a map is practically impossible. To create more accurate cirrus maps, we decided to use the \texttt{DS9} contour tool \citep{DS9_2003ASPC..295..489J}. We generated contours at $29$~mag arcsec$^{-2}$ in the $r$ band, using a smoothness parameter of $4$, in images rescaled to $2143 \times 2143$ pixels. The choice of this isophote level is based on its use in our previous study \citep{Cirrus_2023MNRAS.519.4735S}. We manually selected those contours that correspond to the cirrus filaments. Then, in the selected filaments, we manually removed the brightest sources, such as stars and galaxies, as well as artifacts.
Although this approach is more time-consuming than the semi-automatic method used in our previous work, it creates more accurate and detailed cirrus maps. Models trained on such maps yield larger IoU metric (see Section~\ref{sub_sec:nn_results}).

Using the method described above, we prepared cirrus maps for $70$ randomly selected fields out of a total of $613$.
Although this is five times fewer than in our previous work, it proved sufficient for training and testing neural network models, which produced acceptable cirrus map quality (see Section~\ref{sec:auto_segment}).
The cirrus maps for each field are stored in corresponding FITS files, where a pixel value of $0$ represents the background, and $1$ indicates cirrus. The fraction of pixels that denote cirrus in a field or region is referred to as the \textit{cirrus fraction}.

As a final remark, it is worth noting that the original HSC-SSP data is affected by sky over-subtraction. This can be clearly seen in Fig.~\ref{fig:f1053_example}. Therefore, any pipeline that uses this data will inherit this flaw, including our manual cirrus maps. An important point is that our ultimate goal is to identify cirrus in contaminated data. Therefore, our neural network models should be trained on such data without the need for any additional sky over-subtraction correction. We further discuss sky over-subtraction in Section~\ref{sub_sec:over-subtraction}.

%

\section{Automatic Cirrus Segmentation}
\label{sec:auto_segment}
Since the neural network architectures in our previous work \citep[see Section 3.2 in][]{Cirrus_2023MNRAS.519.4735S} effectively solved the problem of cirrus segmentation (cirrus map generation), we apply the same architectures to the HSC-SSP DR3 data. Improvements to the cirrus segmentation approach involve enhancements in data preprocessing and the use of model ensembles. In this section, we describe the dataset for neural network training, the network architecture, training methods, and the analysis of the trained models.\\

\subsection{Training data for neural networks}
\label{sub_sec:data_for_nn}

In Section~\ref{sec:data}, we described the manual identification process for cirrus filaments. Here, we further process the segmentation data to train a neural network. Due to size limitations of neural network models, we use smaller square image fragments (windows) instead of the original large field images ($10715 \times 10715$ pixels) during the training. This approach reduces computational time, memory usage, and the amount of manually annotated data required for neural network training.
The 3- or 4-channel input images for the neural network are formed from the windows in the $g$, $r$, $i$ bands and the windows from the cleaned images in the $r$ band (fourth channel).
Training a neural network requires dividing the data into three subsets: training, validation, and test sets. To form these subsets, we randomly selected three separate groups of fields, consisting of $37$, $13$ and $20$ fields, respectively. Below, we outline the key steps for preprocessing the training and validation data.
\begin{enumerate}
    \item We calculate the common $99.9$th percentile and $0.1$th percentile values for each band used ($g$, $r$, $i$ and, if necessary, cleaned images in the $r$ band) separately for all training and validation fields ($50$ fields). We then apply corresponding clipping \citep{berry2005handbook, Howell_2006, Gonzalez_2006_handbook} to moderate the effects of extremely bright pixels and over-subtraction near extended sources, which reduces image contrast. This clipping moderately increases the IoU of trained models (improvement is from $0.05$ to $0.1$ for different model parameters).
    Notably, in contrast to our previous work \citet{Cirrus_2023MNRAS.519.4735S}, we use $0.1$th percentile clipping instead of negative value clipping. This adjustment is necessary because over-subtraction occurs in areas near cirrus in HSC-SSP images, an artifact we will demonstrate in Section~\ref{sub_sec:over-subtraction}.
    \item Next, we apply a natural logarithm transformation.
    \item Then, we randomly choose $n_{\mathrm{tr}}$ square windows ($w \times w$ pixels) for each field in the training subset and $n_{\mathrm{val}}$ for each field in the validation subset. After that, we resize each window to match the spatial shape of the input tensor ($w_{\mathrm{in}}, w_{\mathrm{in}}$), using \texttt{cv2.resize} method with \texttt{cv2.INTER\_LINEAR} interpolation from OpenCV library \citep{opencv_library}.
    \item The corresponding windows with the cirrus map are obtained from the files with the cirrus map and then resized, using \texttt{cv2.INTER\_AREA} interpolation.
    \item Finally, during the creation of the input tensor, we apply min-max normalization to the $[-1:1]$ range and augment the data by a symmetry group of the square. This group consists of $\pi/2$ rotations, reflections, and their compositions (eight elements). As a result, this procedure increases the number of windows by a factor of 8.
\end{enumerate}

\subsection{Network architecture and training methods}
\label{sub_sec:nn_arch_and_methods}

\begin{figure}
\begin{center}
\includegraphics[width=0.95\linewidth]{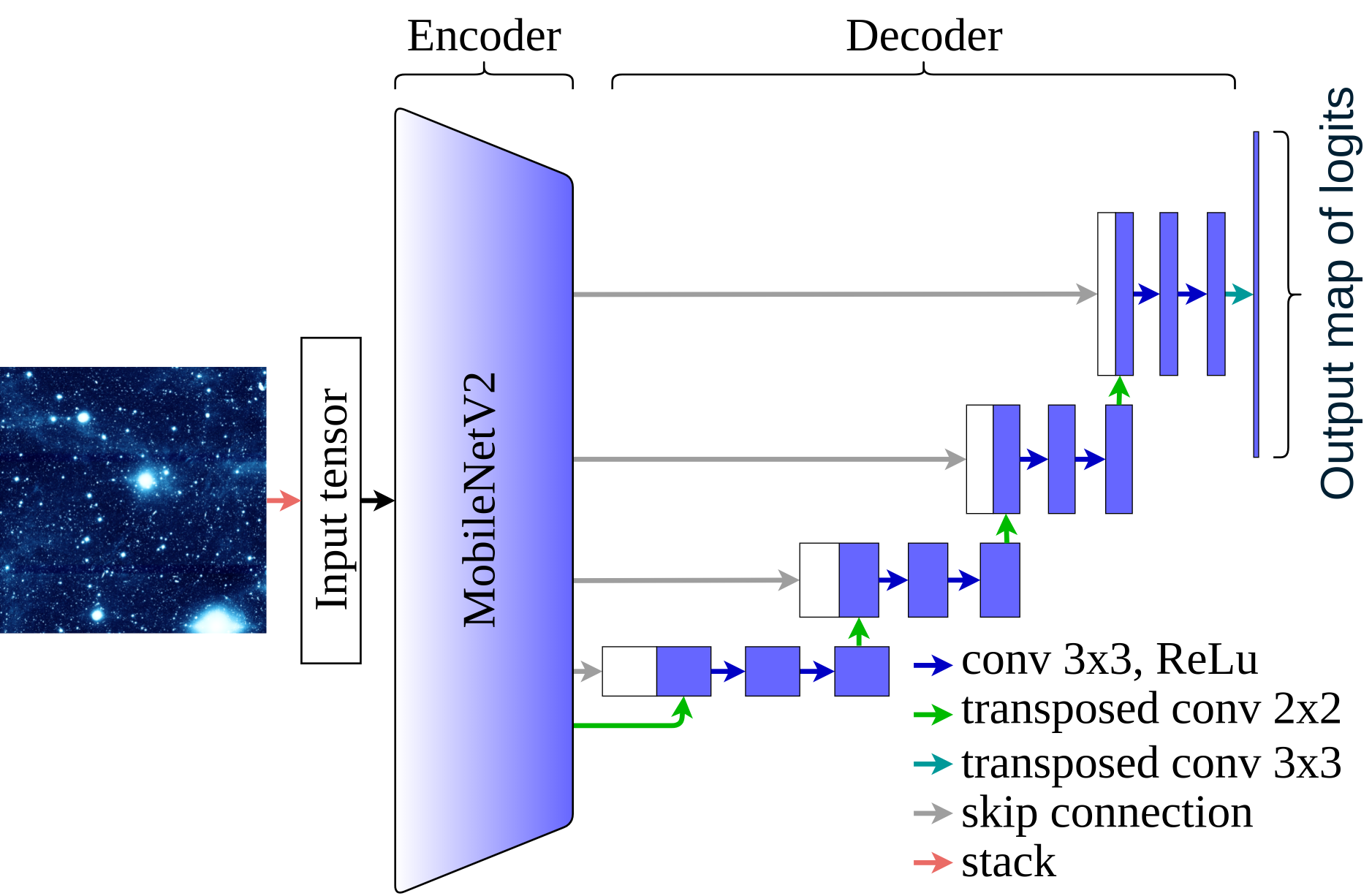}
    \caption{The encoder-decoder architecture used for creating cirrus maps.}
    \label{fig:nn_architecture}
\end{center}
\end{figure}

To generate cirrus maps, we employed the same network architecture used in our previous study on cirrus in SDSS Stripe~82 \citep{Cirrus_2023MNRAS.519.4735S}. This architecture is based on the encoder-decoder U-Net network \citep{U-Net_2015arXiv150504597R}.
Fig.~\ref{fig:nn_architecture} shows a visual representation of the neural network architecture used.
As an encoder, we used MobileNetV2 \citep{MobileNetV2_2018}, as the models with it showed the highest IoU.

For each model under consideration, we used a sparse categorical cross-entropy loss function derived from the logits output tensor. For loss function optimization, we utilized the Adam optimizer \citep{Kingma2014AdamAM} with different learning rates $r$.

During the training experiments, we varied some parameters that influence the optimization process and the final model performance: the number of input channels $n_{\mathrm{ch}}$ (three or four), the scale factor $s$ between the window size $w$ and the input tensor spatial size $w_{\mathrm{in}}$ ($w = s w_{\mathrm{in}}$), class weights $\overline{\omega_{\mathrm{c}}}$, etc.

To train our network models, we used a single NVIDIA GeForce RTX 3060 GPU. The batches consisted of six windows for all models. To estimate a sufficient number of learning epochs, we trained $12$ 4-channel models for $100$ epochs. The validation loss curves for these models are presented in \ref{app_sec:loss_curves}. The average value of the epoch in which the minimum validation loss was achieved was $12.2$, with a standard deviation of $4.8$ and a maximum of $19$. Thus, to save time on training models, we decided to use 20 epochs for the remaining models.
We used $w_{\mathrm{in}} = 448$ for all the models considered. Since the size of the dataset is small, we used a different number of random windows from each field in training and validation datasets ($n_{\mathrm{tr}}$, $n_{\mathrm{val}}$) for different values of $s$.
For scale factors $s = 2$ and $4$, we used $n_{\mathrm{tr}}=200$ and $n_{\mathrm{val}}=250$. For $s = 8$, we used $n_{\mathrm{tr}}=80$ and $n_{\mathrm{val}}=100$.
The training time for a single model ranged from 6 to 15 hours.

The code used to train the segmentation models and produce the cirrus maps is available at \url{https://gitlab.com/polyakovdmi93/cirrus_narrow_segmentation}\footnote{The best way to use and modify this project is to create a fork after gaining access upon request.}.

\subsection{Experimental results analysis}
\label{sub_sec:nn_results}

In this section, we describe experiments aimed at finding a more effective solution for generating cirrus maps. To assess the performance of different models, we employ the IoU metric,
First, we compare the effectiveness of 3-channel and 4-channel models.

As can be seen from Tables~\ref{tab:3D_nn_metrics} and \ref{tab:4D_nn_metrics}, the metrics of models with identical parameters may differ by more than 0.1.
This is probably due to the small size of the dataset. To compare 3-channel and 4-channel models, we compared their efficiency with intermediate class weights $(1.0,~15.0)$. This allowed us to reduce the training time for 3-channel models. The average IoU value for 12 3-channel models, presented in Table~\ref{tab:3D_nn_metrics}, is $0.3606$ and the average IoU value for  corresponding 4-channel models from Table~\ref{tab:4D_nn_metrics} is $0.363$. Such a negligible difference in relation to the variation of the IoU of models with the same parameters indicates a similar quality of the 3-channel and 4-channel models.

Given the limited size of the training dataset, we explored the fine-tuning \citep{Fine-tuning2017} of models trained in \citet{Cirrus_2023MNRAS.519.4735S} for cirrus map generation in SDSS Stripe~82 data. These models use 3-channel images. Each channel corresponds to an image in the $g$, $r$, $i$ bands.
Since the HSC-SSP survey includes the same optical bands as SDSS Stripe~82, we hypothesized that fine-tuning these models might allow us to create cirrus maps for the HSC-SSP survey data.

As a basic model for fine-tuning, we employed the most efficient model from \citet{Cirrus_2023MNRAS.519.4735S}, which demonstrated an IoU of 0.576. This model creates a cirrus map using 3-channel images, sliding window size $w = 448$, scale factor $s = 1$ between the sliding window size $w$ and the input tensor spatial size $w_{\mathrm{in}}$ $(w = s w_{\mathrm{in}})$. The pixel scale in SDSS Stripe~82 is $0.396$~arcsec, which differs from the HSC-SSP pixel scale of $0.168$~arcsec. To address this difference in pixel scale, we conducted several fine-tuning experiments using scale factors of $s=2$ and $s=3$ (see Table~\ref{tab:fine-tuning_nn_metrics}). Unfortunately, as can be seen from Tables~\ref{tab:4D_nn_metrics} and \ref{tab:fine-tuning_nn_metrics}, fine-tuned models did not achieve the same high IoU metrics as models that are trained from scratch. This suggests a difference in the appearance of cirrus between SDSS Stripe~82 and HSC-SSP, which hinders effective fine-tuning. We infer that differences in image depth and data processing between the surveys (Sect.~\ref{sub_sec:hsc_ssp_images}) are the primary factors contributing to the dissimilarity in cirrus appearance. These differences are demonstrated in Section~\ref{sub_sec:Stripe82_vs_HSC_SSP}.

In our previous work on cirrus in SDSS Stripe~82, the best model achieved an IoU of $0.576$. This value is significantly higher than that of the best model in the current work (see Table~\ref{tab:4D_nn_metrics}).
To improve the results, we tried using ensembles of several top 4-channel models available. Each model in the ensemble produces its own independent cirrus map. These maps are combined into a final map using direct pixel-by-pixel majority voting.
The use of an ensemble of models is a highly effective method to improve the performance of machine learning algorithms \citep{Breiman1996BaggingP, Boosting2000, RandomForest2001, Ren2016EnsembleCA, EnsembleLearning2022}. This approach is used to solve problems related to classification, recognition, detection, and segmentation in various scientific fields. Ensembles are widely employed in medical research \citep{QuratulAin2010ClassificationAS, Fraz2012AnEC, Moradi2023DeepEL, Tang2023MobileNetV2ES}, astronomy \citep{nun2016ENSEMBLELM, Priyadarshini2021ACN, Marchuk_etal_2022_B/PS, Pagliaro2023ApplicationOM, Savchenko_etal_2024, Zeraatgari2024MachineLP}, and other fields \citep{Cyganek2012OneClassSV, Stork2012AudiobasedHA, 2013IJRS...34.5166G, Kinattukara2014ANE, Nanni2023ExploringTP}. We tested ensembles of three, five, seven and nine top 4-channel models (see Table~\ref{tab:4D_nn_ensemble_metrics}). All considered ensembles of models show higher IoU compared to individual models. The best ensemble, consisting of nine models, achieves an IoU of 0.48 and is used for cirrus map generation further in this work.

\begin{table}
  \centering
  \begin{tabular}[ht]{c l l l}
    \hline \hline
    $N_{\mathrm{models}}$ & IoU & precision & recall \\
    \hline
        3   &$0.479$ &$0.687$ &$0.612$  \\
        5   &$0.477$ &$0.7$   &$0.599$  \\
        7   &$0.475$ &$0.679$ &$0.612$  \\
        9   &$\mathbf{0.48}$  &$0.686$ &$0.616$  \\
  \end{tabular}
  \caption{The metrics of the ensembles used. It lists IoU, precision, and recall for all test fields for the cirrus class. The first column contains the number of the best models in each ensemble. The maximum value of the IoU is highlighted in bold.}
  \label{tab:4D_nn_ensemble_metrics}
\end{table}

We summarize the improvements to our cirrus map generation approach and the results of model training experiments as follows.

\begin{figure*}[ht]
\begin{center}
\includegraphics[width=0.82\linewidth]{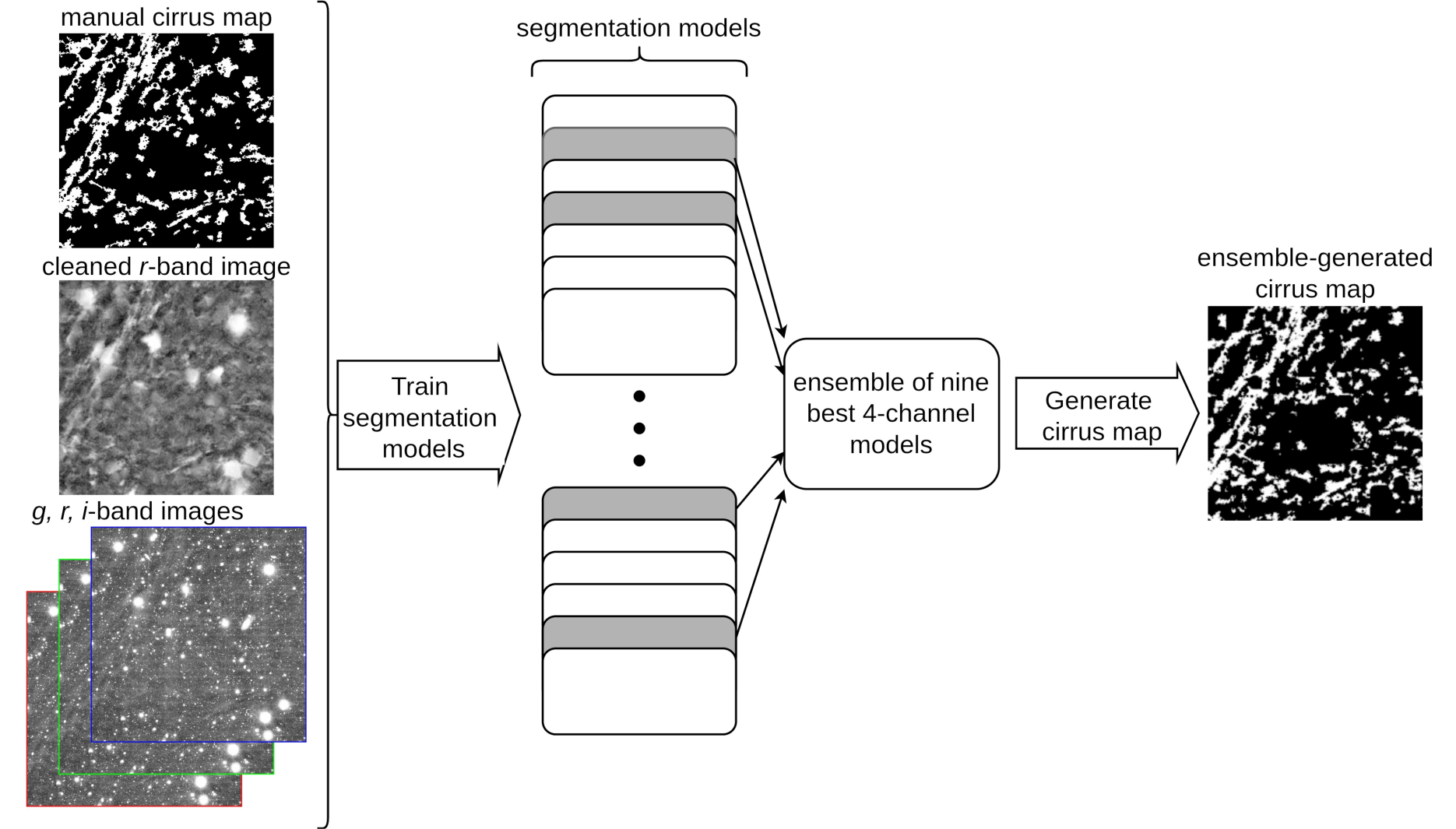}
    \caption{Diagram of the prediction of the cirrus map by the ensemble of nine best $4$-channel models. In the left part of the diagram, the data used for model training is shown. It consists of ground-truth cirrus map created using manually selected and corrected contours in \texttt{DS9} (manual cirrus map) and images from which the input tensor is formed (cleaned $r$-band image and images in $g$, $r$, $i$ bands). Rounded rectangles in the central part of the diagram denote trained models. Gray rounded rectangles denote the best nine models selected for the ensemble. The arrow in the right part of the diagram represents the generation of a cirrus map through direct pixel-by-pixel majority voting by ensemble models.}
    \label{fig:diagram_ensemble}
\end{center}
\end{figure*}

\begin{enumerate}
    \item To adapt our previous approach to the new HSC-SSP data, we incorporated 0.1th percentile clipping in image preprocessing, which allows us to increase contrast in fields rich with cirrus.
    \item We added the capability to train cirrus map generation models using 4-channel images. A comparison between 3-channel and 4-channel models revealed that, given the current problem and dataset size, both models exhibit similar IoU metrics. To better assess this similarity, a larger training sample is required.
    \item We conducted several experiments to find a more effective model. As one can see in Table~\ref{tab:4D_nn_metrics}, the highest IoU of a single model is $0.436$. It is also worth noting that the reliability of these estimates is low due to the small test sample, which consists of only 20 fields.
    \item We also conducted experiments to fine-tune the best model from \citet{Cirrus_2023MNRAS.519.4735S}. Unfortunately, as shown in Tables~\ref{tab:4D_nn_metrics} and \ref{tab:fine-tuning_nn_metrics}, the fine-tuned models demonstrated lower performance than models trained from scratch.
    \item To improve the quality of cirrus map generation, we utilized an ensemble of nine best 4-channel models, achieving an IoU of $0.48$ for cirrus.
\end{enumerate}
In the following sections, we use the cirrus maps obtained by using an ensemble of nine best $4$-channel models. A diagram of the prediction of the cirrus map by this ensemble is shown in Fig.~\ref{fig:diagram_ensemble}.

\section{Results}
\label{sec:results}

In this section, we describe the identified cirrus in each of the HSC-SSP fields under consideration. The key questions we address are: What is the total area covered by cirrus? and What are its physical properties, such as surface brightness and typical sizes?

\subsection{Cirrus maps post processing}
\label{sub_sec:maps_postprocessing}
\begin{figure*}
    \centering
    \includegraphics[width=0.4\linewidth]{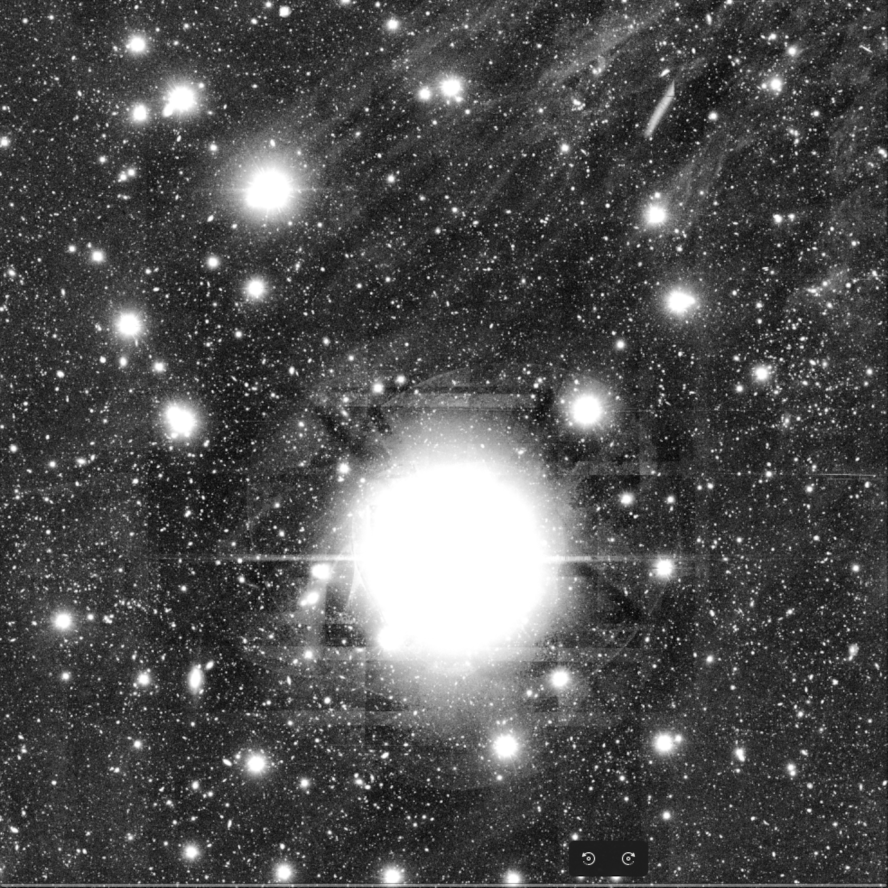} ~~
    \includegraphics[width=0.4\linewidth]{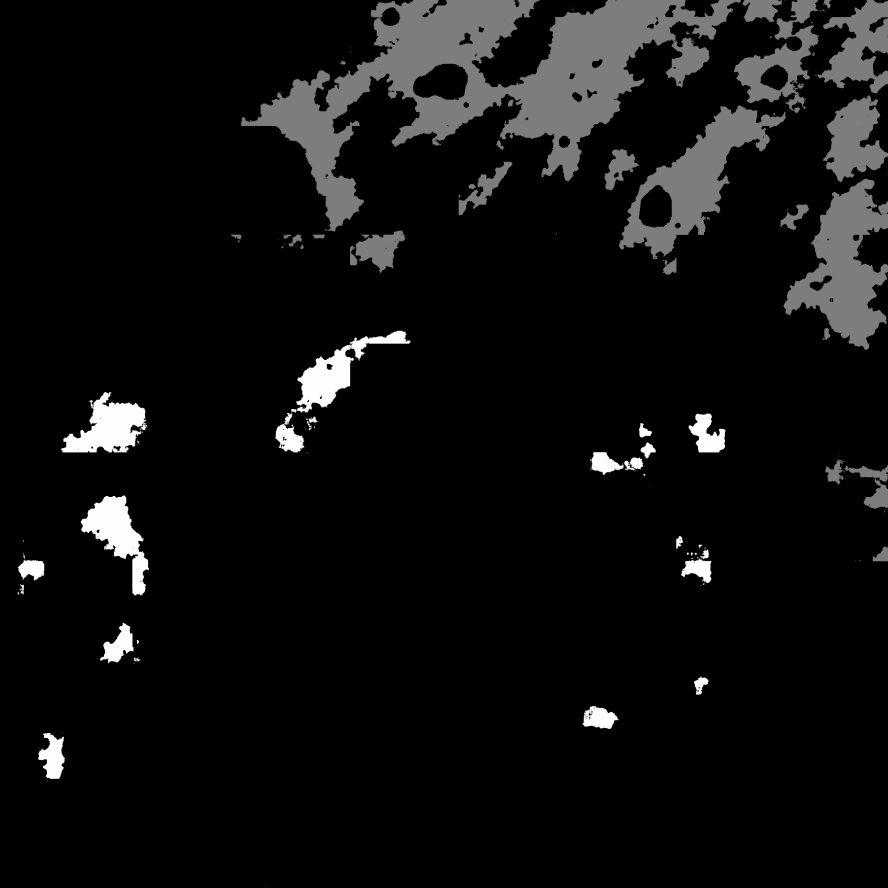}
\caption{The left panel shows a part of the Field f0865 in the $r$ band. The corresponding cirrus map with false cirrus detections is shown in the right panel. White pixels represent removed false cirrus detections. Gray pixels indicate remaining cirrus filaments.}
\label{fig:f0865}
\end{figure*}

Although the ensemble method produces higher-quality cirrus maps compared to individual models, visual inspection reveals some false cirrus detections. Many of these false detections in HSC-SSP images are due to artifacts. An example of such an artifact is the complex pattern of scattered light, which is clearly visible in Fig.~\ref{fig:f0865}.
To rectify the ensemble-generated cirrus maps, we manually removed individual artifacts. If a cirrus filament is partially overlapped by an artifact, the intersection is removed. We performed this procedure on fields where the cirrus fraction exceeded $0.5\%$ (1689 out of 4331 fields).
These fields account for $95.1\%$ of the cirrus area annotated by the ensemble. False cirrus detections were found in 513 out of the 1689 fields considered.

The cirrus fraction in the ensemble-generated maps is 1.985\%.
After removing false cirrus detections, it decreased to 1.884\%, indicating that false cirrus accounted for 0.101\% of the total area of the considered fields.
The cirrus fractions before and after rectification in each region of the HSC-SSP are presented in Table~\ref{tab:cirrus_frac}. Despite the small fraction of false cirrus, removing false detections allowed us to obtain more reliable cirrus properties. Therefore, in all subsequent sections, we use rectified cirrus maps.

\begin{table*}[ht]
  \centering
  \begin{tabular}{l c c c c}
    \hline \hline
    Region name & $f_{\mathrm{ens}}$ & $f_{\mathrm{rect}}$ & $f_{\mathrm{false}}$ & $f_{\mathrm{false}}$ / $f_{\mathrm{ens}}$  \\
    \hline
        Intersection region &$0.0434$ &$0.0408$ &$\mathbf{0.0026}$  &$0.06$   \\
        Fall plus region    &$\mathbf{0.0463}$ &$\mathbf{0.0454}$  &$0.0008$ &$\mathit{0.016}$  \\
        Fall minus region   &$0.0046$ &$0.0041$ & $\mathit{0.0004}$   &$0.09$   \\
    \hline
        Spring region       &$0.0109$ &$0.0101$  &$0.0008$ &$0.07$   \\
        North region        &$\mathit{0.0034}$ &$\mathit{0.0025}$ &$0.0008$ &$\mathbf{0.247}$  \\
  \end{tabular}
  \caption{Cirrus fractions and false cirrus fractions in HSC-SSP regions. The first column contains the names of the regions. Cirrus fractions in ensemble-generated maps for regions are given in the $f_\mathrm{ens}$ column. The $f_{\mathrm{rect}}$ column shows cirrus fractions after false cirrus removing, fractions of false cirrus are contained in the $f_{\mathrm{false}}$ column ($f_{\mathrm{false}} \equiv f_\mathrm{ens} - f_{\mathrm{rect}}$). The last column shows the proportion between false cirrus and cirrus annotated by the ensemble. The maximum and minimum values in the columns are highlighted in bold and italic, respectively.}
  \label{tab:cirrus_frac}
\end{table*}

The rectified cirrus maps that we generated may be useful to other researchers, so we have placed them in cloud storage. A detailed description of the data format and a link to the storage are available in our \texttt{GitLab} repository\footnote{\url{https://gitlab.com/polyakovdmi93/cirrus_narrow_segmentation}}.

\subsection{Cirrus maps in Intersection region}
\label{sub_sec:Stripe82_vs_HSC_SSP}
\begin{figure*}[h!]
\begin{center}
\includegraphics[width=0.65\linewidth]{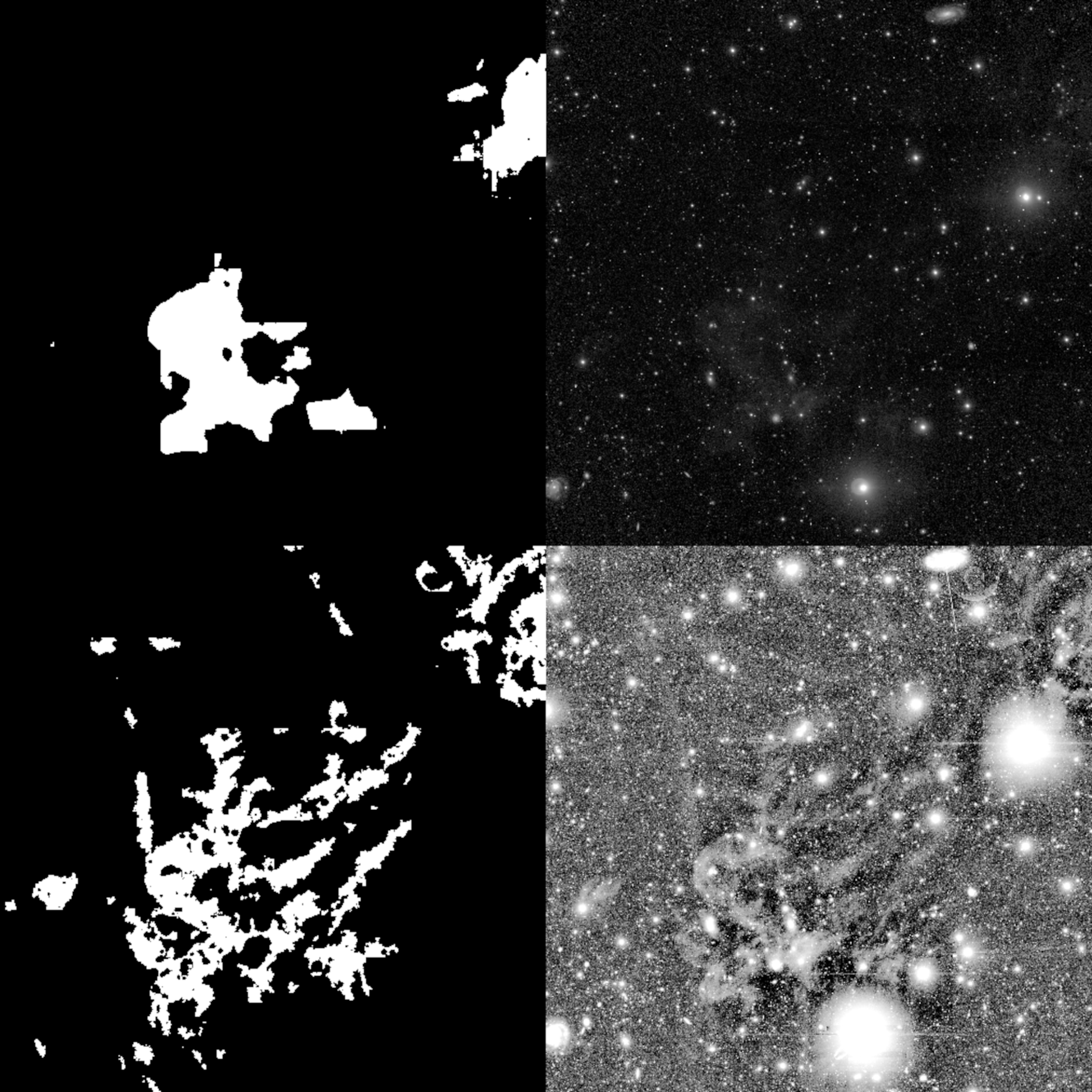}
    \caption{Cirrus maps and images in the $r$ band for the f1082 field. The top row contains the cirrus map on the left and the image from the SDSS Stripe~82 on the right. The bottom row contains the cirrus map on the left and the image from the HSC-SSP on the right.}
    \label{fig:f1082}
\end{center}
\end{figure*}

A cirrus map of the Intersection region was previously generated in our work ~\citep{Cirrus_2023MNRAS.519.4735S} using Stripe~82 data. Here, we examine the consistency between that map and a newly created HSC-SSP-based map from the present study. According to Table~\ref{tab:cirrus_frac}, the cirrus fraction in the HSC-SSP map for the Intersection region is $f_{\mathrm{HSC-SSP}}=0.0408$, whereas in the SDSS Stripe~82 map, it is $f_{\mathrm{S82}} = 0.0092$.
This advantage of HSC-SSP over SDSS Stripe~82 in the detected cirrus fraction can be attributed to the greater depth of HSC-SSP data compared to SDSS Stripe~82 data, allowing for the detection of fainter cirrus structures.
The difference in the appearance of cirrus and the fraction of cirrus detected in HSC-SSP data and SDSS Stripe~82 data can be seen in Fig.~\ref{fig:f1082}.
However, the consistency between these cirrus maps can be assessed using their IoU $=0.105$. While this value may seem low, it aligns with the theoretical maximum, which is determined by the ratio of the cirrus fractions in the two maps: $\mbox{IoU}_{\mathrm{max}} = f_{\mathrm{S82}} / f_{\mathrm{HSC-SSP}} = 0.225$. $\mbox{IoU}_{\mathrm{max}}$ is achieved when all cirrus detected in the Stripe~82 map are fully contained within the cirrus regions identified in the HSC-SSP map.
We also compare the actual IoU with random IoU $\left(\mbox{IoU}_{\mathrm{rand}}\right)$. Random IoU is the IoU for maps consisting of $10^{4}$ pairs of random non-equal fields (for any pair $\mbox{Field}_i, \mbox{Field}_j$, $i \neq j$) from the Intersection region. The first map consists of SDSS Sripe~82 cirrus maps for the first fields in each pair, while the second map contains HSC-SSP cirrus maps for the second fields in each pair.
The IoU$_{\mathrm{rand}}$ corresponds to the degree of similarity between random maps from SDSS Stripe~82 and HSC-SSP. Since IoU$_{\mathrm{rand}}$ depends on choosing random pairs of maps, we repeated the calculation $100$ times to get a sample distribution of IoU$_{\mathrm{rand}}$.
The average sample value $\langle \mbox{IoU}_{\mathrm{rand}}\rangle = 0.0076$ and the sample standard deviation $\mbox{sd}\left(\mbox{IoU}_{\mathrm{rand}}\right) = 0.0004$. The $\langle \mbox{IoU}_{\mathrm{rand}}\rangle$ is significantly lower than the actual IoU, indicating a strong agreement between the cirrus map from \citet{Cirrus_2023MNRAS.519.4735S} and the newly generated map.

\begin{figure*}[h!]
    \centering
    \includegraphics[width=0.95\linewidth]{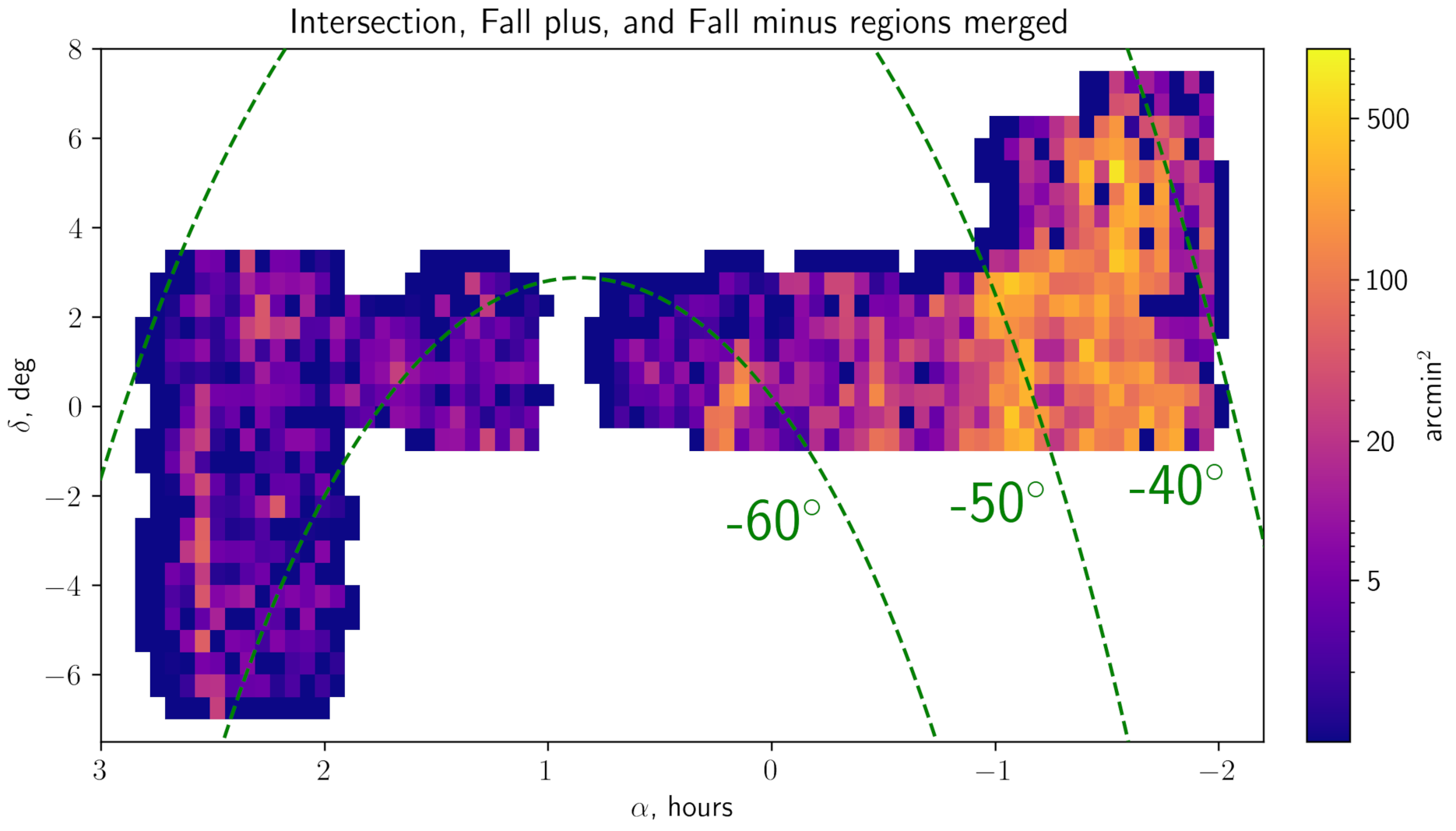}
    \includegraphics[width=0.95\linewidth]{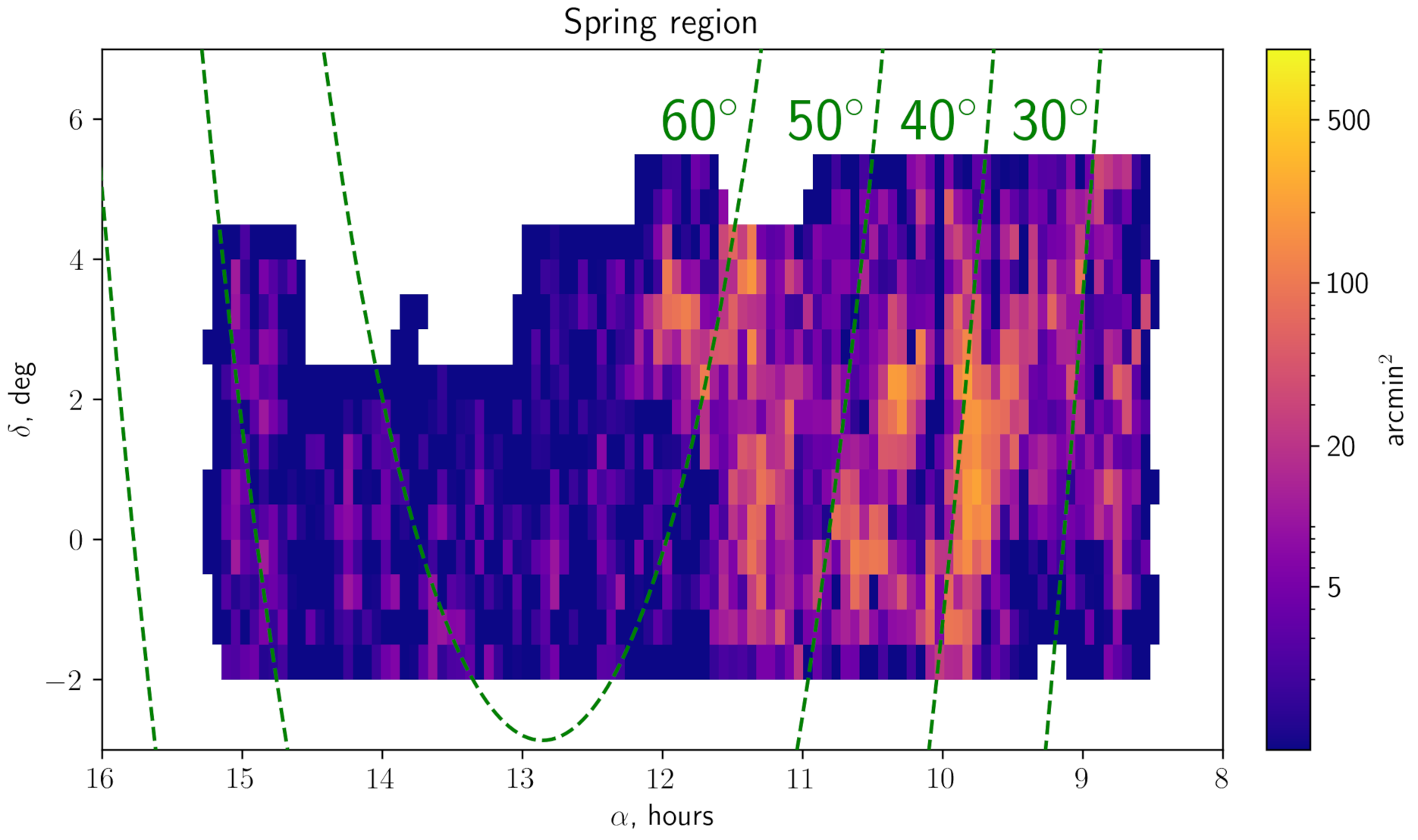} \\

    \caption{Cirrus maps obtained in the present work for the Intersection, Fall plus, Fall minus regions (\textit{top}) and Spring region (\textit{bottom}). Each rectangle in these maps corresponds to a bin of $1$ deg in right ascension and $0.5$ deg in declination. In each subplot, green lines indicate the Galactic latitude contours. The color of each bin corresponds to the total area covered by cirrus within a bin of $1$ deg in right ascension and $0.5$ deg in declination.}
    \label{fig:map1}
\end{figure*}

\begin{figure*}[h!]
\centering
\includegraphics[width=0.68\linewidth]{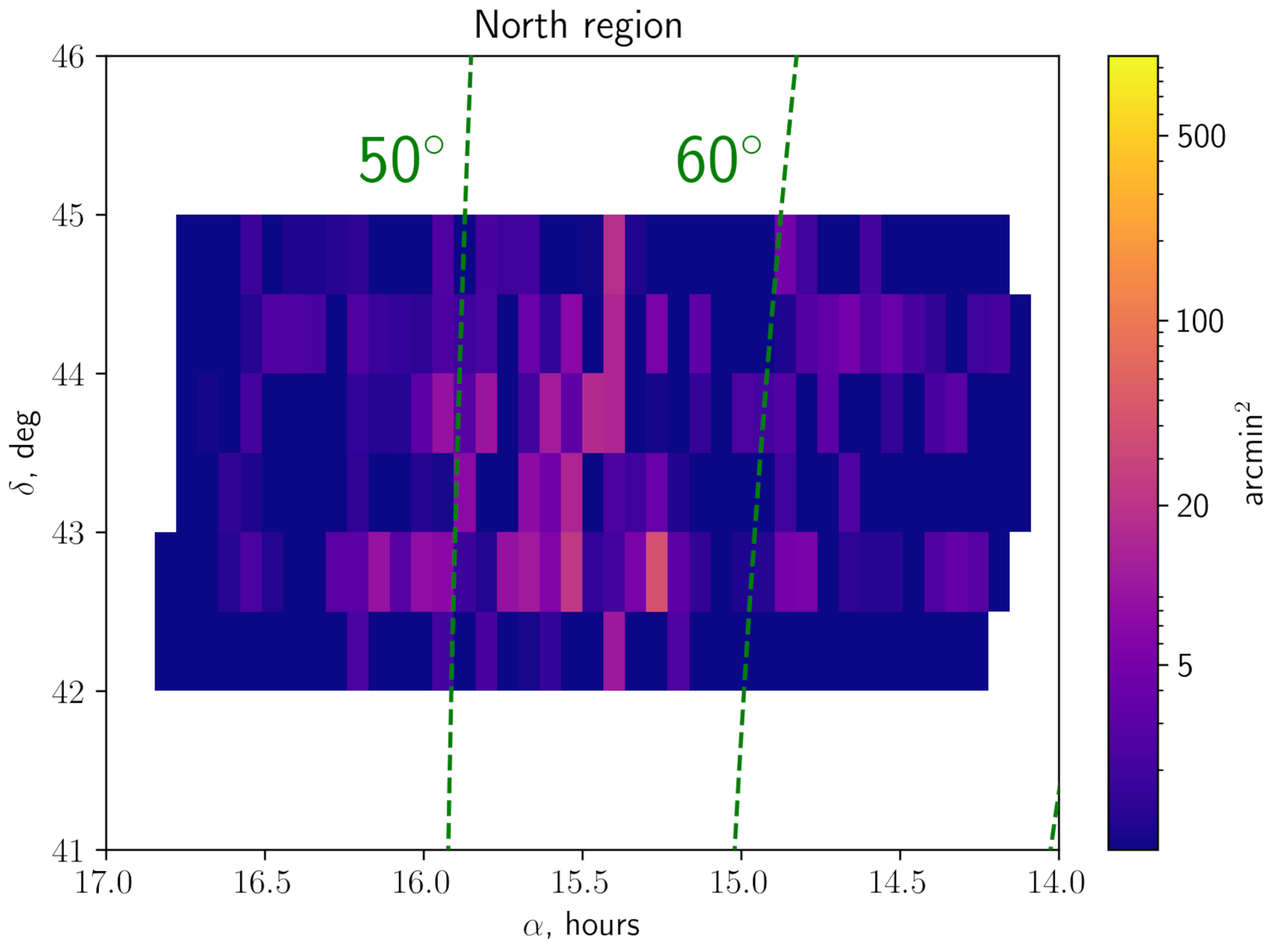}
    \caption{A cirrus map obtained in the present work for the North region. Each rectangle in this map corresponds to a bin of $1$ deg in right ascension and $0.5$ deg in declination. Green lines indicate the Galactic latitude contours. The color of each bin corresponds to the total area covered by cirrus within a bin of $1$ deg in right ascension and $0.5$ deg in declination.}
    \label{fig:map2}
\end{figure*}

The spatial distribution of cirrus filaments over each HSC-SSP region after rebinning to a bin of $1$ degree in right ascension and $0.5$ degree in declination is presented in Figs.~\ref{fig:map1} and~\ref{fig:map2}.
The maps were obtained by summing the areas covered by cirrus in the corresponding fields and then rebinning the resulting values into bins of 1 deg in right ascension and 0.5 deg in declination, assuming that the cirrus should be assigned to a bin if the center of its field lies within that bin. First of all, as evident from these maps, the cirrus distribution across the sky is highly uneven (note the logarithmic scale of the colorbar). In the Intersection and the Fall regions, a large cirrus cloud is present on the right side of the region, covering $\alpha$ from $-2^h$ to $0^h$ and $\delta$ from $0^{\circ}$ to $7^{\circ}$, with the highest cirrus concentration around $\alpha \sim -1.5^{h}$ and $\delta \sim 1^{\circ}$. We should note that while this cloud does appear to be a large, coherent structure, it was not the case in~\citet{Stripe82_2020A&A...644A..42R} and~\citet{Cirrus_2023MNRAS.519.4735S}, where only Stripe~82 data was used. In~\citet{Stripe82_2020A&A...644A..42R}, this cloud was divided into several smaller parts (see Fields 2-3 and Fields 6-10 in Table 2 of that work). Since Stripe~82 data is limited by $\delta=\pm1.25^{\circ}$, the extension of this cloud toward higher declination values was not previously studied in deep optical images. Our study demonstrates that this is now possible using HSC-SSP data. In contrast, the left part of the Intersection region is almost devoid of cirrus, though a thin filamentary structure is still present. In the Spring region, a similar pattern is observed: the right part of the region contains numerous cirrus filaments, while the left part is nearly devoid of them.
Also, note that the cirrus in this region appears as a collection of inclined stripes, revealing the characteristic filamentary structure of cirrus across multiple stacked HSC-SSP fields. We highlight that this is the first study of the global structure of cirrus clouds in this region of the sky using deep optical imaging, as Stripe~82 does not cover this area at all. In contrast, the North region is less noteworthy, as it contains almost no cirrus. We identified only a few small cirrus patches, with no clear evidence of a larger-scale dust structure.

In each subplot, we also provide contours for Galactic latitude. As seen in the data, most of the identified cirrus clouds are located between $b=30^{\circ}$ and $b=60^{\circ}$, as well as $b=-60^{\circ}$ and $b=-40^{\circ}$.

\subsection{Cirrus properties}

\begin{figure*}
    \centering
    \includegraphics[width=0.3\linewidth]{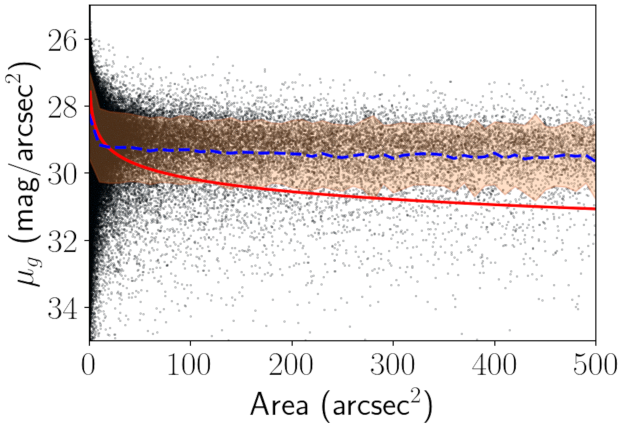}
    \includegraphics[width=0.3\linewidth]{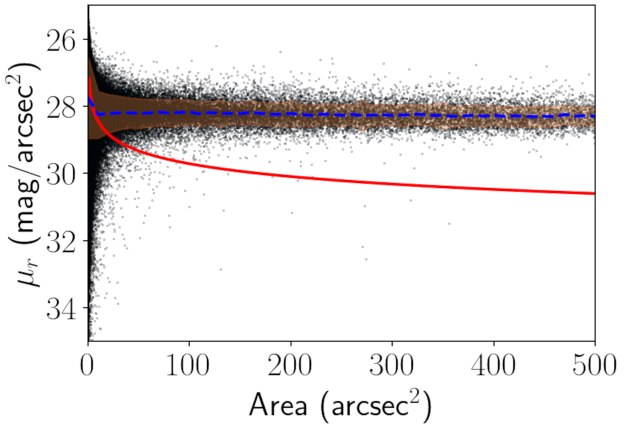}
    \includegraphics[width=0.3\linewidth]{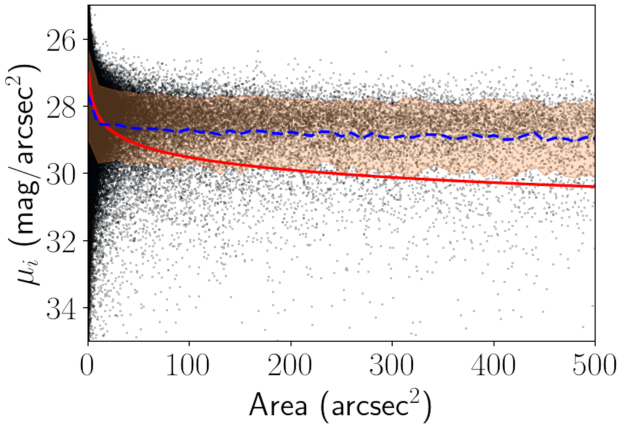}
    \caption{Surface brightness of filaments in the Spring region depending on the area in the $g$, $r$, and $i$ optical bands (from left to right). A red line in each subplot marks the $3$-sigma limit for the corresponding band. A blue dashed line shows the average value while the shaded area corresponds to $1$-sigma limits.}
    \label{fig:noise}
\end{figure*}

\begin{figure*}
    \centering
    \includegraphics[width=0.42\linewidth]{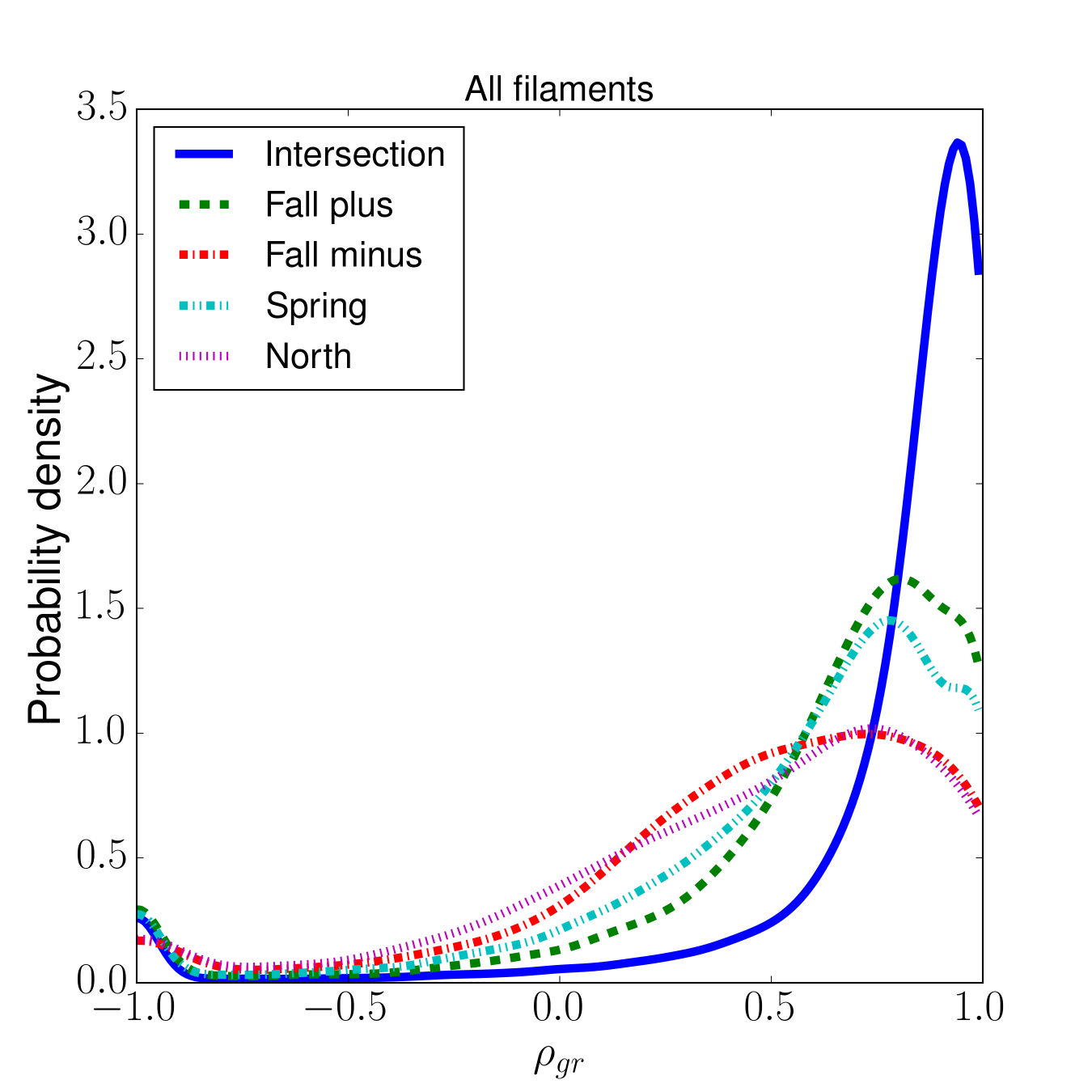}
    \includegraphics[width=0.42\linewidth]{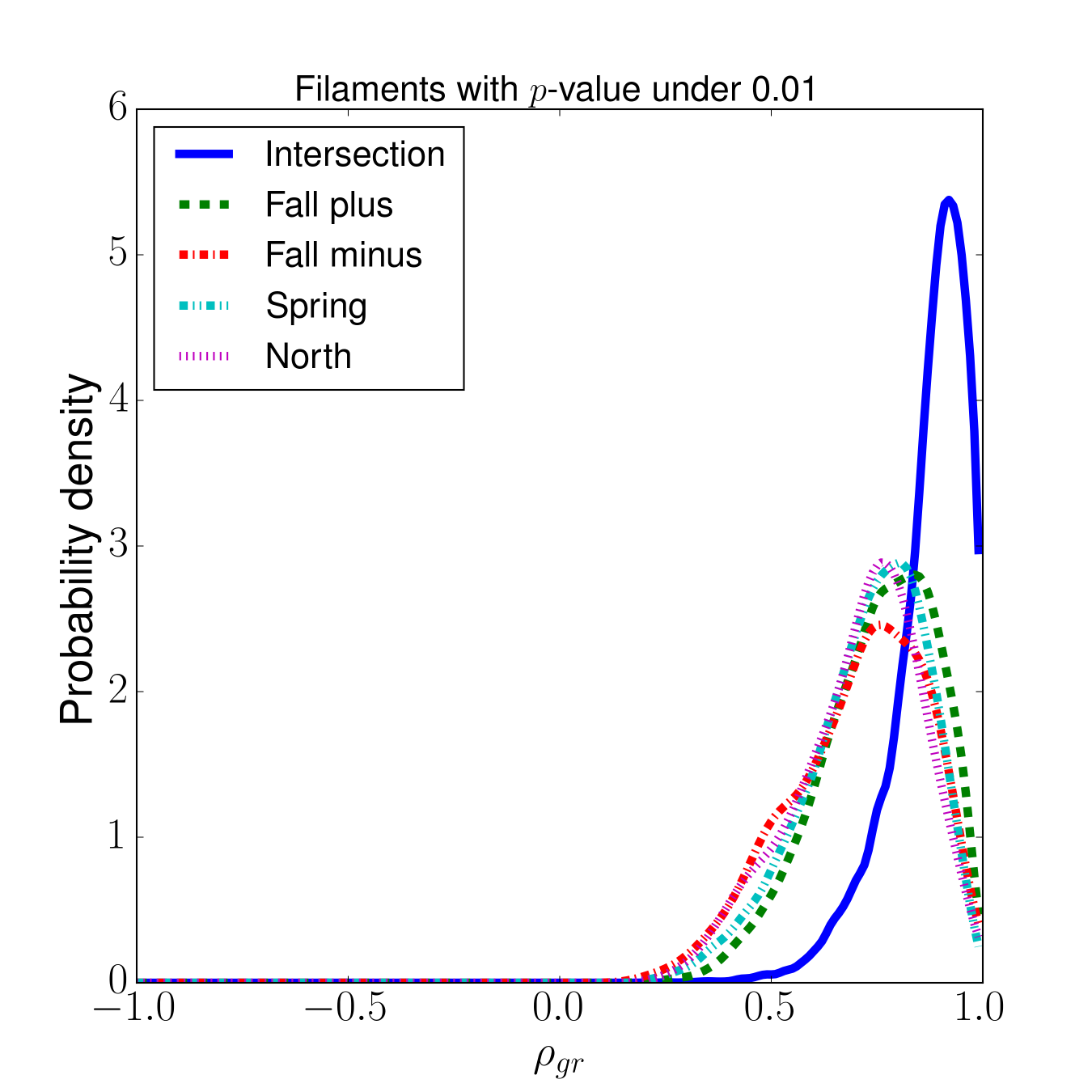}
    \\%
     \includegraphics[width=0.42\linewidth]{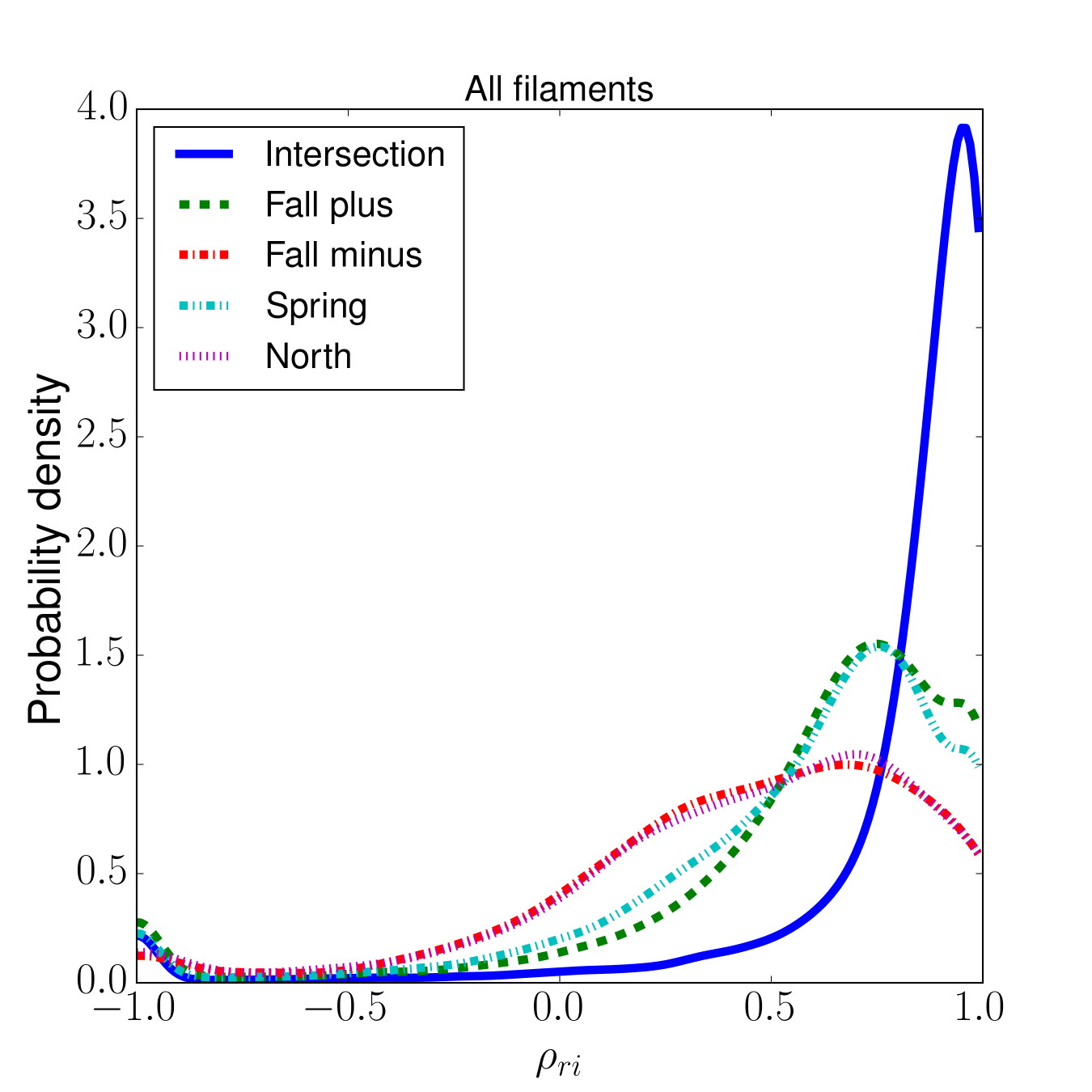}%
    \includegraphics[width=0.42\linewidth]{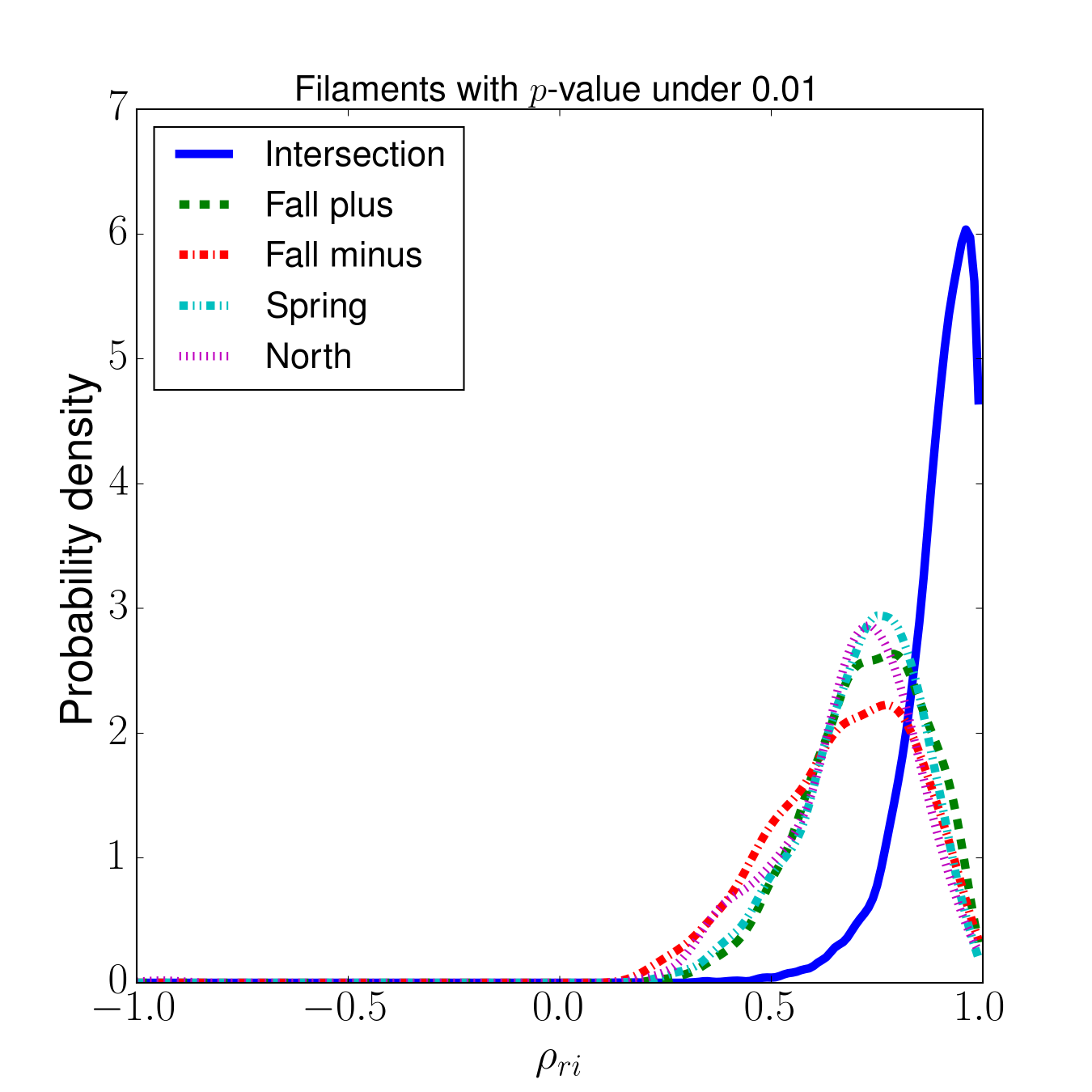}
    \caption{Distributions of Pearson correlation coefficients between different optical bands for cirrus filaments in HSC-SSP regions. Top row: $g$ and $r$ bands; Bottom row: $r$ and $i$ bands; Left column: for all filaments; Right column: only for statistically significant correlation with $p$-value under 0.01.}
    \label{fig:corr}
\end{figure*}

In this section, we analyze how bright the cirrus is compared to background noise and how cirrus fluxes in different bands correlate with each other.
First, we assess how well the filaments stand out from the background noise. Fig.~\ref{fig:noise} shows the surface brightness of the filaments identified in the Spring region as a function of their size (total filament area). The surface brightness of a given filament is measured in the the manner as it is usually measured for extended objects: $\mu=-2.5\log10 \,(F/S) + m_0$, where $S$ is an area of a given filament, $F$ is the total flux inside the filament's area, and $m_0=27$ mag is the photometric zero-point. We also plot the $3$-sigma noise limit as a function of filament area. The exact value of the limit for different areas was determined using a sample of randomly placed squares which had a certain area value. After the sigma value was calculated for each square, it was averaged across all squares. As can be seen, most filaments lie well above the limits, indicating that they are bright enough to be clearly distinguished from the background noise. However, we should also note that there is a tail of small filaments extending below the $3$-sigma limits. This is naturally explained by the fact that the total flux in small filaments is more susceptible to disruptions from noise and over-subtraction by the background algorithm (see Section~\ref{sub_sec:over-subtraction}). Nevertheless, the neural network identifies these small filaments as cirrus based on their morphology, even though they cannot be distinguished from noise using the signal-to-noise ratio alone.

Secondly, compared to the $g$ and $i$ bands, the flux distribution in the $r$ band is more concentrated above the 3-sigma limit across different area values. In the $g$ and $i$ bands, filaments more frequently appear below the $3$-sigma limits. This difference between bands may be attributed to the fact that the training sample of filaments was prepared using only $r$-band images, while the cirrus mask remains unchanged for other optical bands.
Although we focused on the Spring region in this discussion, the results are highly similar across other HSC-SSP regions. Fig.~\ref{fig:app_noise} in \ref{app_sec:noise} shows the plots for these regions.

Following our previous work~\citep{Cirrus_2023MNRAS.519.4735S}, we calculated Pearson correlation coefficients between different optical bands for each filament:
\begin{equation}
    \rho_{xy} = \displaystyle\frac{\sum_{i=1}^n (x_i-\bar{x}) (y_i-\bar{y})}{\sqrt{\sum_{i=1}^n (x_i-\bar{x})^2} \sqrt{\sum_{i=1}^n (y_i-\bar{y})^2}},
\end{equation}
where $x_i$ and $y_i$ are the fluxes of the $i$-th pixel of the filament in $x$ and $y$ bands, respectively, and $\bar{x}$ and $\bar{y}$ are the mean fluxes found by averaging over all pixels contained in the filament area. It is worth noting that a combined mask was applied to the cirrus map before the correlation coefficients were calculated. To estimate the correlations, we also rebinned the data (the original $g,\,r,$ and $i$ images, as well as the cirrus maps and combined masks) to a 6-arcsecond resolution. This also facilitates direct comparison with the results of~\citet{Cirrus_2023MNRAS.519.4735S}, which were obtained at this exact spatial resolution.

Fig.~\ref{fig:corr} presents the distributions of correlation coefficients for all filaments and for those with a statistically significant correlation ($p$-value~$< 0.01$).
For these calculations we used \texttt{pearsonr} function from \texttt{scipy}. It is worth noting that this test requires a normality test, but we did not perform it. For some filaments, this condition may not be met, but we prefer to filter out filaments with a certain degree of error rather than not filtering them at all.
As shown in the figure, the correlation coefficients exhibit a wide distribution, with some filaments even displaying negative correlations. Such cases arise when a filament contains a small number of pixels, making its total flux highly susceptible to disruptions from noise or other sources of error, such as proximity to bright sources, background subtraction issues, or flat-fielding artifacts. We effectively excluded these filaments from further analysis by filtering them based on the $p$-value of their correlation: \mbox{$p$-value~$< 0.01$}. As can be seen, applying this criterion shows that most filaments with statistically significant correlation exhibit positive correlation coefficients: $\rho_{gr}>0.5$ and $\rho_{ri} > 0.5$. A significant correlation is generally expected for any source; however, since all compact sources have been masked, the observed correlation is likely associated with either cirrus filaments or other components of diffuse Galactic light.
Thus, Fig.~\ref{fig:corr} demonstrates that our cirrus identification methods successfully detect well-correlated cirrus filaments at a spatial scale of 6 arcsec.

\begin{figure*}[h!]
    \centering
    \includegraphics[width=0.45\linewidth]{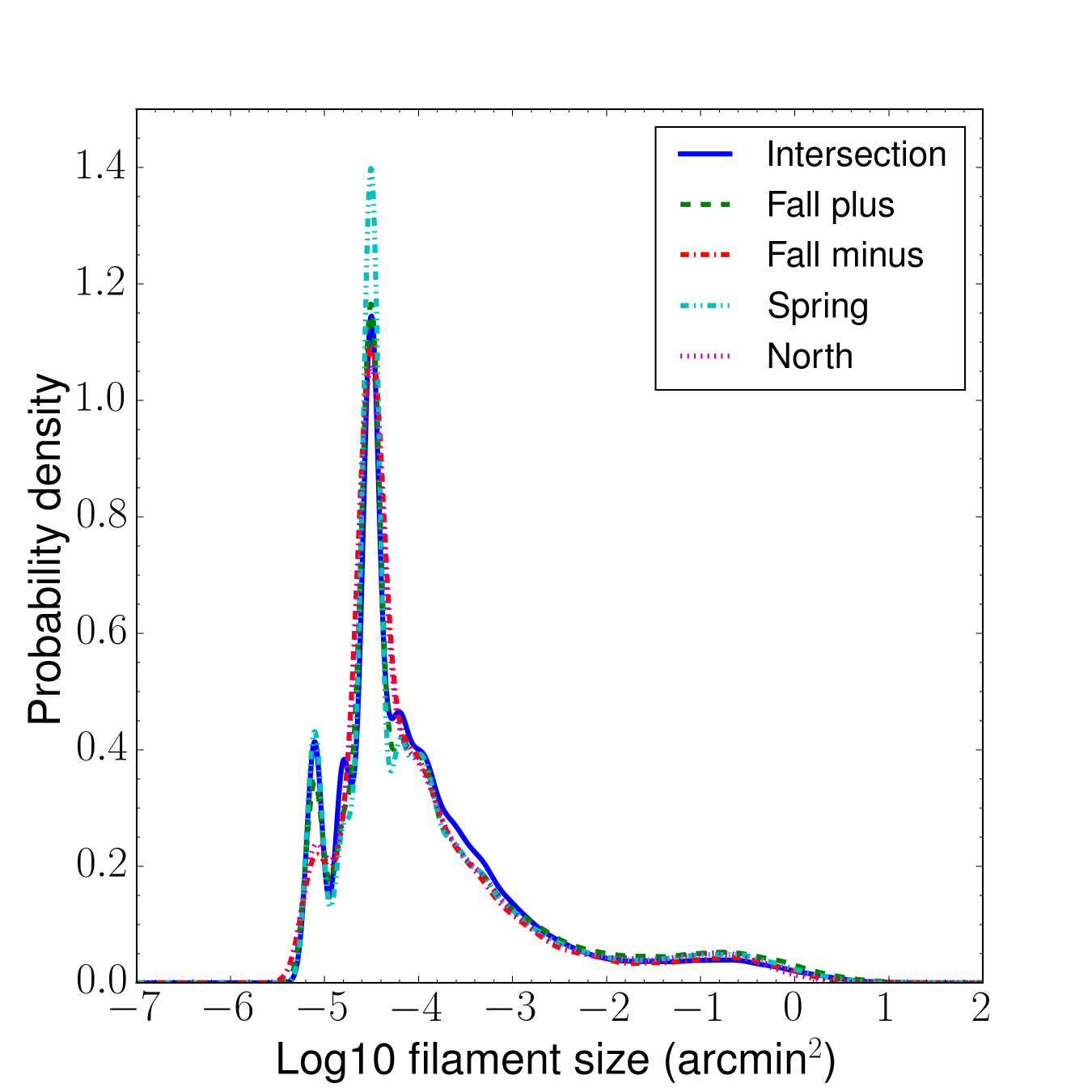}%
    \includegraphics[width=0.45\linewidth]{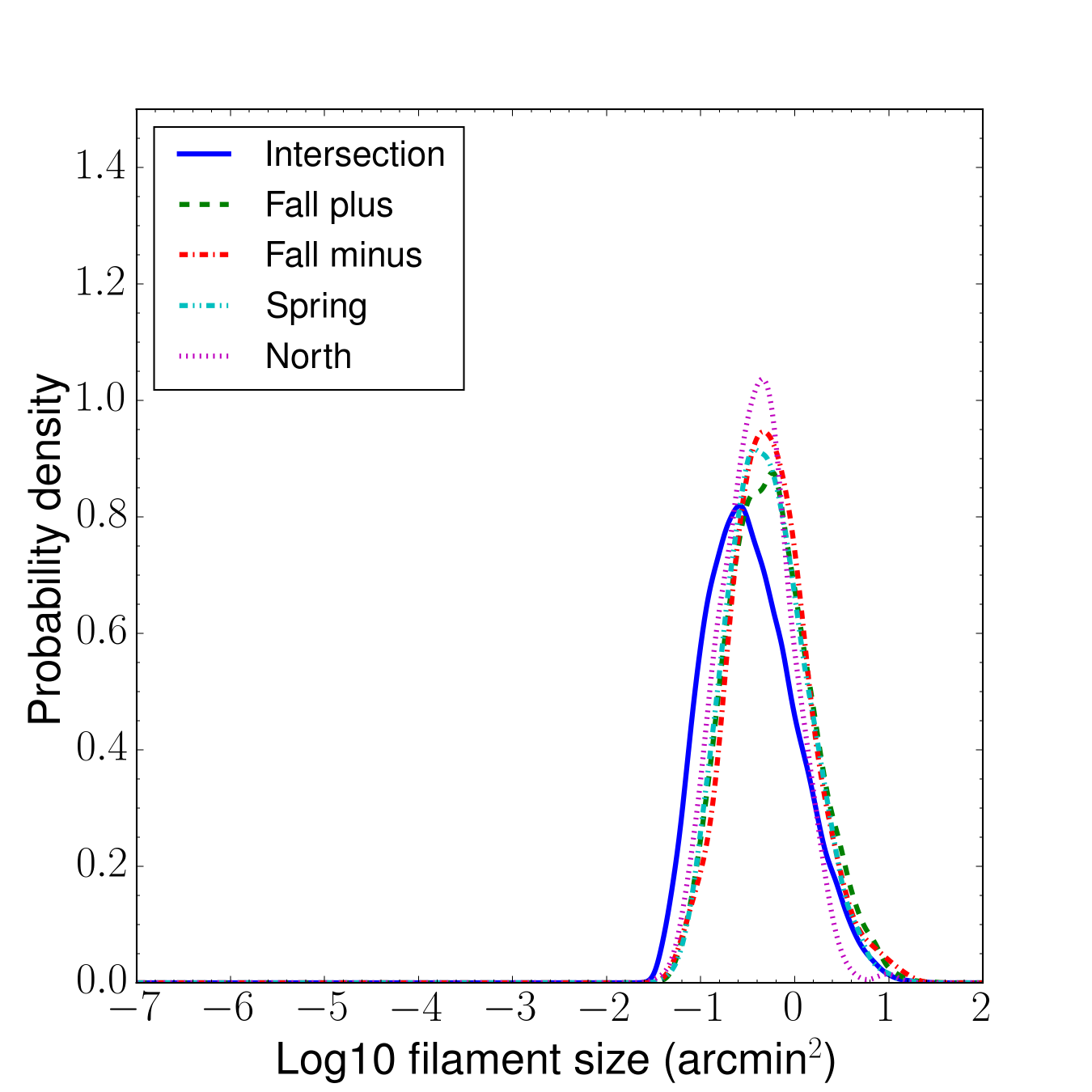}%
    \caption{Distribution of sizes of individual filaments in HSC-SSP regions for the original resolution of HSC-SSP 0.168 arcsec (\textit{left}) and for the data rebinned to the resolution of 6 arcsec (\textit{right}).}
    \label{fig:sizes}
\end{figure*}

\begin{figure*}[h!]
    \centering
    \includegraphics[width=0.88\linewidth]{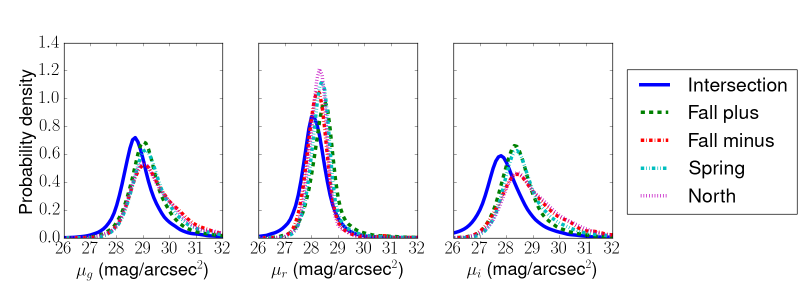}%
    \caption{Distributions of the surface brightnesses of individual filaments in the HSC-SSP regions, in $g$, $r$, and $i$ bands (from left to right, respectively), for data rebinned to a resolution of 6 arcsec.}
    \label{fig:mu_collage}
\end{figure*}

Fig.~\ref{fig:sizes} presents the size distribution of filaments at two different resolutions: the original HSC-SSP resolution of $0.168$ arcsec and the 6-arcsecond resolution, considering only filaments with a statistically significant correlation ($p$-value~$< 0.01$). As before, a combined mask was applied to the cirrus map before calculating the size distribution of filaments. It is also worth noting that filaments touching along the boundaries of fields are considered to be separate filaments, as in \citet{Cirrus_2023MNRAS.519.4735S}.
We present the size distribution at two different resolutions because examining both cases provides valuable insights into how filament properties change with spatial scale. For the original resolution, we see that the sample is dominated by very small filaments with a typical size of about 3--4~pixels (0.5--0.6~arcsec). A small peak to the left corresponds to the area of a single HSC-SSP pixel ($\approx0.03$~arcsec$^2$). At the same time, large filaments ($\sim$ several arcmin$^2$) --- comparable in size to some galaxies --- are present across all HSC-SSP regions.
\par
At the 6-arcsecond resolution (right panel of Fig.~\ref{fig:sizes}), the size distributions appear noticeably different. First, since the smallest scale is now 6~arcsec, all filaments smaller than this threshold are effectively filtered out.
Secondly, applying the $p$-value filter causes the distribution peaks to shift to a characteristic value higher than the smallest allowed scale of 6 arcsec. This value varies slightly across individual regions but remains approximately 0.5--1.0 arcmin$^2$ for all HSC-SSP regions. This is in good agreement with the findings of \citet{Cirrus_2023MNRAS.519.4735S} for 6-arcsecond resolution, where the typical filament size in Stripe~82 data was found to be approximately $1$ arcmin$^2$. It is also important to note that, although the size distributions at the original resolution and 6-arcsecond resolution appear quite different, the total area covered by cirrus remains nearly the same in both cases: 13.3 deg$^2$ at the original resolution and 12.6 deg$^2$ for filaments with statistically significant correlation at 6-arcsecond resolution. This is because the majority of identified cirrus is contained within large, extended filaments, rather than in small patches with sizes below 6 arcseconds.

Taking into account that only filaments filtered by $p$-value have a strong correlation between fluxes in optical bands, we prepared two versions of cirrus filament catalogs for the entire HSC-SSP.
The first catalog contains all the detected cirrus filaments that remain after rebinning to a 6-arcsecond resolution. Therefore, all filaments with an area greater than 36~arcsec$^{2}$ are contained in this catalog. The second catalog consists of filaments filtered by the $p$-value from the first one. Both catalogs are available on our \texttt{GitLab} repository\footnote{\url{https://gitlab.com/polyakovdmi93/cirrus_narrow_segmentation}}. Fig.~\ref{fig:mu_collage} presents the distributions of surface brightnesses of individual filaments in the $g$, $r$, and $i$ bands. These distributions are qualitatively similar across different HSC-SSP regions, with filaments in the Intersection region exhibiting the highest average brightness.

While the colors of filaments are important properties of cirrus, we intentionally do not analyze their distributions here. This is because cirrus fluxes in HSC-SSP data may be affected by background subtraction, potentially leading to incorrect estimations. We discuss this issue in detail in Section~\ref{sec:discussion}.



\section{Discussion}
\label{sec:discussion}

\subsection{Comparison of HSC-SSP data with Stripe~82 data}
\label{sub_sec:comparison_with_Stripe82}
\begin{figure}
    \centering
    \includegraphics[width=0.8\linewidth]{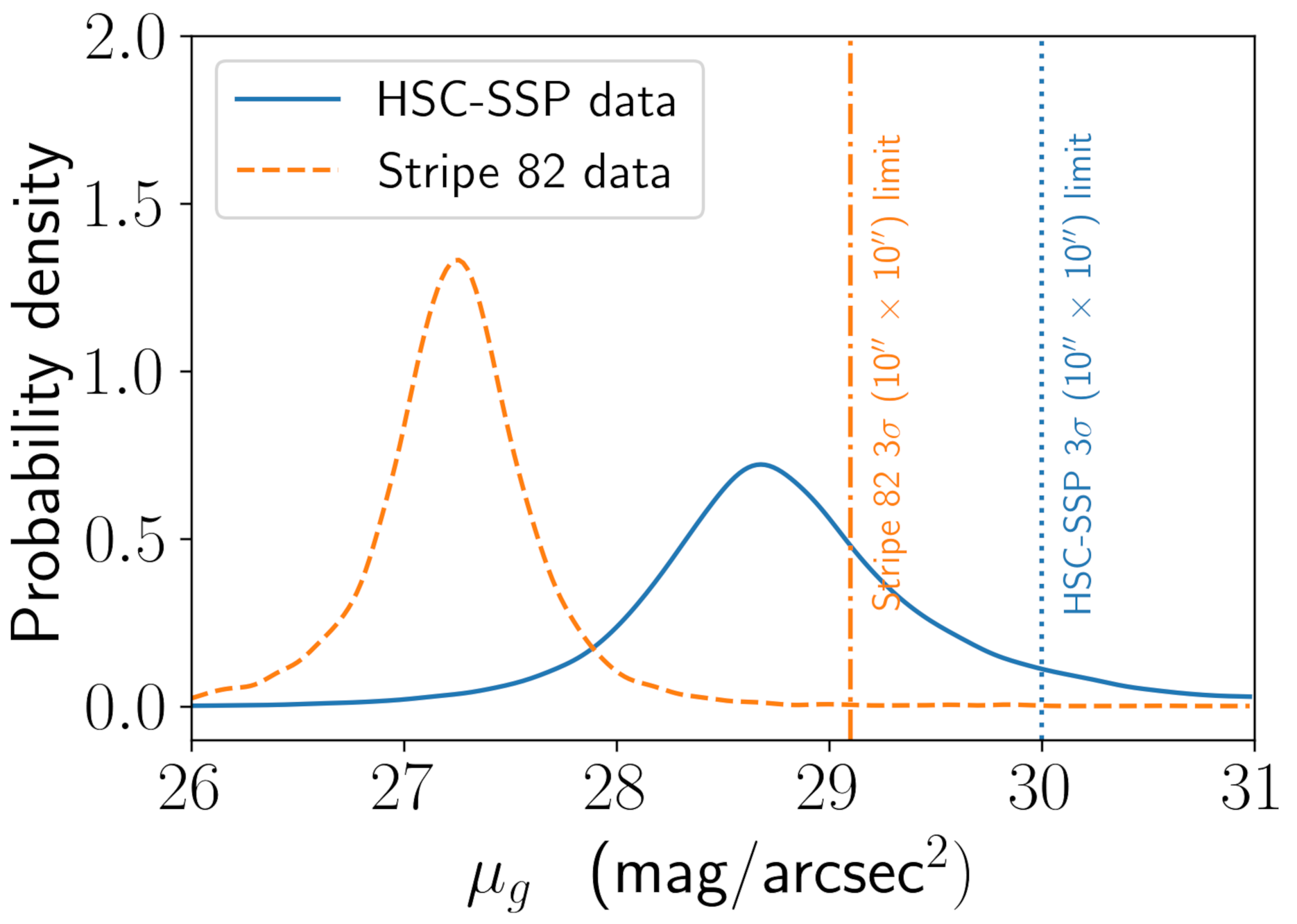}%

    \caption{Comparison of $g$-band surface brightnesses distributions for individual filaments in the Intersection region for Stripe~82 from~\citet{Cirrus_2023MNRAS.519.4735S} and HSC-SSP data (this work). Vertical lines indicate the 3$\sigma$(10$''$ $\times$ 10$''$) surface brightness limits in the $g$ band for the corresponding surveys.}
    \label{fig:sdss_vs_hscssp}
\end{figure}

\begin{figure*}
    \centering
    \includegraphics[width=0.44\linewidth]{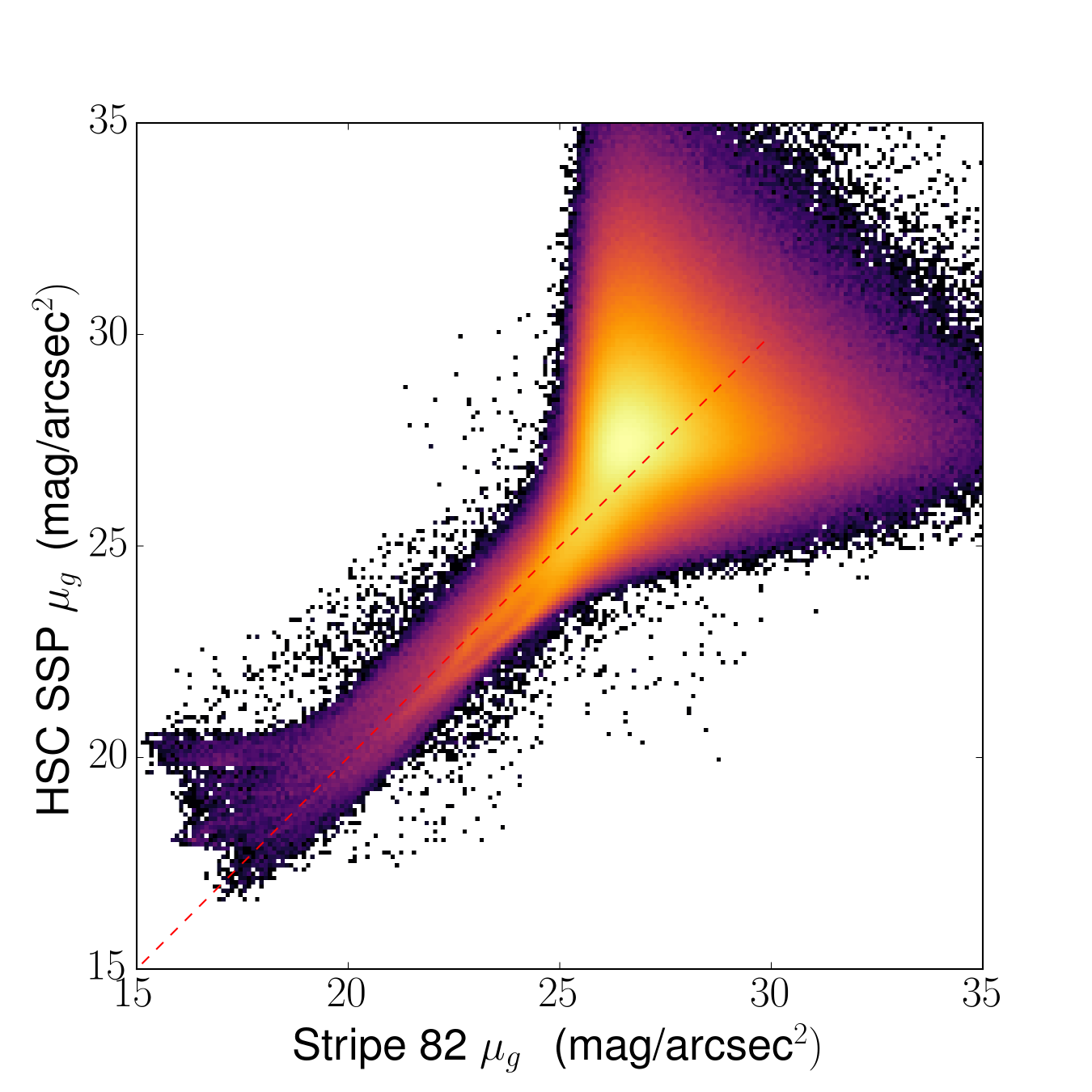}%
    \includegraphics[width=0.44\linewidth]{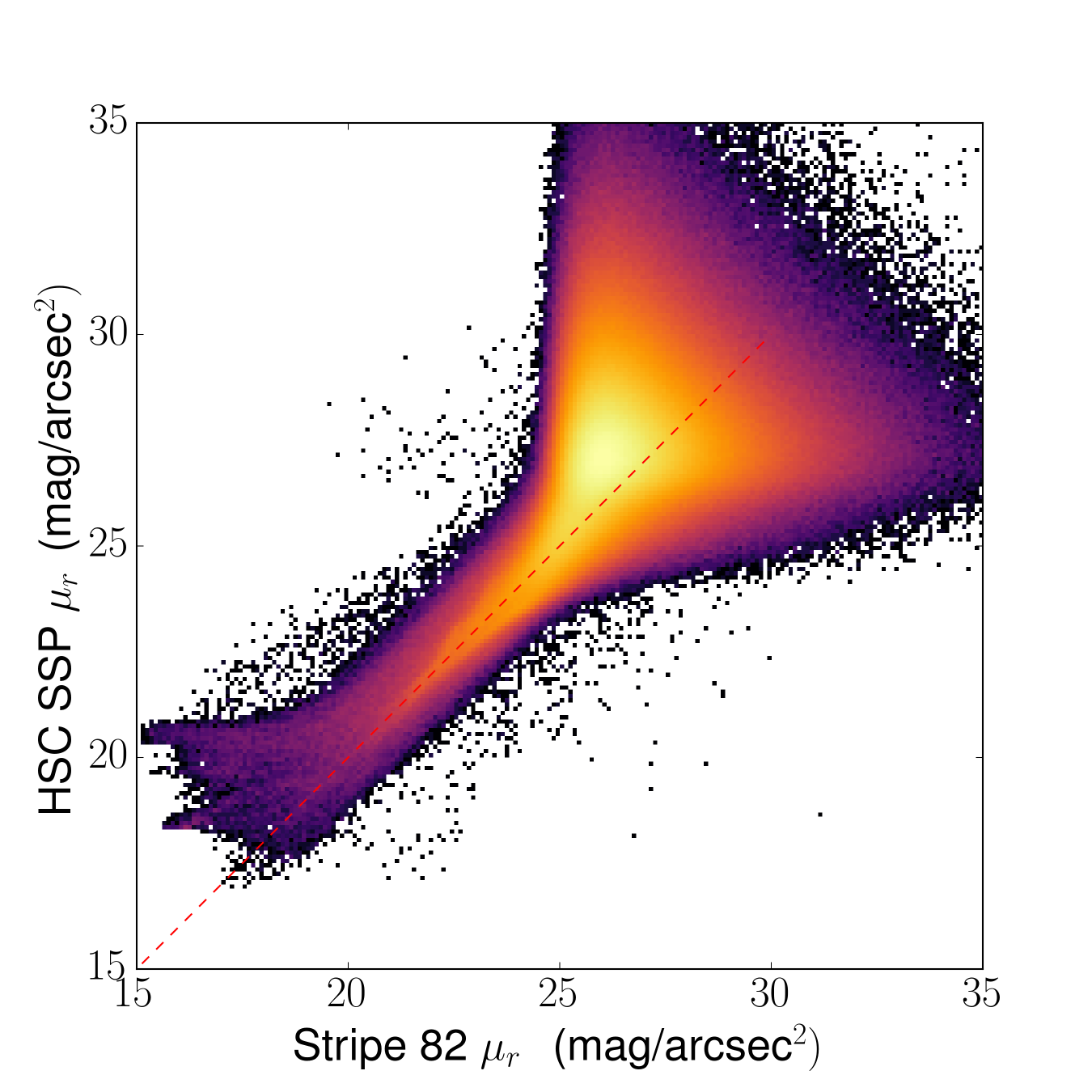}%
    \\
    \includegraphics[width=0.72\linewidth]{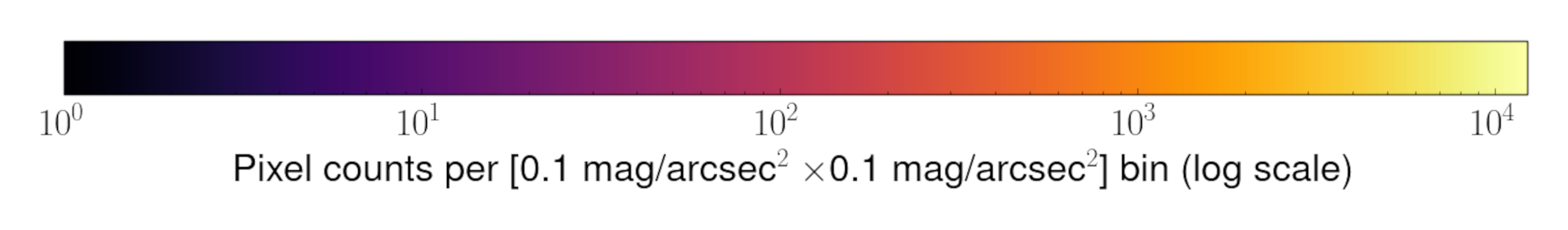}%

    \caption{Pixel by pixel comparison of surface brightnesses in Stripe~82 and HSC-SSP data for the Field f1053 in the $g$ (left) and $r$ (right) optical bands.}
    \label{fig:ST_vs_SSP_f1053}
\end{figure*}

Fig.~\ref{fig:sdss_vs_hscssp} compares the probability density distribution of the surface brightness of cirrus filaments identified in \citet{Cirrus_2023MNRAS.519.4735S} from Stripe~82 data at a resolution of 6 arcsec with those found in the Intersection region in this study (at the same 6-arcsecond resolution). It is important to note that HSC-SSP covers only a portion of the original Stripe~82 data; therefore, we consider only the common fields from both datasets in this analysis. In Fig.~\ref{fig:sdss_vs_hscssp}, we expected the distributions to resemble each other. However, this is clearly not the case. The distributions are shifted relative to one another, indicating that the filaments in HSC-SSP appear dimmer, with the distribution mode shifted by approximately $\Delta \mu_g \sim 2$ mag arcsec$^{-2}$. Also, the distribution from the original Stripe~82 data is slightly sharper and narrower. \par

To identify the root of the descrepancy, we performed a pixel-by-pixel comparison in the $g$ and $r$ bands for selected fields in Stripe~82 and HSC-SSP data. Fig.~\ref{fig:ST_vs_SSP_f1053} shows this comparison for Field f1053, which is very rich in cirrus. Since HSC-SSP and Stripe~82 data have different native resolutions, we rebinned the HSC-SSP image to match the SDSS resolution of 0.396 arcsec/pixel. Note that the comparison includes all pixels from the original fields. In other words, no masking was done, and no cirrus was selected when we carried out this comparison. This is done so to demonstrate that, for bright sources, the fluxes are nearly the same. The figure indeed shows a strong 1:1 correlation up to a certain surface brightness level in both bands. However, beyond approximately 26–27 mag arcsec$^{-2}$, this correlation is no more. HSC SSP flux values are on average smaller (magnitudes are larger) and there is a large spread of values from the 1:1 line.
 Figs.~\ref{fig:sdss_vs_hscssp} and~\ref{fig:ST_vs_SSP_f1053} suggest that cirrus in the HSC-SSP data may be affected by background over-subtraction. We examine this issue in detail in the next section.



\subsection{Over-subtraction areas in HSC-SSP}
\label{sub_sec:over-subtraction}

In the previous section, we demonstrated that cirrus in HSC-SSP data may be affected by background over-subtraction. To assess its impact, we examined the properties of various layers adjacent to cirrus filaments.

The boundary layer $L_{d, t}$ in a field is defined by the distance $d$ from the cirrus and its thickness $t$. We calculate this layer using mathematical morphology applied to field's cirrus maps, as described below.

\begin{equation} \label{eq:boundary_layer}
	L_{d, t} = \begin{cases} {\left(F \oplus K_{\mathrm{out}}\right) \setminus \left(F \oplus K_{\mathrm{in}}\right) \setminus M_{\mathrm{c}},} & {d > 0}
    \cr {\left(F \oplus K_{\mathrm{out}}\right) \setminus F \setminus M_{\mathrm{c}},} & {d \equiv 0} \end{cases},
\end{equation}
where $F$ is the field cirrus map, $\oplus$ denotes a dilation operation, $\setminus$ denotes a set difference operation, $M_{\mathrm{c}}$ is a combined mask. We use the combined mask to exclude non-cirrus sources from analysis. $K_{\mathrm{out}}$ and $K_{\mathrm{in}}$ are circle structuring elements defined using the \texttt{OpenCV} library.
{\scriptsize \begin{verbatim}
    k_out = cv2.getStructuringElement(cv2.MORPH_ELLIPSE, (2 * (t + d) + 1, 2 * (t + d) + 1))
    k_in = cv2.getStructuringElement(cv2.MORPH_ELLIPSE, (2 * d + 1, 2 * d + 1))
\end{verbatim}}

\begin{figure*}[h!]
\begin{center}
    \begin{minipage}[b]{0.265\textwidth}
        \includegraphics[width=\textwidth]{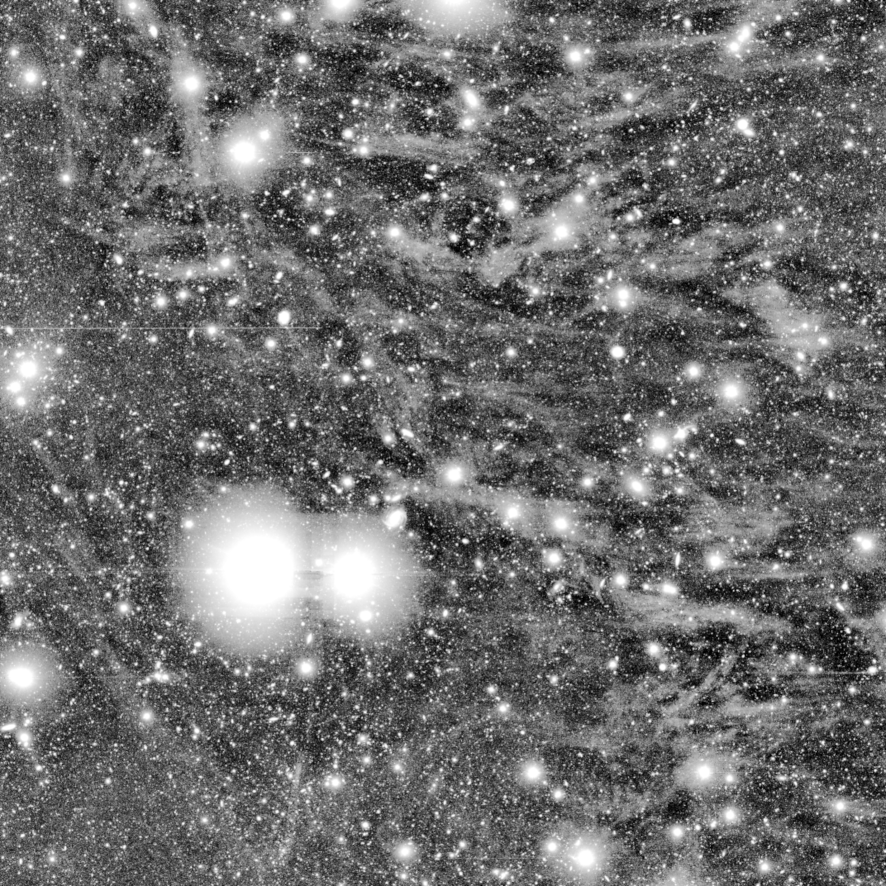}
    \end{minipage}
     \begin{minipage}[b]{0.265\textwidth}
        \includegraphics[width=\textwidth]{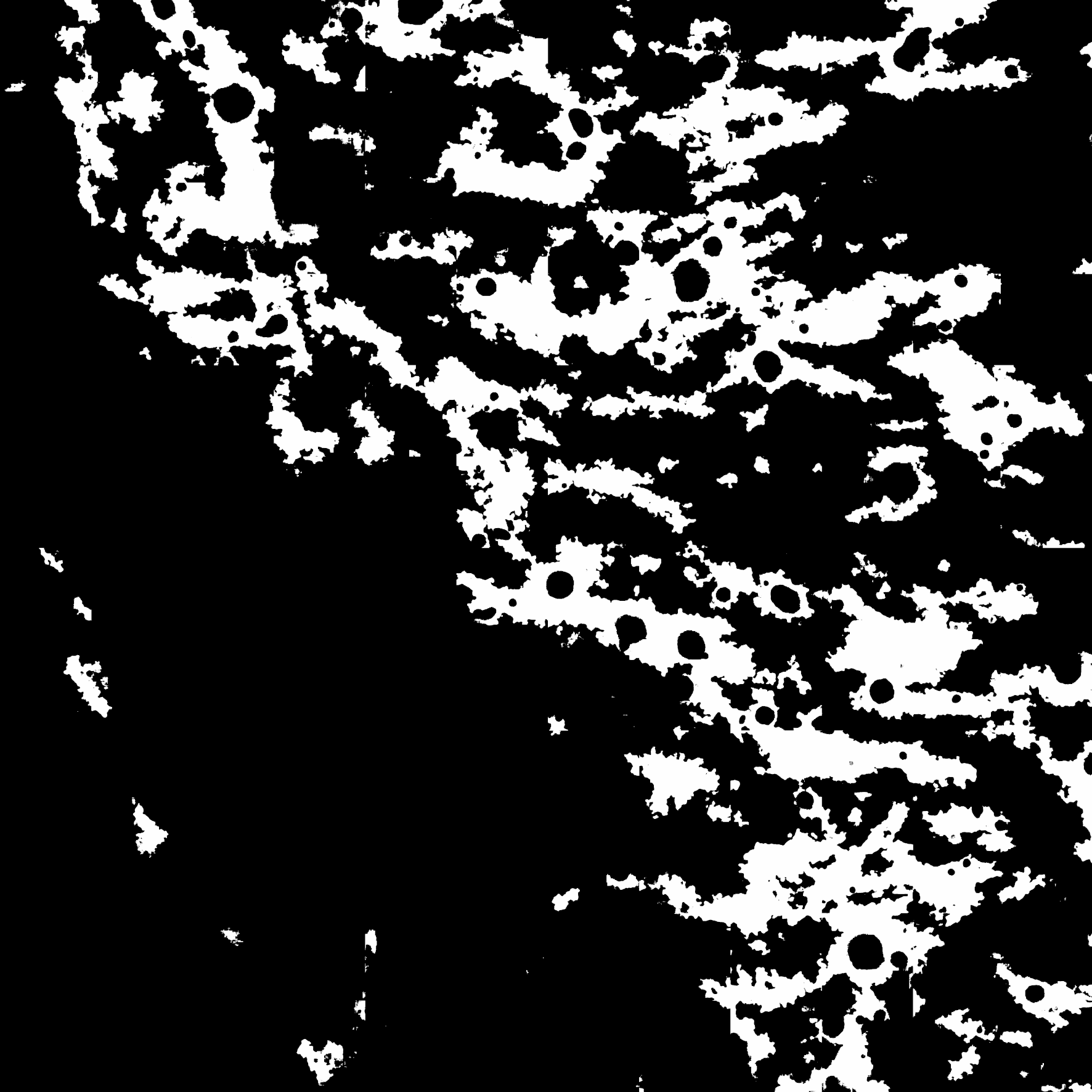}
    \end{minipage}
    \begin{minipage}[b]{0.265\textwidth}
        \includegraphics[width=\textwidth]{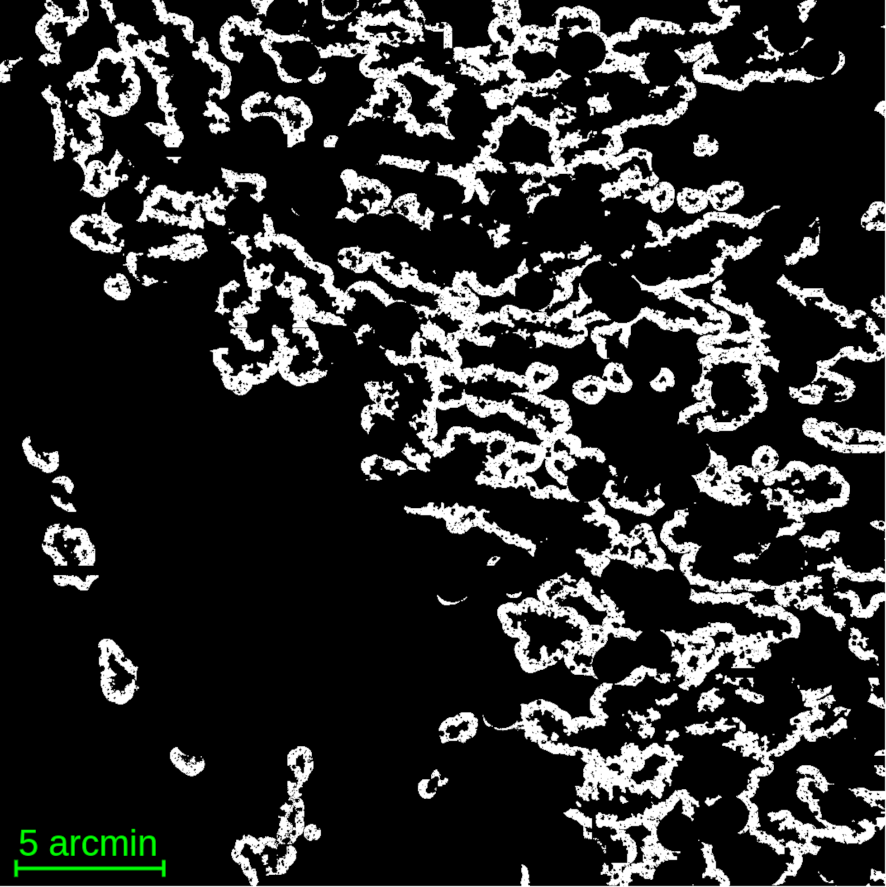}
    \end{minipage}
\end{center}
\caption{Example of boundary layer $L_{0, 100}$ for field f1053. The field itself in the $r$ band is shown in the left panel. The central panel shows the ensemble-generated cirrus map. The boundary layer is demonstrated in the right panel.}
\label{fig:boundary_layer}
\end{figure*}

The properties of boundary layers were analyzed for fields in the Intersection region.
As an example, Fig.~\ref{fig:boundary_layer} illustrates the boundary layer for Field f1053.

The simplest way to estimate the characteristic value of over-subtraction in a boundary layer is to compare the median pixel value within the layer to the median value in the background. Alternatively, for fields with a significant presence of cirrus filaments, one can compare the corresponding fractions of negative pixels.
To start, we considered the boundary layer with a thickness of 100 pixels adjacent to cirrus, denoted as $L_{0, 100}$, in fields where the cirrus fraction exceeds $0.1$. There are 98 such fields in the Intersection region. For each field in the $r$ band, we calculated the median pixel value of $L_{0, 100}$, denoted as $m(L_{0, 100})$. For the same fields, we also computed the fraction of negative pixels, $f_{-0}(L_{0, 100})$.
Next, we calculated the median values of $m(L_{0, 100})$ and $f_{-0}(L_{0, 100})$: $\langle m(L_{0, 100}) \rangle = -1.49 \times 10^{-3}$, $\langle f_{-0}(L_{0, 100}) \rangle = 0.512$. Then, we computed similar values for the background\footnote{The background is defined as the area where the combined mask is 0: $B \equiv \overline{M_{\mathrm{c}}}$} pixels in the same fields. Also note that some cirrus and boundary layer pixels are inevitably included in $B$. The median values for the background are $\langle m(B) \rangle = 4.67 \times 10^{-4}$ and $\langle f_{-0}(B) \rangle = 0.496$. Thus, the median value in pixels of the boundary layer is less than the corresponding value in the background, then this indicates an over-subtraction of background in the boundary layer.
To demonstrate the statistical significance of this conclusion, we performed a Wilcoxon signed-rank test with a ``less'' alternative for the pairs of samples $\left(m(L_{0, 100}), m(B) \right)$ and $\left(f_{-0}(B), f_{-0}(L_{0, 100})\right)$. These tests yielded $p$-values of $4.9\times10^{-18}$ and $4.7\times10^{-18}$, respectively. Therefore, the null hypotheses can be reliably rejected.

\begin{figure*}[h!]
    \centering
    \includegraphics[width=0.265\linewidth]{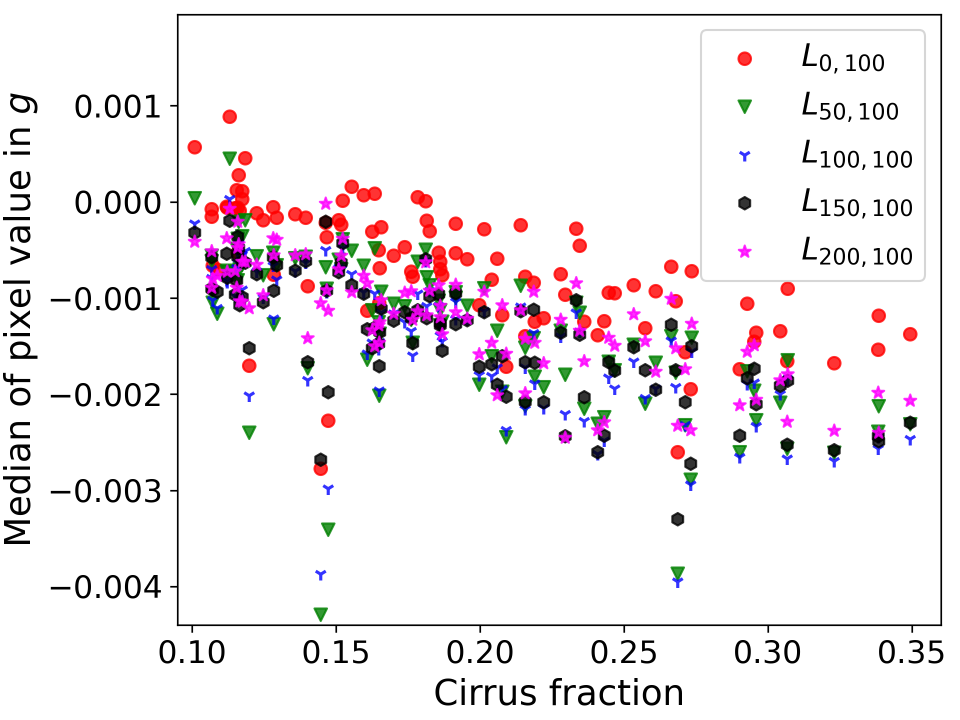}%
    \includegraphics[width=0.265\linewidth]{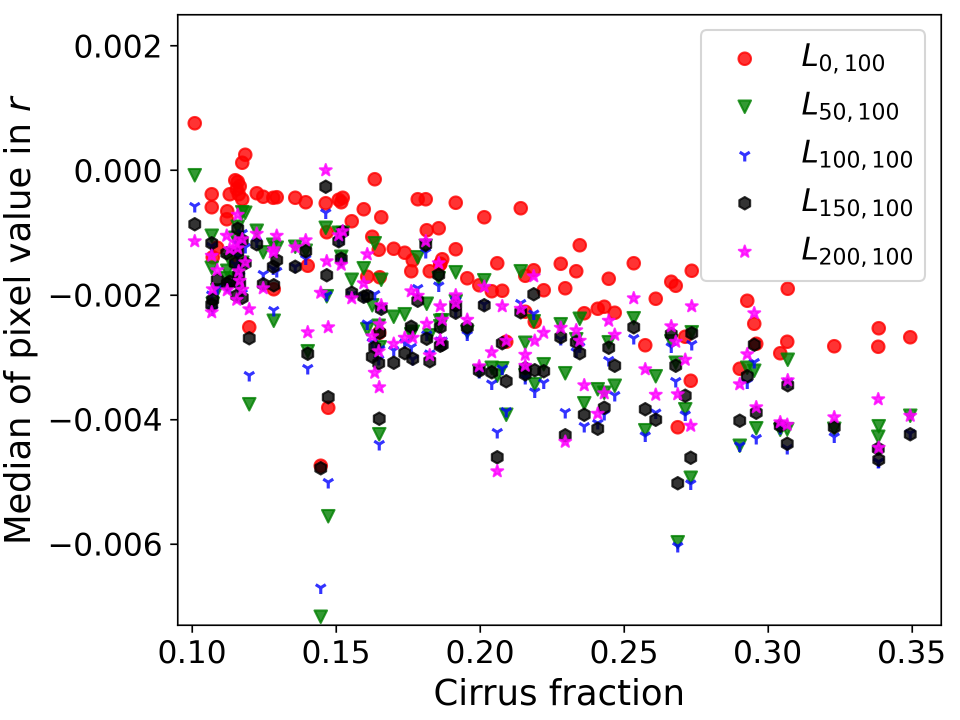}
    \includegraphics[width=0.265\linewidth]{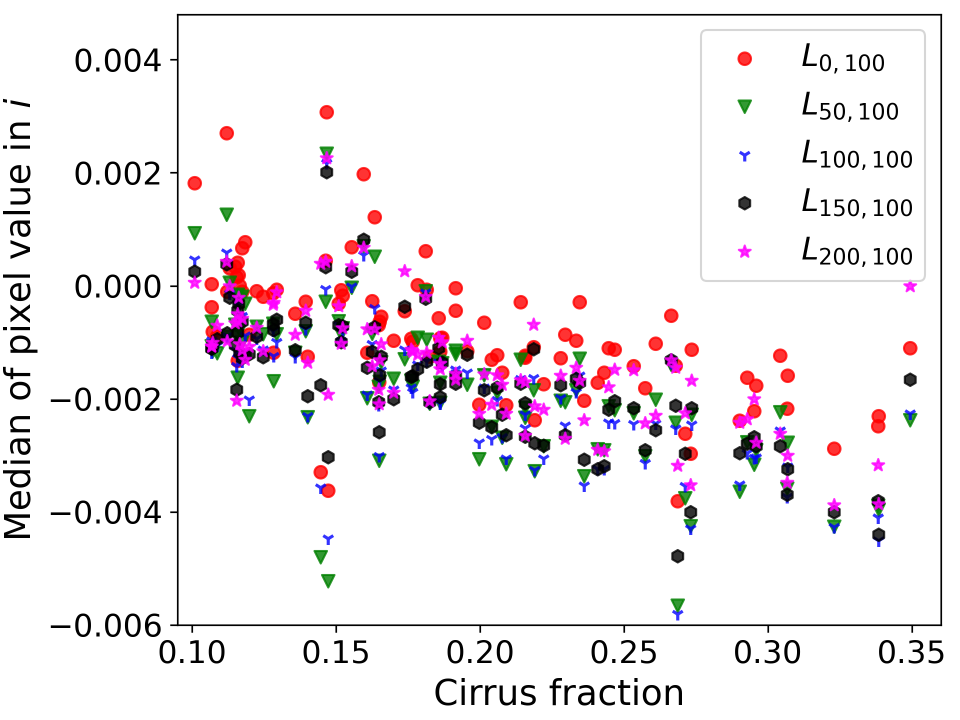} \\

    \includegraphics[width=0.265\linewidth]{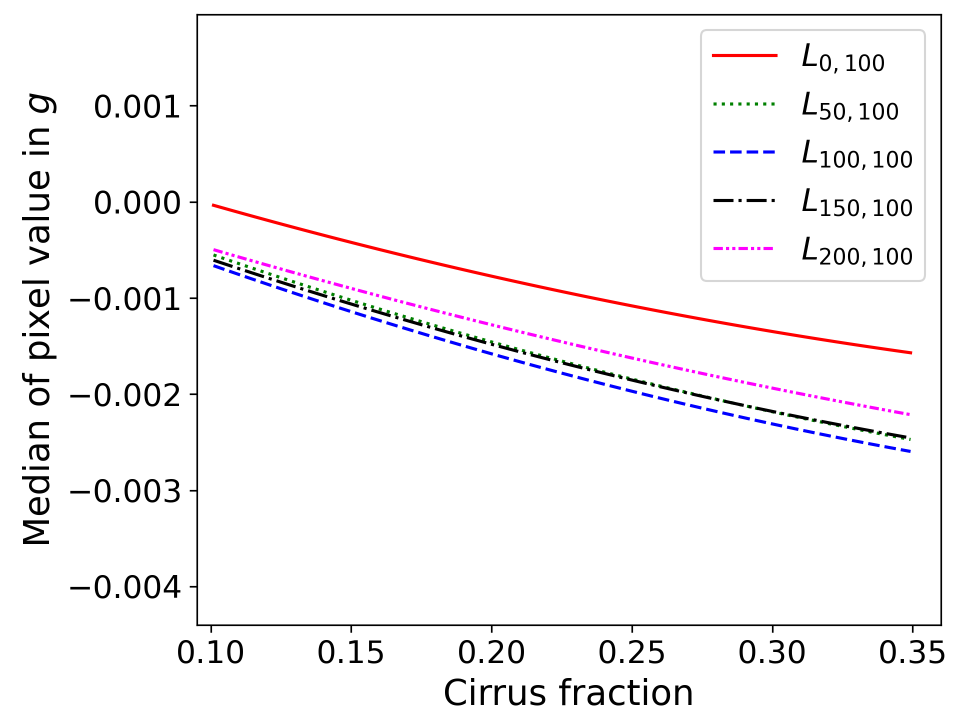}%
    \includegraphics[width=0.265\linewidth]{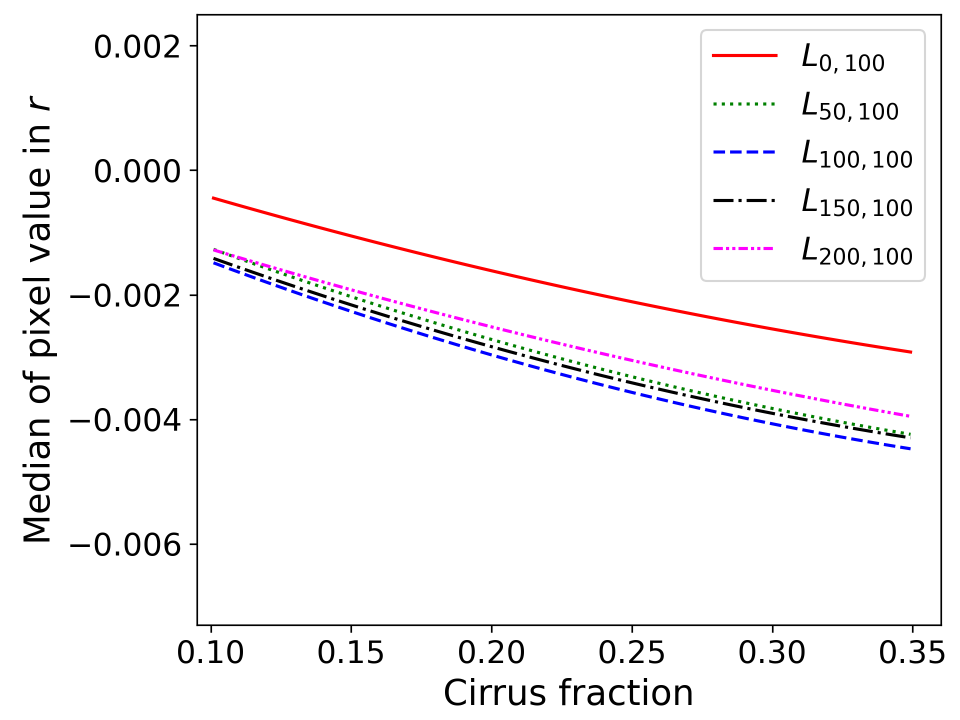}
    \includegraphics[width=0.265\linewidth]{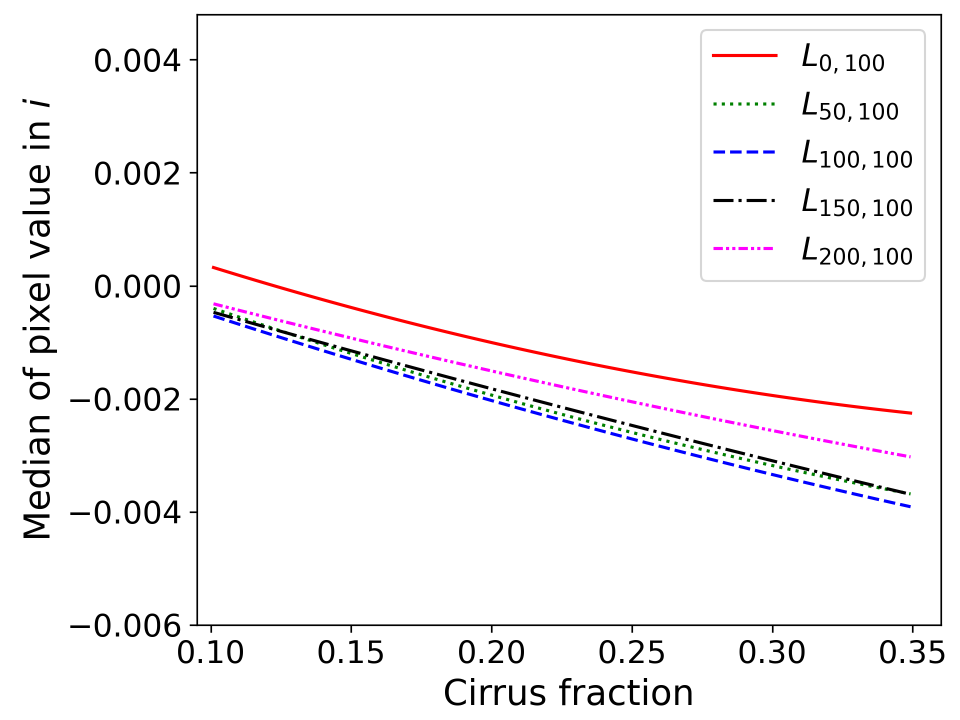} \\
\caption{The top panel contains scatter plots of the median values in the boundary layers with a thickness of 100 pixels ($L_{0, 100}$, $L_{50, 100}$, $L_{100, 100}$, $L_{150, 100}$, $L_{200, 100}$) in the $g$, $r$ and $i$ optical bands (from left to right) for fields in the Intersection region, where the fraction of cirrus is greater than 0.1. The bottom panel contains corresponding approximations by polynomials of the second degree.}
\label{fig:scatterplots_and_approx_dist}
\end{figure*}

\begin{figure*}[h!]
    \centering
    \includegraphics[width=0.265\linewidth]{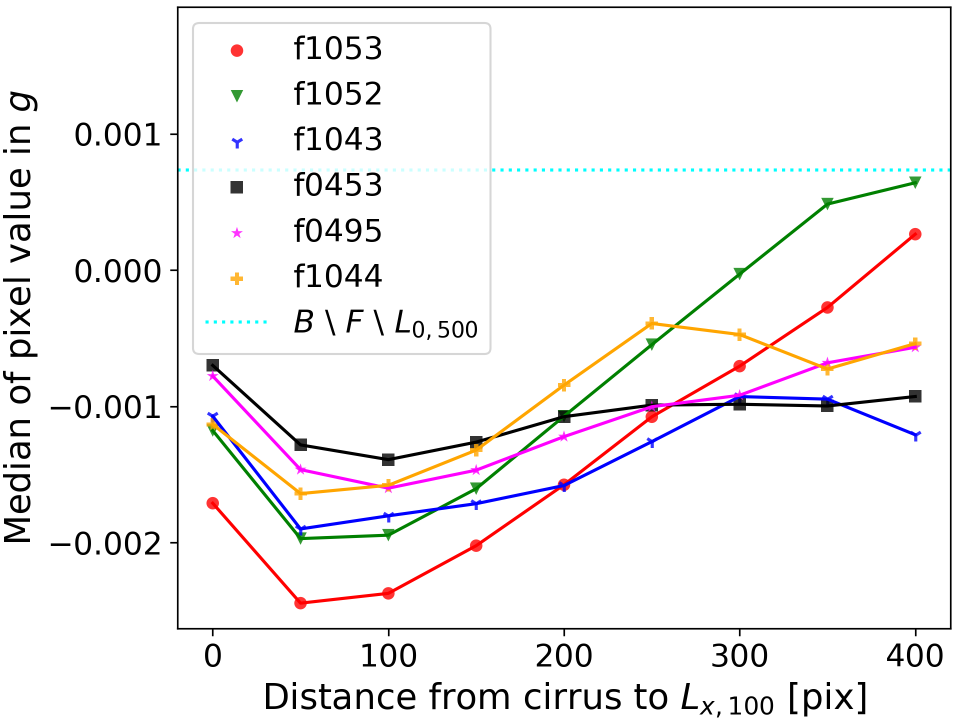}%
    \includegraphics[width=0.265\linewidth]{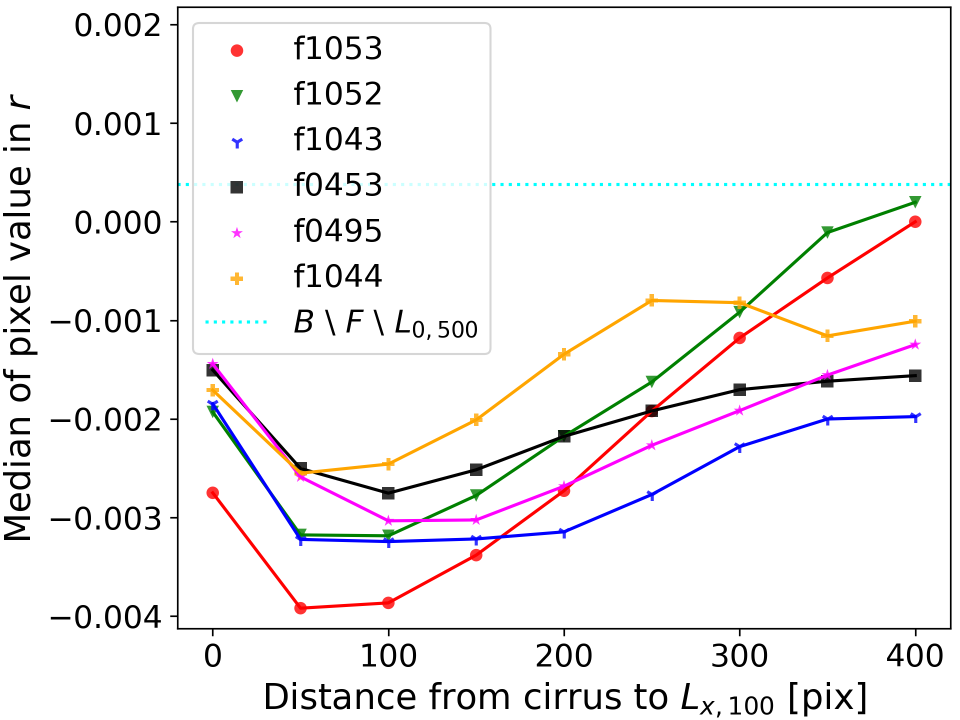}
    \includegraphics[width=0.265\linewidth]{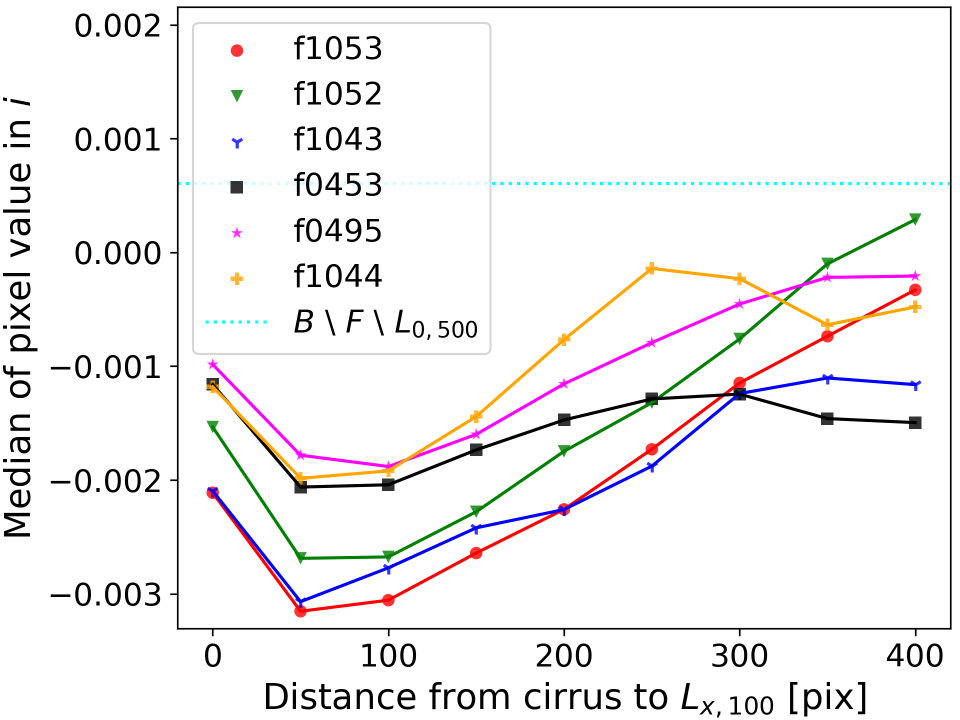}
\caption{Median values in the boundary layers $L_{x, 100}$ in the $g$, $r$ and $i$ optical bands (from left to right) for six fields f1053, f1052, f1043, f0453, f0495, and f1044 are represented by different colored dots. The horizontal dotted line denotes the average of median values of the background without the cirrus and the layer $L_{0, 500}$ for the six fields considered.}
\label{fig:6_fields_layers}
\end{figure*}

Next, we examined the relationship between the fraction of cirrus in a field and the degree of over-subtraction in the boundary layer, represented by $m(L_{d, t})$. For the 98 previously analyzed fields, we calculated the median pixel value in the $g$, $r$ and $i$ bands for boundary layers with a thickness of 100 pixels: $L_{0, 100}$, $L_{50, 100}$, $L_{100, 100}$, $L_{150, 100}$, $L_{200, 100}$. The scatter plots of these values are presented in the top panel of Fig.~\ref{fig:scatterplots_and_approx_dist}. The corresponding second-degree polynomial approximations are shown in the bottom panel. As evident from these graphs, the regions experiencing the most significant over-subtraction are concentrated approximately 100–200 pixels away from cirrus filaments in each band. Additionally, it is noteworthy that a higher cirrus fraction anti correlates with an increase in $m(L_{d, t})$ across all analyzed layers.
To further illustrate the presence of an over-subtracted region near cirrus filaments, we analyzed the median values in layers $L_{x, 100}$ for six cirrus-rich fields: f1053, f1052, f1043, f0453, f0495, and f1044. The cirrus fraction decreases from $0.209$ in f1053 to $0.161$ in f1044. These median values are presented in Fig.~\ref{fig:6_fields_layers}. For comparison, a horizontal reference line is included in the graph, representing the average of the median background values (excluding cirrus and the layer $L_{0, 500}$) across the six selected fields. The graphs indicate that the most over-subtracted layers are those located 50--150 pixels away from cirrus filaments in these fields.

\begin{figure*}[h!]
    \centering
    \includegraphics[width=0.265\linewidth]{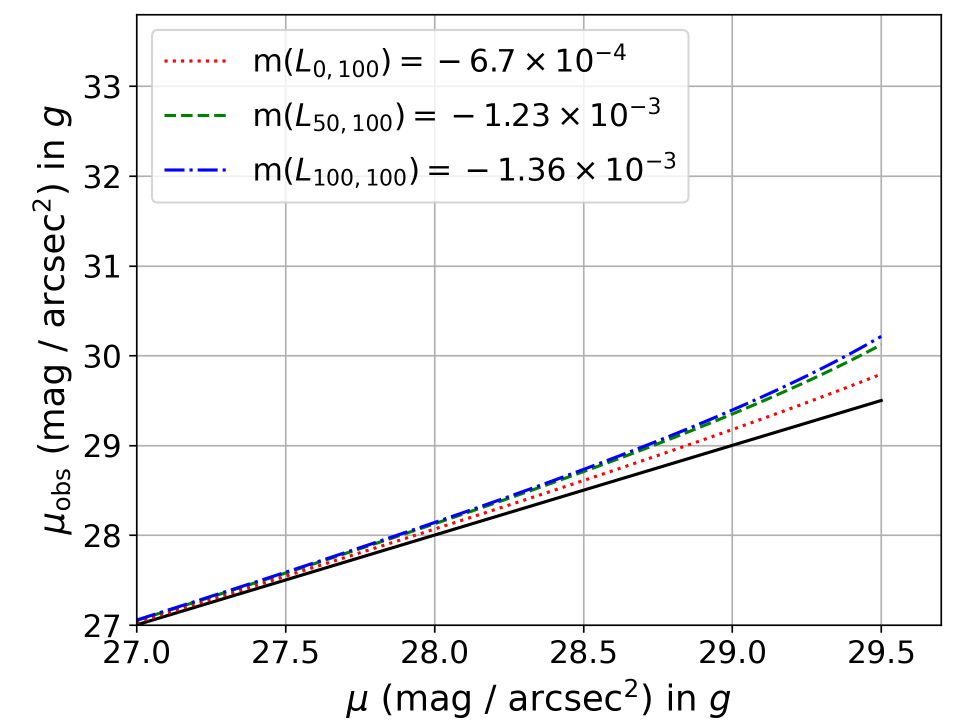}%
    \includegraphics[width=0.265\linewidth]{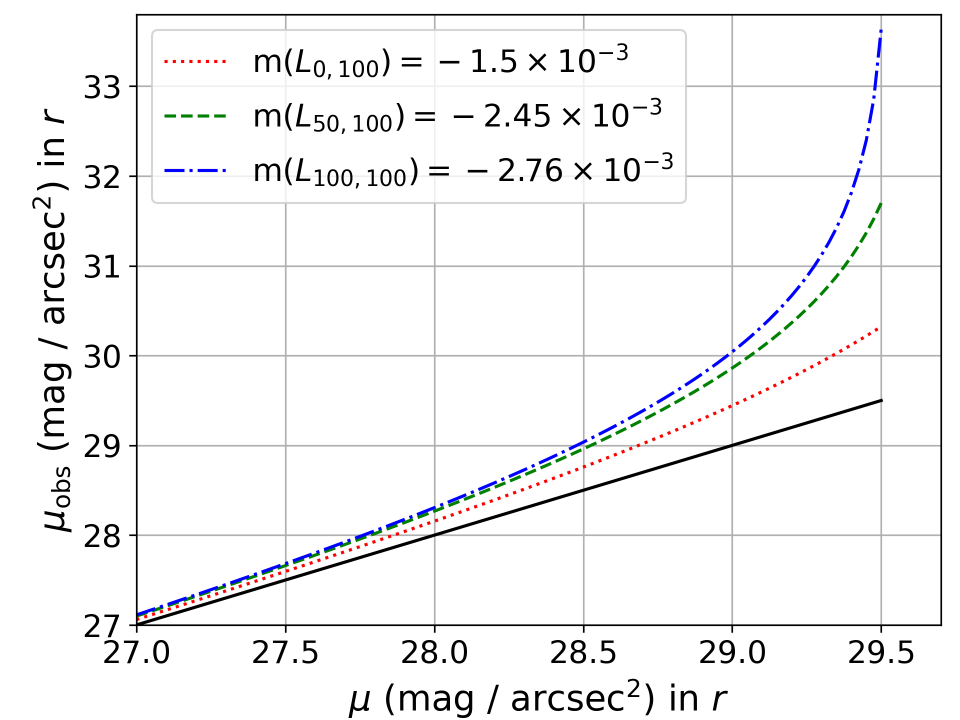}
    \includegraphics[width=0.265\linewidth]{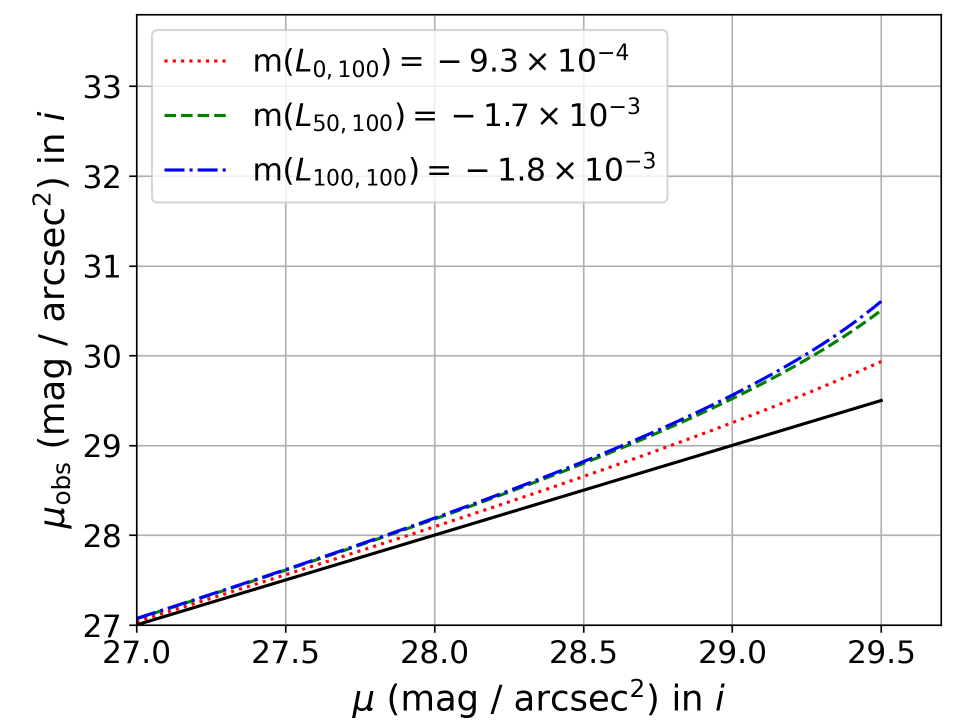}
\caption{Bias in the observed surface brightness of the source due to over-subtraction of various values in the $g$, $r$ and $i$ optical bands (from left to right). The bias values correspond to the medians of the pixel value for different boundary layers: $L_{0,100}$, $L_{50,100}$ and $L_{100,100}$. The ground-truth surface brightness of the source is plotted along the x-axis. Biased observed surface brightness is plotted along the y-axis. Solid lines correspond to the 1:1 line.}
\label{fig:surface_mag_corrections}
\end{figure*}

Thus, the magnitude of the over-subtraction depends on both the cirrus fraction and its distance, and we assume it is also influenced by the intrinsic properties of the cirrus itself.
To evaluate the impact of over-subtraction, we estimate how much the surface brightness of a source would change, assuming that the over-subtraction is uniform and equal to the median of pixel value calculated for the 98 fields discussed above for the various $L_{x, 100}$ layers. This effect is demonstrated in Fig.~\ref{fig:surface_mag_corrections} in the $g$, $r$ and $i$ bands.
As evident from the graph for the $r$ band, even when assessing over-subtraction using the lowest median values of $m(L_{0, 100})$, it results in a surface brightness bias of $0.5$ magnitudes for objects with a surface brightness of $m = 29~\mbox{mag~arcsec}^{-2}$in the $r$ band.
It is worth mentioned that medians of pixel value in $g$ and $i$ bands are sufficiently larger than for the $r$ band. This difference may be attributed to the circumstance that the cirrus maps are more precisely matched to the cirrus filaments in the $r$ band because the training sample of the filaments was prepared using only $r$-band images.
Although these estimates are approximate, it is evident that large cirrus filaments significantly impact the accuracy of sky background subtraction. Therefore, their presence should be taken into account in algorithms for sky background estimation. The cirrus maps we have generated can aid in refining these techniques for both existing and upcoming deep optical surveys, such as Euclid \citep{Euclid_2011} or the Vera C. Rubin Observatory \citep[former LSST,][]{LSST_2009}.

Unfortunately, sky background over-subtraction can also affect the visible boundaries of cirrus filaments, which can affect the reliability of cirrus maps. It is difficult to determine how much this influence is, as well as the value of sky background over-subtraction. However, the background of the sky in the HSC-SSP DR3 is taken into account on a large scale for ``global-sky'' coadd images ($8\mbox{K} \times 8\mbox{K}$ superpixels was used to estimate it). Therefore, the value of sky background over-subtraction also varies on a large scale. If this scale exceeds the flux change scale at the boundaries of filaments, sky over-subtraction does not significantly affect the location of the visible boundaries of cirrus clouds.

An indirect way to verify the reliability of finding the boundaries of cirrus filaments is to compare them with data from other surveys. However, it is worth noting that the surface brightness limit of survey images is very important for determining the visible position of filament boundaries, as well as for the boundaries of galaxies. The difference in the cirrus fraction in the Intersection region obtained from HSC-SSP DR3 and SDSS Stripe~82 ($f_{\mathrm{HSC-SSP}}=0.0408$, $f_{\mathrm{S82}} = 0.0092$) confirms the importance of the surface brightness limit of images. Thus, the best comparison would be with a survey of the same surface brightness limit.

\section{Summary and conclusions}
\label{sec:summary}
In this study, we have analyzed cirri in the HSC-SSP DR3 data as a continuation of  \citet{Cirrus_2023MNRAS.519.4735S}, which investigated the color properties of cirrus using Stripe~82 data. In \citet{Cirrus_2023MNRAS.519.4735S}, a catalog of cirrus filaments was created using a neural network trained on a dataset of semi-automatically generated cirrus maps, with the best model achieving an IoU of $0.576$ for cirrus. Here, we adopted a similar neural network architecture to identify cirrus filaments in the new HSC-SSP data.

Since individual models exhibited a lower IoU for cirrus than in \citet{Cirrus_2023MNRAS.519.4735S} ($\mbox{IoU} \leq 0.436$), we employed an ensemble of the nine best 4-channel models to generate the cirrus maps and catalog. Using this ensemble improved the IoU to $0.48$.
After compiling the catalog of cirrus filaments based on the HSC-SSP data, we compared our results with those from \citet{Cirrus_2023MNRAS.519.4735S}.
For 613 fields in the Intersection region, we obtained cirrus maps from both the Stripe~82 and HSC-SSP datasets. These maps demonstrate high consistency, despite the deeper HSC-SSP dataset revealing $4.5$ times more cirrus clouds. The obtained cirrus maps for various regions of HSC-SSP are presented in Fig.~\ref{fig:map1} and Fig.~\ref{fig:map2}. The cirrus filaments in the Spring, Fall plus, Fall minus, and North regions are studied for the first time in the present work. For all regions, the typical size of identified filaments is about 0.5--1.0 arcmin. The value is in agreement with the results of our previous work, where only Stripe~82 data was considered~\citep{Cirrus_2023MNRAS.519.4735S}. The total area of identified cirri, summed across all the HSC-SSP regions analyzed, is approximately 13 deg$^2$.

The cirrus filament catalogs and the rectified cirrus maps that we have created are publicly available in our \texttt{GitLab} repository\footnote{\url{https://gitlab.com/polyakovdmi93/cirrus_narrow_segmentation}}.

Through several tests, we revealed a discrepancy in the surface brightness of filaments between Stripe~82 and HSC-SSP data. This suggests that the background subtraction procedure in HSC-SSP results in the over-subtraction of cirrus fluxes, and possibly other faint sources. Using our cirrus maps, we examined the surface brightness in regions adjacent to the filaments and provided a rough estimate of the over-subtraction effect.


This study has validated the cirrus map creation approach proposed in \citet{Cirrus_2023MNRAS.519.4735S} and enhanced it through ensemble learning. We confirm the importance of accounting for cirrus in background estimation algorithms. Moving forward, we plan to refine our method for generating cirrus maps and extend its application to upcoming deep optical surveys.



\section*{Acknowledgements}
We thank the anonymous referees by their detailed revision of the manuscript that helped to improve its quality.

We acknowledge financial support from the Russian Science Foundation (grant no. 20-72-10052).


The Hyper Suprime-Cam (HSC) collaboration includes the astronomical communities of Japan and Taiwan, and Princeton University. The HSC instrumentation and software were developed by the National Astronomical Observatory of Japan (NAOJ), the Kavli Institute for the Physics and Mathematics of the Universe (Kavli IPMU), the University of Tokyo, the High Energy Accelerator Research Organization (KEK), the Academia Sinica Institute for Astronomy and Astrophysics in Taiwan (ASIAA), and Princeton University. Funding was contributed by the FIRST program from the Japanese Cabinet Office, the Ministry of Education, Culture, Sports, Science and Technology (MEXT), the Japan Society for the Promotion of Science (JSPS), the Japan Science and Technology Agency(JST), the Toray Science Foundation, NAOJ, Kavli IPMU, KEK, ASIAA, and Princeton University. This paper makes use of software developed for the Large Synoptic Survey Telescope. We thank the LSST Project for making their code available as free software at http://dm.lsst.org. The Pan-STARRS1 Surveys have been made possible through contributions of the Institute for Astronomy, the University of Hawaii, the Pan-STARRS Project Office, the Max-Planck Society and its participating institutes, the Max Planck Institute for Astronomy, Heidelberg and the Max Planck Institute for Extraterrestrial Physics, Garching, The Johns Hopkins University, Durham University, the University of Edinburgh, Queen’s University Belfast, the Harvard-Smithsonian Center for Astrophysics, the Las Cumbres Observatory Global Telescope Network Incorporated, the National Central University of Taiwan, the Space Telescope Science Institute, the National Aeronautics and Space Administration under Grant No. NNX08AR22G issued through the Planetary Science Division of the NASA Science Mission Directorate, the National Science Foundation under Grant No. AST-1238877, the University of Maryland, Eotvos Lorand University (ELTE), and the Los Alamos National Laboratory. This paper is based on data collected at the Subaru Telescope and retrieved from the HSC data archive system, which is operated by the Subaru Telescope and Astronomy Data Center at NAOJ.



\appendix
\section{Surface brightness limits of HSC-SSP}
\label{app_sec:sb_limit}
There, the process of obtaining surface brightness limits of HSC-SSP coadd images is described. In this paper, we define this value in the same way as \citet{Stripe82_2020A&A...644A..42R} did. The surface brightness limit at $3 \sigma$ level for a source with an angular size of $10'' \times 10''$ can be defined as follows:
$$
    \mu_{\mathrm{lim}}\left(3 \sigma; 10'' \times 10''\right) = m_{0} - 2.5 \lg(3 \sigma),
$$
where $\sigma$ is the standard deviation of the background flux in a random box with an angular size of $10'' \times 10''$, and $m_{0}$ is the photometric zero-point of the data in the optical band under consideration. The coadd images have a common photometric zero-point of $27.0$ in all bands in the HSC-SSP.

To compute $ \mu_{\mathrm{lim}}\left(3 \sigma; 10'' \times 10''\right)$ we directly estimate $\sigma$. For this procedure, we use all fields that we consider ($4331$ fields). For each field, we place $10 000$ random boxes with a size of $10'' \times 10''$, which do not intersect with the combined mask. Then, for each box, we calculate the standard deviation of the flux. This value is recalculated into stellar magnitudes using the photometric zero-point $m_0$. Averaging this value over all $10000$ boxes gives the surface brightness limit for the field. Then, averaging this value for all $4331$ fields, we obtain the surface brightness limit for the HSC-SSP in the optical band under consideration.

\section{Validation loss curves across 100 training epochs}
\label{app_sec:loss_curves}
One of the main stages of this work is training artificial neural networks to create cirrus maps. Finding optimal parameters for the network architecture and learning process requires repeatedly training various models. Since our computing resources were limited to a single NVIDIA GeForce RTX 3060 GPU graphics card, we needed to speed up the learning process of our models.

\begin{figure*}[h!]
    \centering
    \includegraphics[width=0.45\linewidth]{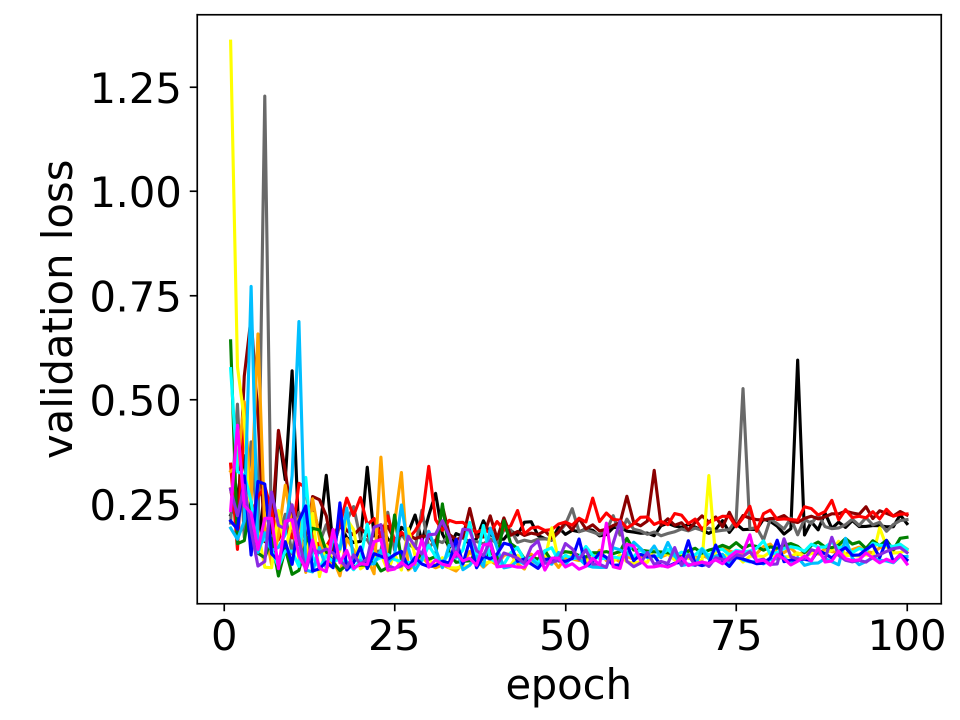} ~~
    \includegraphics[width=0.45\linewidth]{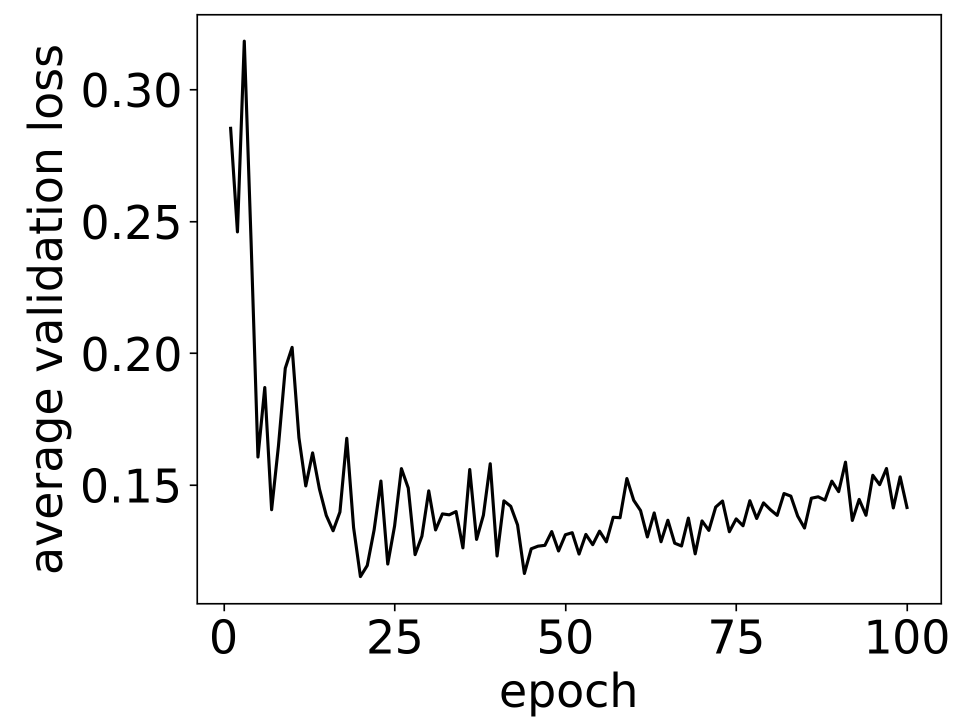}
\caption{The left panel shows validation loss curves across $100$ epochs for $12$ $4$-channel models. Validation loss curves of different colors correspond to different models. The corresponding average loss curve is presented in the right panel.}
\label{fig:loss_curves}
\end{figure*}

The simplest way, without affecting the network architecture or optimization process, is to reduce the number of training epochs. To make sure that reducing the number of training epochs does not significantly affect the final result, we trained 12 4-channel models. There are presented validation loss curves for these models across 100 training epochs. As can be clearly seen from Fig.~\ref{fig:loss_curves}, the loss curves reach a plateau by the 20th epoch. This allows us to train models for 20 epochs in order to save time.

\section{Metrics and parameters of cirrus segmentation models}
\label{app_sec:models_metrics}

Section~\ref{sub_sec:nn_results} discusses the results of experiments on training neural network models to create cirrus maps.

The main results of the network training are presented here in the form of tables. Each model in these tables is represented by three parameters of the learning process: class weights ($\overline{\omega_{\mathrm{c}}}$), the spatial size of the input tensor (shape), the ratio of the window size to the spatial size of the input tensor (scale), the learning rate ($r$) and metrics on the test sample: intersection over union \citep[IoU,][]{IoU1901}, precision and recall for cirrus class. IoU metric measures the extent of correspondence between the predicted and actual cirrus map.

The metrics presented in Tables~\ref{tab:3D_nn_metrics} and \ref{tab:4D_nn_metrics} allow one to compare the quality of 3-channel and 4-channel models. Table~\ref{tab:fine-tuning_nn_metrics} shows the results of an experiment on fine-tuning the best models from \citet{Cirrus_2023MNRAS.519.4735S}.

\begin{table*}[ht]
	\caption{The results of the training experiments conducted on 3-channel models. It lists class weights for background and cirrus ($\overline{\omega_{\mathrm{c}}}$), input tensor spatial shape, scale factor, learning rate ($r$) and IoU, precision, recall for all test fields for cirrus class.}
	\centering
    \begin{tabular}{c c c c c c }
        \hline \hline
        $\overline{\omega_{\mathrm{c}}}$  & shape / scale  & $r$ & IoU & precision & recall \\
\hline
$(1.0, 15.0)$  & $448\times448$ / 2&$0.0005$&$0.381$ &$0.573$ &$0.532$  \\
$(1.0, 15.0)$  & $448\times448$ / 2&$0.0005$&$0.399$ &$0.614$ &$0.533$  \\
$(1.0, 15.0)$  & $448\times448$ / 2&$0.001$ &$0.344$ &$0.607$ &$0.442$  \\
$(1.0, 15.0)$  & $448\times448$ / 2&$0.001$ &$0.416$ &$0.678$ &$0.518$  \\
\hline
$(1.0, 15.0)$  & $448\times448$ / 4&$0.0005$&$0.389$ &$0.502$ &$0.634$  \\
$(1.0, 15.0)$  & $448\times448$ / 4&$0.0005$&$0.385$ &$0.67$  &$0.475$  \\
$(1.0, 15.0)$  & $448\times448$ / 4&$0.001$ &$0.355$ &$0.537$ &$0.512$  \\
$(1.0, 15.0)$  & $448\times448$ / 4&$0.001$ &$0.397$ &$0.582$ &$0.556$  \\
\hline
$(1.0, 15.0)$  & $448\times448$ / 8&$0.0005$&$0.348$ &$0.432$ &$0.641$  \\
$(1.0, 15.0)$  & $448\times448$ / 8&$0.0005$&$0.281$ &$0.385$ &$0.509$  \\
$(1.0, 15.0)$  & $448\times448$ / 8&$0.001$ &$0.298$ &$0.536$ &$0.402$  \\
$(1.0, 15.0)$  & $448\times448$ / 8&$0.001$ &$0.334$ &$0.645$ &$0.409$  \\
    \end{tabular}
    \label{tab:3D_nn_metrics}
\end{table*}

\begin{table*}[ht]
	\caption{The results of the training experiments conducted on 4-channel models. It lists class weights for background and cirrus ($\overline{\omega_{\mathrm{c}}}$), input tensor spatial shape, scale factor, learning rate ($r$) and IoU, precision, recall for all tests fields for cirrus class. The last column shows the rank of the model, if it is among the top 9 4-channel models based on the IoU metric.}
	\centering
    \begin{tabular}{c c c c c c c}
        \hline \hline
        $\overline{\omega_{\mathrm{c}}}$  & shape / scale  & $r$ & IoU & precision & recall & Rank among the top 9 models \\
\hline
$(1.0, 10.0)$  & $448\times448$ / 2&$0.0005$&$0.317$ &$0.641$ &$0.385$ &      \\
$(1.0, 10.0)$  & $448\times448$ / 2&$0.0005$&$0.29$  &$0.665$ &$0.34$  &      \\
$(1.0, 10.0)$  & $448\times448$ / 2&$0.001$ &$0.436$ &$0.569$ &$0.651$ &$1$    \\
$(1.0, 10.0)$  & $448\times448$ / 2&$0.001$ &$0.322$ &$0.791$ &$0.352$ &      \\
\hline
$(1.0, 10.0)$  & $448\times448$ / 4&$0.0005$&$0.345$ &$0.653$ &$0.423$ &      \\
$(1.0, 10.0)$  & $448\times448$ / 4&$0.0005$&$0.395$ &$0.634$ &$0.511$ &      \\
$(1.0, 10.0)$  & $448\times448$ / 4&$0.001$ &$0.362$ &$0.555$ &$0.51$  &      \\
$(1.0, 10.0)$  & $448\times448$ / 4&$0.001$ &$0.424$ &$0.672$ &$0.535$ &$2$    \\
\hline
$(1.0, 10.0)$  & $448\times448$ / 8&$0.0005$&$0.379$ &$0.543$ &$0.556$ &      \\
$(1.0, 10.0)$  & $448\times448$ / 8&$0.0005$&$0.391$ &$0.554$ &$0.57$  &      \\
$(1.0, 10.0)$  & $448\times448$ / 8&$0.001$ &$0.386$ &$0.566$ &$0.549$ &      \\
$(1.0, 10.0)$  & $448\times448$ / 8&$0.001$ &$0.356$ &$0.477$ &$0.584$ &      \\
$(1.0, 10.0)$  & $448\times448$ / 8&$0.005$ &$0.424$ &$0.565$ &$0.629$ &$3$    \\
$(1.0, 10.0)$  & $448\times448$ / 8&$0.005$ &$0.373$ &$0.547$ &$0.54$  &      \\
\hline
\hline
$(1.0, 15.0)$  & $448\times448$ / 2&$0.0005$&$0.404$ &$0.51$  &$0.66$  &$7$    \\
$(1.0, 15.0)$  & $448\times448$ / 2&$0.0005$&$0.391$ &$0.649$ &$0.496$ &$ $      \\
$(1.0, 15.0)$  & $448\times448$ / 2&$0.001$ &$0.383$ &$0.527$ &$0.583$ &$ $      \\
$(1.0, 15.0)$  & $448\times448$ / 2&$0.001$ &$0.275$ &$0.828$ &$0.292$ &$ $      \\
\hline
$(1.0, 15.0)$  & $448\times448$ / 4&$0.0005$&$0.375$ &$0.574$ &$0.519$ &$ $      \\
$(1.0, 15.0)$  & $448\times448$ / 4&$0.0005$&$0.4$   &$0.613$ &$0.536$ &$ $      \\
$(1.0, 15.0)$  & $448\times448$ / 4&$0.001$ &$0.386$ &$0.563$ &$0.552$ &$ $      \\
$(1.0, 15.0)$  & $448\times448$ / 4&$0.001$ &$0.403$ &$0.549$ &$0.604$ &$8$    \\
\hline
$(1.0, 15.0)$  & $448\times448$ / 8&$0.0005$&$0.347$ &$0.566$ &$0.472$ &$ $      \\
$(1.0, 15.0)$  & $448\times448$ / 8&$0.0005$&$0.304$ &$0.511$ &$0.429$ &$ $      \\
$(1.0, 15.0)$  & $448\times448$ / 8&$0.001$ &$0.35$  &$0.679$ &$0.42$  &$ $      \\
$(1.0, 15.0)$  & $448\times448$ / 8&$0.001$ &$0.338$ &$0.547$ &$0.47$  &$ $      \\
\hline
$(1.0, 20.0)$  & $448\times448$ / 2&$0.0005$&$0.422$ &$0.653$ &$0.545$ &$4$    \\
$(1.0, 20.0)$  & $448\times448$ / 2&$0.0005$&$0.413$ &$0.571$ &$0.598$ &$6$    \\
$(1.0, 20.0)$  & $448\times448$ / 2&$0.001$ &$0.394$ &$0.527$ &$0.609$ &      \\
$(1.0, 20.0)$  & $448\times448$ / 2&$0.001$ &$0.337$ &$0.421$ &$0.627$ &      \\
\hline
$(1.0, 20.0)$  & $448\times448$ / 4&$0.0005$&$0.244$ &$0.604$ &$0.291$ &      \\
$(1.0, 20.0)$  & $448\times448$ / 4&$0.0005$&$0.375$ &$0.504$ &$0.593$ &      \\
$(1.0, 20.0)$  & $448\times448$ / 4&$0.001$ &$0.312$ &$0.53$  &$0.431$ &      \\
$(1.0, 20.0)$  & $448\times448$ / 4&$0.001$ &$0.33$  &$0.613$ &$0.417$ &      \\
\hline
$(1.0, 20.0)$  & $448\times448$ / 8&$0.0005$&$0.323$ &$0.468$ &$0.51$  &      \\
$(1.0, 20.0)$  & $448\times448$ / 8&$0.0005$&$0.401$ &$0.504$ &$0.666$ &$9$    \\
$(1.0, 20.0)$  & $448\times448$ / 8&$0.001$ &$0.419$ &$0.562$ &$0.621$ &$5$    \\
$(1.0, 20.0)$  & $448\times448$ / 8&$0.001$ &$0.363$ &$0.479$ &$0.601$ &      \\
    \end{tabular}
    \label{tab:4D_nn_metrics}
\end{table*}

\begin{table*}[ht]
	\caption{The results of fine-tuning the best model from \citet{Cirrus_2023MNRAS.519.4735S} using HSC-SSP data. It lists class weights for background and cirrus ($\overline{\omega_{\mathrm{c}}}$), input tensor spatial shape, scale factor, learning rate ($r$) and IoU, precision, recall for all test fields for cirrus class.}
	\centering
    \begin{tabular}{c c c c c c }
        \hline \hline
        $\overline{\omega_{\mathrm{c}}}$  & shape / scale  & $r$ & IoU & precision & recall \\
\hline
$(1.0, 15.0)$  &$448\times448$ / 2&$0.0001$&$0.364$ &$0.583$ &$0.492$  \\
$(1.0, 15.0)$  &$448\times448$ / 2&$0.0001$&$0.383$ &$0.496$ &$0.628$  \\
$(1.0, 15.0)$  &$448\times448$ / 2&$0.00025$   &$0.348$ &$0.525$ &$0.508$  \\
$(1.0, 15.0)$  &$448\times448$ / 2&$0.00025$   &$0.359$ &$0.521$ &$0.536$  \\
$(1.0, 15.0)$  &$448\times448$ / 2&$0.0005$&$0.377$ &$0.589$ &$0.512$  \\
$(1.0, 15.0)$  &$448\times448$ / 2&$0.0005$&$0.388$ &$0.67$  &$0.48$   \\
$(1.0, 15.0)$  &$448\times448$ / 2&$0.001$ &$0.338$ &$0.56$  &$0.46$   \\
$(1.0, 15.0)$  &$448\times448$ / 2&$0.001$ &$0.372$ &$0.437$ &$0.716$  \\
\hline
$(1.0, 15.0)$  &$448\times448$ / 3&$0.0001$&$0.342$ &$0.564$ &$0.465$  \\
$(1.0, 15.0)$  &$448\times448$ / 3&$0.0001$&$0.349$ &$0.539$ &$0.498$  \\
$(1.0, 15.0)$  &$448\times448$ / 3&$0.00025$   &$0.325$ &$0.487$ &$0.495$  \\
$(1.0, 15.0)$  &$448\times448$ / 3&$0.00025$   &$0.384$ &$0.494$ &$0.633$  \\
$(1.0, 15.0)$  &$448\times448$ / 3&$0.0005$&$0.301$ &$0.498$ &$0.432$  \\
$(1.0, 15.0)$  &$448\times448$ / 3&$0.0005$&$0.377$ &$0.557$ &$0.539$  \\
$(1.0, 15.0)$  &$448\times448$ / 3&$0.001$ &$0.287$ &$0.582$ &$0.362$  \\
$(1.0, 15.0)$  &$448\times448$ / 3&$0.001$ &$0.317$ &$0.471$ &$0.493$  \\
    \end{tabular}
    \label{tab:fine-tuning_nn_metrics}
\end{table*}


\section{Surface brightness of filaments depending on the area}
\label{app_sec:noise}
\begin{figure*}[ht]
    \centering
    \includegraphics[width=0.265\linewidth]{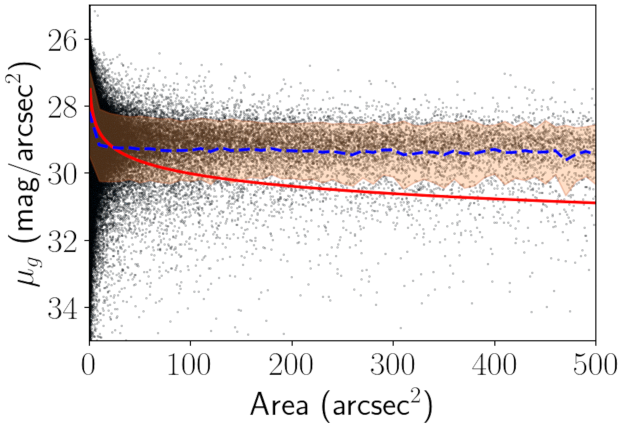}%
    \includegraphics[width=0.265\linewidth]{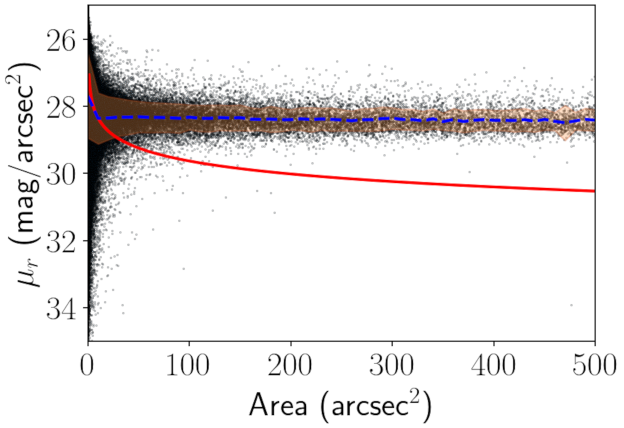}
    \includegraphics[width=0.265\linewidth]{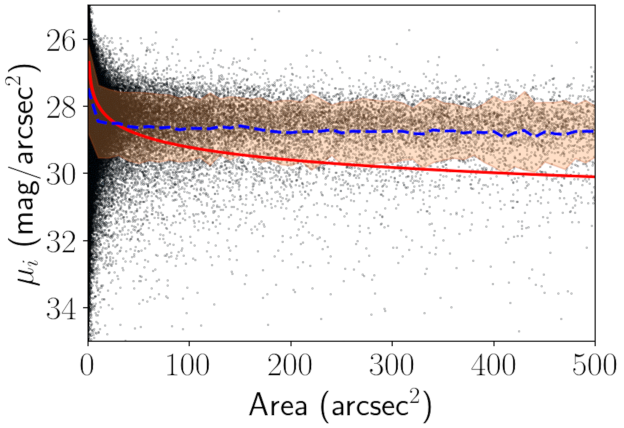} \\

    \includegraphics[width=0.265\linewidth]{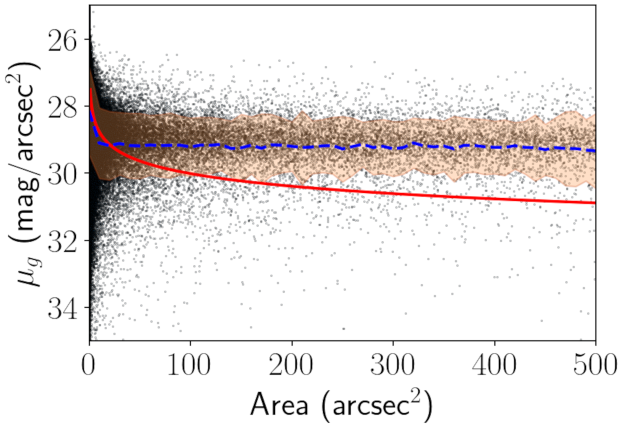}%
    \includegraphics[width=0.265\linewidth]{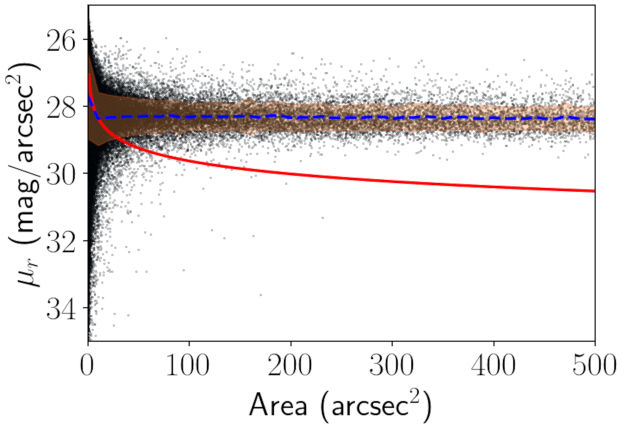}
    \includegraphics[width=0.265\linewidth]{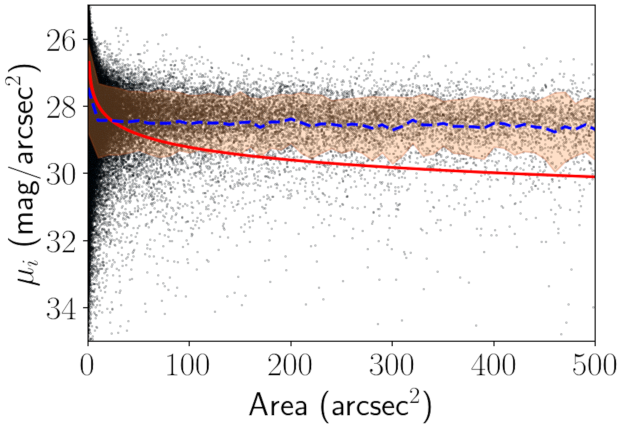} \\

    \includegraphics[width=0.265\linewidth]{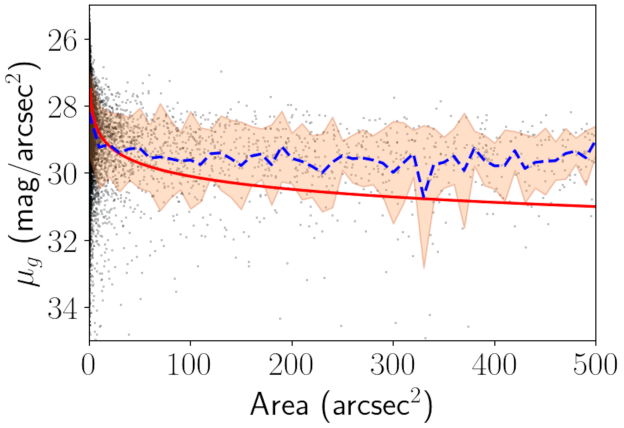}%
    \includegraphics[width=0.265\linewidth]{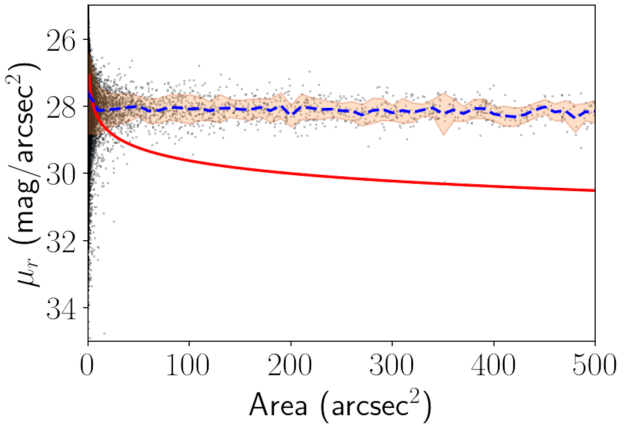}
    \includegraphics[width=0.265\linewidth]{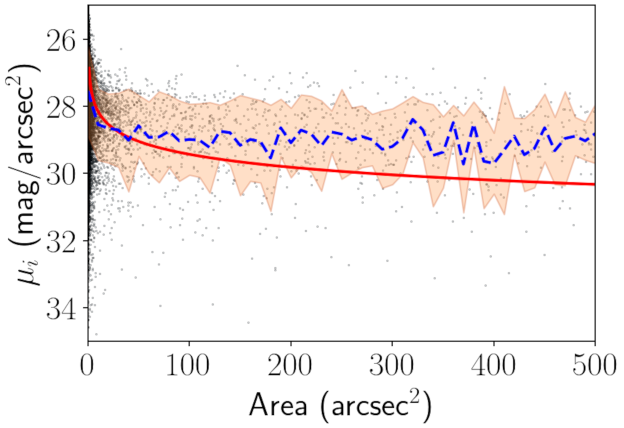} \\

    \includegraphics[width=0.265\linewidth]{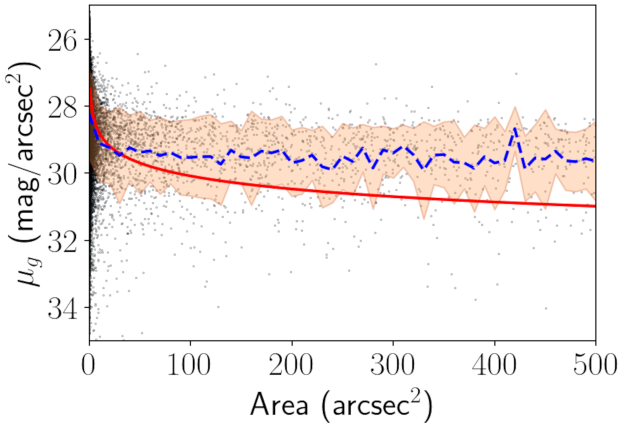}%
    \includegraphics[width=0.265\linewidth]{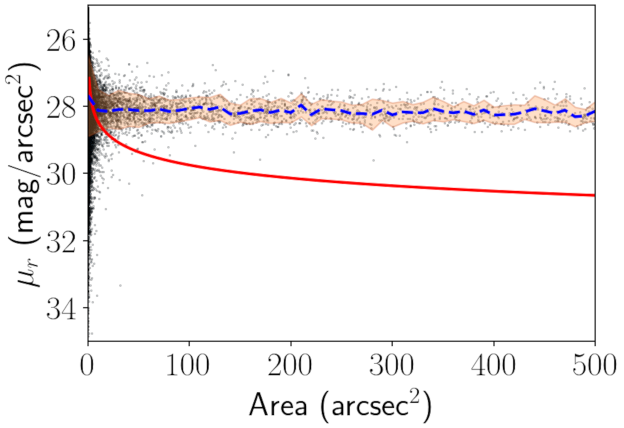}
    \includegraphics[width=0.265\linewidth]{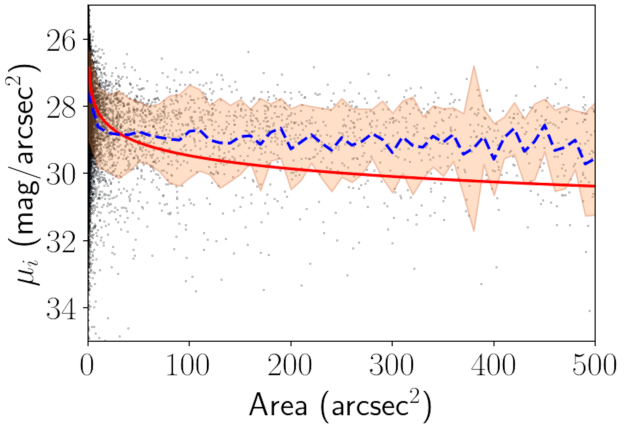}\\
\caption{Surface brightness of filaments in the Intersection, Fall plus, Fall minus, and North regions (from top to bottom) depending on the area in the $g$, $r$, and $i$ optical bands (from left to right). A red line in each subplot marks the $3$-sigma limit for the corresponding band. A blue dashed line shows the average value while the shaded area corresponds to $1$-sigma limits.}    
\label{fig:app_noise}
\end{figure*}

\clearpage

\bibliographystyle{elsarticle-harv} 
\bibliography{main}

@PREAMBLE{ {\providecommand{\noopsort}[1]{}} }

@ARTICLE{HSC-SSP_DR1_2018PASJ...70S...8A,
       author = {{Aihara}, Hiroaki and {Armstrong}, Robert and {Bickerton}, Steven and {Bosch}, James and {Coupon}, Jean and {Furusawa}, Hisanori and {Hayashi}, Yusuke and {Ikeda}, Hiroyuki and {Kamata}, Yukiko and {Karoji}, Hiroshi and {Kawanomoto}, Satoshi and {Koike}, Michitaro and {Komiyama}, Yutaka and {Lang}, Dustin and {Lupton}, Robert H. and {Mineo}, Sogo and {Miyatake}, Hironao and {Miyazaki}, Satoshi and {Morokuma}, Tomoki and {Obuchi}, Yoshiyuki and {Oishi}, Yukie and {Okura}, Yuki and {Price}, Paul A. and {Takata}, Tadafumi and {Tanaka}, Manobu M. and {Tanaka}, Masayuki and {Tanaka}, Yoko and {Uchida}, Tomohisa and {Uraguchi}, Fumihiro and {Utsumi}, Yousuke and {Wang}, Shiang-Yu and {Yamada}, Yoshihiko and {Yamanoi}, Hitomi and {Yasuda}, Naoki and {Arimoto}, Nobuo and {Chiba}, Masashi and {Finet}, Francois and {Fujimori}, Hiroki and {Fujimoto}, Seiji and {Furusawa}, Junko and {Goto}, Tomotsugu and {Goulding}, Andy and {Gunn}, James E. and {Harikane}, Yuichi and {Hattori}, Takashi and {Hayashi}, Masao and {He{\l}miniak}, Krzysztof G. and {Higuchi}, Ryo and {Hikage}, Chiaki and {Ho}, Paul T.~P. and {Hsieh}, Bau-Ching and {Huang}, Kuiyun and {Huang}, Song and {Imanishi}, Masatoshi and {Iwata}, Ikuru and {Jaelani}, Anton T. and {Jian}, Hung-Yu and {Kashikawa}, Nobunari and {Katayama}, Nobuhiko and {Kojima}, Takashi and {Konno}, Akira and {Koshida}, Shintaro and {Kusakabe}, Haruka and {Leauthaud}, Alexie and {Lee}, Chien-Hsiu and {Lin}, Lihwai and {Lin}, Yen-Ting and {Mandelbaum}, Rachel and {Matsuoka}, Yoshiki and {Medezinski}, Elinor and {Miyama}, Shoken and {Momose}, Rieko and {More}, Anupreeta and {More}, Surhud and {Mukae}, Shiro and {Murata}, Ryoma and {Murayama}, Hitoshi and {Nagao}, Tohru and {Nakata}, Fumiaki and {Niida}, Mana and {Niikura}, Hiroko and {Nishizawa}, Atsushi J. and {Oguri}, Masamune and {Okabe}, Nobuhiro and {Ono}, Yoshiaki and {Onodera}, Masato and {Onoue}, Masafusa and {Ouchi}, Masami and {Pyo}, Tae-Soo and {Shibuya}, Takatoshi and {Shimasaku}, Kazuhiro and {Simet}, Melanie and {Speagle}, Joshua and {Spergel}, David N. and {Strauss}, Michael A. and {Sugahara}, Yuma and {Sugiyama}, Naoshi and {Suto}, Yasushi and {Suzuki}, Nao and {Tait}, Philip J. and {Takada}, Masahiro and {Terai}, Tsuyoshi and {Toba}, Yoshiki and {Turner}, Edwin L. and {Uchiyama}, Hisakazu and {Umetsu}, Keiichi and {Urata}, Yuji and {Usuda}, Tomonori and {Yeh}, Sherry and {Yuma}, Suraphong},
        title = "{First data release of the Hyper Suprime-Cam Subaru Strategic Program}",
      journal = {\pasj},
         year = 2018,
        month = jan,
       volume = {70},
          eid = {S8},
        pages = {S8},
          doi = {10.1093/pasj/psx081},
archivePrefix = {arXiv},
       eprint = {1702.08449},
 primaryClass = {astro-ph.IM},
       adsurl = {https://ui.adsabs.harvard.edu/abs/2018PASJ...70S...8A}
}

@ARTICLE{HSC-SSP_DR3_2022PASJ...74..247A,
       author = {{Aihara}, Hiroaki and {AlSayyad}, Yusra and {Ando}, Makoto and {Armstrong}, Robert and {Bosch}, James and {Egami}, Eiichi and {Furusawa}, Hisanori and {Furusawa}, Junko and {Harasawa}, Sumiko and {Harikane}, Yuichi and {Hsieh}, Bau-Ching and {Ikeda}, Hiroyuki and {Ito}, Kei and {Iwata}, Ikuru and {Kodama}, Tadayuki and {Koike}, Michitaro and {Kokubo}, Mitsuru and {Komiyama}, Yutaka and {Li}, Xiangchong and {Liang}, Yongming and {Lin}, Yen-Ting and {Lupton}, Robert H. and {Lust}, Nate B. and {MacArthur}, Lauren A. and {Mawatari}, Ken and {Mineo}, Sogo and {Miyatake}, Hironao and {Miyazaki}, Satoshi and {More}, Surhud and {Morishima}, Takahiro and {Murayama}, Hitoshi and {Nakajima}, Kimihiko and {Nakata}, Fumiaki and {Nishizawa}, Atsushi J. and {Oguri}, Masamune and {Okabe}, Nobuhiro and {Okura}, Yuki and {Ono}, Yoshiaki and {Osato}, Ken and {Ouchi}, Masami and {Pan}, Yen-Chen and {Plazas Malag{\'o}n}, Andr{\'e}s A. and {Price}, Paul A. and {Reed}, Sophie L. and {Rykoff}, Eli S. and {Shibuya}, Takatoshi and {Simunovic}, Mirko and {Strauss}, Michael A. and {Sugimori}, Kanako and {Suto}, Yasushi and {Suzuki}, Nao and {Takada}, Masahiro and {Takagi}, Yuhei and {Takata}, Tadafumi and {Takita}, Satoshi and {Tanaka}, Masayuki and {Tang}, Shenli and {Taranu}, Dan S. and {Terai}, Tsuyoshi and {Toba}, Yoshiki and {Turner}, Edwin L. and {Uchiyama}, Hisakazu and {Vijarnwannaluk}, Bovornpratch and {Waters}, Christopher Z. and {Yamada}, Yoshihiko and {Yamamoto}, Naoaki and {Yamashita}, Takuji},
        title = "{Third data release of the Hyper Suprime-Cam Subaru Strategic Program}",
      journal = {\pasj},
         year = 2022,
        month = apr,
       volume = {74},
       number = {2},
        pages = {247-272},
          doi = {10.1093/pasj/psab122},
archivePrefix = {arXiv},
       eprint = {2108.13045},
 primaryClass = {astro-ph.IM},
       adsurl = {https://ui.adsabs.harvard.edu/abs/2022PASJ...74..247A}
}

@ARTICLE{HSC-SSP_DR2_2019PASJ...71..114A,
       author = {{Aihara}, Hiroaki and {AlSayyad}, Yusra and {Ando}, Makoto and {Armstrong}, Robert and {Bosch}, James and {Egami}, Eiichi and {Furusawa}, Hisanori and {Furusawa}, Junko and {Goulding}, Andy and {Harikane}, Yuichi and {Hikage}, Chiaki and {Ho}, Paul T.~P. and {Hsieh}, Bau-Ching and {Huang}, Song and {Ikeda}, Hiroyuki and {Imanishi}, Masatoshi and {Ito}, Kei and {Iwata}, Ikuru and {Jaelani}, Anton T. and {Kakuma}, Ryota and {Kawana}, Kojiro and {Kikuta}, Satoshi and {Kobayashi}, Umi and {Koike}, Michitaro and {Komiyama}, Yutaka and {Li}, Xiangchong and {Liang}, Yongming and {Lin}, Yen-Ting and {Luo}, Wentao and {Lupton}, Robert and {Lust}, Nate B. and {MacArthur}, Lauren A. and {Matsuoka}, Yoshiki and {Mineo}, Sogo and {Miyatake}, Hironao and {Miyazaki}, Satoshi and {More}, Surhud and {Murata}, Ryoma and {Namiki}, Shigeru V. and {Nishizawa}, Atsushi J. and {Oguri}, Masamune and {Okabe}, Nobuhiro and {Okamoto}, Sakurako and {Okura}, Yuki and {Ono}, Yoshiaki and {Onodera}, Masato and {Onoue}, Masafusa and {Osato}, Ken and {Ouchi}, Masami and {Shibuya}, Takatoshi and {Strauss}, Michael A. and {Sugiyama}, Naoshi and {Suto}, Yasushi and {Takada}, Masahiro and {Takagi}, Yuhei and {Takata}, Tadafumi and {Takita}, Satoshi and {Tanaka}, Masayuki and {Terai}, Tsuyoshi and {Toba}, Yoshiki and {Uchiyama}, Hisakazu and {Utsumi}, Yousuke and {Wang}, Shiang-Yu and {Wang}, Wenting and {Yamada}, Yoshihiko},
        title = "{Second data release of the Hyper Suprime-Cam Subaru Strategic Program}",
      journal = {\pasj},
         year = 2019,
        month = dec,
       volume = {71},
       number = {6},
          eid = {114},
        pages = {114},
          doi = {10.1093/pasj/psz103},
archivePrefix = {arXiv},
       eprint = {1905.12221},
 primaryClass = {astro-ph.IM},
       adsurl = {https://ui.adsabs.harvard.edu/abs/2019PASJ...71..114A}
}

@ARTICLE{IAC_Stripe_82_2016MNRAS.456.1359F,
       author = {{Fliri}, J{\"u}rgen and {Trujillo}, Ignacio},
        title = "{The IAC Stripe 82 Legacy Project: a wide-area survey for faint surface brightness astronomy}",
      journal = {\mnras},
         year = 2016,
        month = feb,
       volume = {456},
       number = {2},
        pages = {1359-1373},
          doi = {10.1093/mnras/stv2686},
archivePrefix = {arXiv},
       eprint = {1603.04474},
 primaryClass = {astro-ph.GA},
       adsurl = {https://ui.adsabs.harvard.edu/abs/2016MNRAS.456.1359F}
}

@ARTICLE{SDSS_DR7_2009ApJS..182..543A,
       author = {{Abazajian}, Kevork N. and {Adelman-McCarthy}, Jennifer K. and {Ag{\"u}eros}, Marcel A. and {Allam}, Sahar S. and {Allende Prieto}, Carlos and {An}, Deokkeun and {Anderson}, Kurt S.~J. and {Anderson}, Scott F. and {Annis}, James and {Bahcall}, Neta A. and {Bailer-Jones}, C.~A.~L. and {Barentine}, J.~C. and {Bassett}, Bruce A. and {Becker}, Andrew C. and {Beers}, Timothy C. and {Bell}, Eric F. and {Belokurov}, Vasily and {Berlind}, Andreas A. and {Berman}, Eileen F. and {Bernardi}, Mariangela and {Bickerton}, Steven J. and {Bizyaev}, Dmitry and {Blakeslee}, John P. and {Blanton}, Michael R. and {Bochanski}, John J. and {Boroski}, William N. and {Brewington}, Howard J. and {Brinchmann}, Jarle and {Brinkmann}, J. and {Brunner}, Robert J. and {Budav{\'a}ri}, Tam{\'a}s and {Carey}, Larry N. and {Carliles}, Samuel and {Carr}, Michael A. and {Castander}, Francisco J. and {Cinabro}, David and {Connolly}, A.~J. and {Csabai}, Istv{\'a}n and {Cunha}, Carlos E. and {Czarapata}, Paul C. and {Davenport}, James R.~A. and {de Haas}, Ernst and {Dilday}, Ben and {Doi}, Mamoru and {Eisenstein}, Daniel J. and {Evans}, Michael L. and {Evans}, N.~W. and {Fan}, Xiaohui and {Friedman}, Scott D. and {Frieman}, Joshua A. and {Fukugita}, Masataka and {G{\"a}nsicke}, Boris T. and {Gates}, Evalyn and {Gillespie}, Bruce and {Gilmore}, G. and {Gonzalez}, Belinda and {Gonzalez}, Carlos F. and {Grebel}, Eva K. and {Gunn}, James E. and {Gy{\"o}ry}, Zsuzsanna and {Hall}, Patrick B. and {Harding}, Paul and {Harris}, Frederick H. and {Harvanek}, Michael and {Hawley}, Suzanne L. and {Hayes}, Jeffrey J.~E. and {Heckman}, Timothy M. and {Hendry}, John S. and {Hennessy}, Gregory S. and {Hindsley}, Robert B. and {Hoblitt}, J. and {Hogan}, Craig J. and {Hogg}, David W. and {Holtzman}, Jon A. and {Hyde}, Joseph B. and {Ichikawa}, Shin-ichi and {Ichikawa}, Takashi and {Im}, Myungshin and {Ivezi{\'c}}, {\v{Z}}eljko and {Jester}, Sebastian and {Jiang}, Linhua and {Johnson}, Jennifer A. and {Jorgensen}, Anders M. and {Juri{\'c}}, Mario and {Kent}, Stephen M. and {Kessler}, R. and {Kleinman}, S.~J. and {Knapp}, G.~R. and {Konishi}, Kohki and {Kron}, Richard G. and {Krzesinski}, Jurek and {Kuropatkin}, Nikolay and {Lampeitl}, Hubert and {Lebedeva}, Svetlana and {Lee}, Myung Gyoon and {Lee}, Young Sun and {French Leger}, R. and {L{\'e}pine}, S{\'e}bastien and {Li}, Nolan and {Lima}, Marcos and {Lin}, Huan and {Long}, Daniel C. and {Loomis}, Craig P. and {Loveday}, Jon and {Lupton}, Robert H. and {Magnier}, Eugene and {Malanushenko}, Olena and {Malanushenko}, Viktor and {Mandelbaum}, Rachel and {Margon}, Bruce and {Marriner}, John P. and {Mart{\'\i}nez-Delgado}, David and {Matsubara}, Takahiko and {McGehee}, Peregrine M. and {McKay}, Timothy A. and {Meiksin}, Avery and {Morrison}, Heather L. and {Mullally}, Fergal and {Munn}, Jeffrey A. and {Murphy}, Tara and {Nash}, Thomas and {Nebot}, Ada and {Neilsen}, Eric H., Jr. and {Newberg}, Heidi Jo and {Newman}, Peter R. and {Nichol}, Robert C. and {Nicinski}, Tom and {Nieto-Santisteban}, Maria and {Nitta}, Atsuko and {Okamura}, Sadanori and {Oravetz}, Daniel J. and {Ostriker}, Jeremiah P. and {Owen}, Russell and {Padmanabhan}, Nikhil and {Pan}, Kaike and {Park}, Changbom and {Pauls}, George and {Peoples}, John, Jr. and {Percival}, Will J. and {Pier}, Jeffrey R. and {Pope}, Adrian C. and {Pourbaix}, Dimitri and {Price}, Paul A. and {Purger}, Norbert and {Quinn}, Thomas and {Raddick}, M. Jordan and {Re Fiorentin}, Paola and {Richards}, Gordon T. and {Richmond}, Michael W. and {Riess}, Adam G. and {Rix}, Hans-Walter and {Rockosi}, Constance M. and {Sako}, Masao and {Schlegel}, David J. and {Schneider}, Donald P. and {Scholz}, Ralf-Dieter and {Schreiber}, Matthias R. and {Schwope}, Axel D. and {Seljak}, Uro{\v{s}} and {Sesar}, Branimir and {Sheldon}, Erin and {Shimasaku}, Kazu and {Sibley}, Valena C. and {Simmons}, A.~E. and {Sivarani}, Thirupathi and {Allyn Smith}, J. and {Smith}, Martin C. and {Smol{\v{c}}i{\'c}}, Vernesa and {Snedden}, Stephanie A. and {Stebbins}, Albert and {Steinmetz}, Matthias and {Stoughton}, Chris and {Strauss}, Michael A. and {SubbaRao}, Mark and {Suto}, Yasushi and {Szalay}, Alexander S. and {Szapudi}, Istv{\'a}n and {Szkody}, Paula and {Tanaka}, Masayuki and {Tegmark}, Max and {Teodoro}, Luis F.~A. and {Thakar}, Aniruddha R. and {Tremonti}, Christy A. and {Tucker}, Douglas L. and {Uomoto}, Alan and {Vanden Berk}, Daniel E. and {Vandenberg}, Jan and {Vidrih}, S. and {Vogeley}, Michael S. and {Voges}, Wolfgang and {Vogt}, Nicole P. and {Wadadekar}, Yogesh and {Watters}, Shannon and {Weinberg}, David H. and {West}, Andrew A. and {White}, Simon D.~M. and {Wilhite}, Brian C. and {Wonders}, Alainna C. and {Yanny}, Brian and {Yocum}, D.~R. and {York}, Donald G. and {Zehavi}, Idit and {Zibetti}, Stefano and {Zucker}, Daniel B.},
        title = "{The Seventh Data Release of the Sloan Digital Sky Survey}",
      journal = {\apjs},
         year = 2009,
        month = jun,
       volume = {182},
       number = {2},
        pages = {543-558},
          doi = {10.1088/0067-0049/182/2/543},
archivePrefix = {arXiv},
       eprint = {0812.0649},
 primaryClass = {astro-ph},
       adsurl = {https://ui.adsabs.harvard.edu/abs/2009ApJS..182..543A}
}

@ARTICLE{Stripe82_2020A&A...644A..42R,
       author = {{Rom{\'a}n}, Javier and {Trujillo}, Ignacio and {Montes}, Mireia},
        title = "{Galactic cirri in deep optical imaging}",
      journal = {\aap},
         year = 2020,
        month = dec,
       volume = {644},
          eid = {A42},
        pages = {A42},
          doi = {10.1051/0004-6361/201936111},
archivePrefix = {arXiv},
       eprint = {1907.00978},
 primaryClass = {astro-ph.GA},
       adsurl = {https://ui.adsabs.harvard.edu/abs/2020A&A...644A..42R}
}

@ARTICLE{Marchuk_etal2021,
       author = {{Marchuk}, Alexander A. and {Smirnov}, Anton A. and {Mosenkov}, Aleksandr V. and {Il'in}, Vladimir B. and {Gontcharov}, George A. and {Savchenko}, Sergey S. and {Rom{\'a}n}, Javier},
        title = "{Fractal dimension of optical cirrus in Stripe82}",
      journal = {\mnras},
         year = 2021,
        month = dec,
       volume = {508},
       number = {4},
        pages = {5825-5841},
          doi = {10.1093/mnras/stab2846},
archivePrefix = {arXiv},
       eprint = {2109.14034},
 primaryClass = {astro-ph.GA},
       adsurl = {https://ui.adsabs.harvard.edu/abs/2021MNRAS.508.5825M}
}

@ARTICLE{Gontcharov_etal_2022_S82_Wolf,
       author = {{Gontcharov}, G.~A. and {Mosenkov}, A.~V. and {Savchenko}, S.~S. and {Il'in}, V.~B. and {Marchuk}, A.~A. and {Smirnov}, A.~A. and {Usachev}, P.~A. and {Polyakov}, D.~M. and {Shakespear}, Z.},
        title = "{Interstellar Extinction in Galactic Cirri in SDSS Stripe 82}",
      journal = {Astronomy Letters},
         year = 2022,
        month = sep,
       volume = {48},
       number = {9},
        pages = {503-516},
          doi = {10.1134/S1063773722090031},
archivePrefix = {arXiv},
       eprint = {2301.10591},
 primaryClass = {astro-ph.GA},
       adsurl = {https://ui.adsabs.harvard.edu/abs/2022AstL...48..503G}
}

@ARTICLE{Cirrus_2023MNRAS.519.4735S,
       author = {{Smirnov}, Anton A. and {Savchenko}, Sergey S. and {Poliakov}, Denis M. and {Marchuk}, Alexander A. and {Mosenkov}, Aleksandr V. and {Il'in}, Vladimir B. and {Gontcharov}, George A. and {Rom{\'a}n}, Javier and {Seguine}, Jonah},
        title = "{Prospects for future studies using deep imaging: analysis of individual Galactic cirrus filaments}",
      journal = {\mnras},
         year = 2023,
        month = mar,
       volume = {519},
       number = {3},
        pages = {4735-4752},
          doi = {10.1093/mnras/stac3765},
archivePrefix = {arXiv},
       eprint = {2301.12410},
 primaryClass = {astro-ph.GA},
       adsurl = {https://ui.adsabs.harvard.edu/abs/2023MNRAS.519.4735S}
}

@ARTICLE{Haigh_etal_2021A&A...645A.107H,
       author = {{Haigh}, Caroline and {Chamba}, Nushkia and {Venhola}, Aku and {Peletier}, Reynier and {Doorenbos}, Lars and {Watkins}, Matthew and {Wilkinson}, Michael H.~F.},
        title = "{Optimising and comparing source-extraction tools using objective segmentation quality criteria}",
      journal = {\aap},
         year = 2021,
        month = jan,
       volume = {645},
          eid = {A107},
        pages = {A107},
          doi = {10.1051/0004-6361/201936561},
archivePrefix = {arXiv},
       eprint = {2009.07586},
 primaryClass = {astro-ph.GA},
       adsurl = {https://ui.adsabs.harvard.edu/abs/2021A&A...645A.107H}
}

@ARTICLE{U-Net_2015arXiv150504597R,
       author = {{Ronneberger}, Olaf and {Fischer}, Philipp and {Brox}, Thomas},
        title = "{U-Net: Convolutional Networks for Biomedical Image Segmentation}",
      journal = {arXiv e-prints},
         year = 2015,
        month = may,
          eid = {arXiv:1505.04597},
        pages = {arXiv:1505.04597},
archivePrefix = {arXiv},
       eprint = {1505.04597},
 primaryClass = {cs.CV},
       adsurl = {https://ui.adsabs.harvard.edu/abs/2015arXiv150504597R}
}

@article{Zhou2018UNetAN,
  title={UNet++: A Nested U-Net Architecture for Medical Image Segmentation},
  author={Zongwei Zhou and Md Mahfuzur Rahman Siddiquee and Nima Tajbakhsh and Jianming Liang},
  journal={Deep Learning in Medical Image Analysis and Multimodal Learning for Clinical Decision Support : 4th International Workshop, DLMIA 2018, and 8th International Workshop, ML-CDS 2018, held in conjunction with MICCAI 2018, Granada, Spain, S...},
  year={2018},
  volume={11045},
  pages={
          3-11
        },
  url={https://api.semanticscholar.org/CorpusID:50786304}
}

@ARTICLE{MobileNetV2_2018,
       author = {{Sandler}, Mark and {Howard}, Andrew and {Zhu}, Menglong and {Zhmoginov}, Andrey and {Chen}, Liang-Chieh},
        title = "{MobileNetV2: Inverted Residuals and Linear Bottlenecks}",
      journal = {arXiv e-prints},
         year = 2018,
        month = jan,
          eid = {arXiv:1801.04381},
        pages = {arXiv:1801.04381},
archivePrefix = {arXiv},
       eprint = {1801.04381},
 primaryClass = {cs.CV},
       adsurl = {https://ui.adsabs.harvard.edu/abs/2018arXiv180104381S}
}

@ARTICLE{MaskR-CNN_2017,
       author = {{He}, Kaiming and {Gkioxari}, Georgia and {Doll{\'a}r}, Piotr and {Girshick}, Ross},
        title = "{Mask R-CNN}",
      journal = {arXiv e-prints},
         year = 2017,
        month = mar,
          eid = {arXiv:1703.06870},
        pages = {arXiv:1703.06870},
          doi = {10.48550/arXiv.1703.06870},
archivePrefix = {arXiv},
       eprint = {1703.06870},
 primaryClass = {cs.CV},
       adsurl = {https://ui.adsabs.harvard.edu/abs/2017arXiv170306870H}
}

@INPROCEEDINGS{Stork2012AudiobasedHA,
  author={Stork, Johannes A. and Spinello, Luciano and Silva, Jens and Arras, Kai O.},
  booktitle={2012 IEEE RO-MAN: The 21st IEEE International Symposium on Robot and Human Interactive Communication}, 
  title={Audio-based human activity recognition using Non-Markovian Ensemble Voting}, 
  year={2012},
  volume={},
  number={},
  pages={509-514},
  doi={10.1109/ROMAN.2012.6343802}
}

@ARTICLE{Fraz2012AnEC,
  author={Fraz, Muhammad Moazam and Remagnino, Paolo and Hoppe, Andreas and Uyyanonvara, Bunyarit and Rudnicka, Alicja R. and Owen, Christopher G. and Barman, Sarah A.},
  journal={IEEE Transactions on Biomedical Engineering}, 
  title={An Ensemble Classification-Based Approach Applied to Retinal Blood Vessel Segmentation}, 
  year={2012},
  volume={59},
  number={9},
  pages={2538-2548},
  doi={10.1109/TBME.2012.2205687}
}

@ARTICLE{2013IJRS...34.5166G,
       author = {{Guan}, Haiyan and {Li}, Jonathan and {Chapman}, Michael and {Deng}, Fei and {Ji}, Zheng and {Yang}, Xu},
        title = "{Integration of orthoimagery and lidar data for object-based urban thematic mapping using random forests}",
      journal = {International Journal of Remote Sensing},
         year = 2013,
        month = jul,
       volume = {34},
       number = {14},
        pages = {5166-5186},
          doi = {10.1080/01431161.2013.788261},
       adsurl = {https://ui.adsabs.harvard.edu/abs/2013IJRS...34.5166G}
}

@inproceedings{QuratulAin2010ClassificationAS,
    author = {Qurat-Ul-Ain, Qurat-Ul-Ain and Latif, Ghazanfar and Kazmi, Sidra Batool and Jaffar, M. Arfan and Mirza, Anwar M.},
    title = {Classification and segmentation of brain tumor using texture analysis},
    year = {2010},
    isbn = {9789604741540},
    publisher = {World Scientific and Engineering Academy and Society (WSEAS)},
    address = {Stevens Point, Wisconsin, USA},
    booktitle = {Proceedings of the 9th WSEAS International Conference on Artificial Intelligence, Knowledge Engineering and Data Bases},
    pages = {147–155},
    numpages = {9},
    location = {UK},
    series = {AIKED'10}
}

@ARTICLE{Cyganek2012OneClassSV,
       author = {{Cyganek}, Bogus{\l}aw},
        title = "{One-Class Support Vector Ensembles for Image Segmentation and Classification}",
      journal = {Journal of Mathematical Imaging and Vision},
         year = 2012,
        month = feb,
       volume = {42},
       number = {2-3},
        pages = {103-117},
          doi = {10.1007/s10851-011-0304-0},
       adsurl = {https://ui.adsabs.harvard.edu/abs/2012JMIV...42..103C}
}

@InProceedings{Kinattukara2014ANE,
    author="Kinattukara, Tejy
    and Verma, Brijesh",
    editor="Loo, Chu Kiong
    and Yap, Keem Siah
    and Wong, Kok Wai
    and Beng Jin, Andrew Teoh
    and Huang, Kaizhu",
    title="A Neural Ensemble Approach for Segmentation and Classification of Road Images",
    booktitle="Neural Information Processing",
    year="2014",
    publisher="Springer International Publishing",
    address="Cham",
    pages="183--193",
    isbn="978-3-319-12643-2"
}

@article{Tang2023MobileNetV2ES,
  title={MobileNetV2 Ensemble Segmentation for Mandibular on Panoramic Radiography},
  author={Peng Tang and Qiaokang Liang and Xintong Yan and Dan Zhang and Gianmarc Coppola and Wei Sun Multi-proportion and Anne M. Alsup and Kelli Fowlds and Michael Cho and Jacob M Luber and BetaBuddy},
  journal={International Journal of Intelligent Engineering and Systems},
  year={2023},
  url={https://api.semanticscholar.org/CorpusID:257211721}
}

@article{Moradi2023DeepEL,
  title={Deep ensemble learning for automated non-advanced AMD classification using optimized retinal layer segmentation and SD-OCT scans},
  author={Mousa Moradi and Yu Chen and Xian Du and Johanna M. Seddon},
  journal={Computers in biology and medicine},
  year={2023},
  volume={154},
  pages={
          106512
        },
  url={https://api.semanticscholar.org/CorpusID:255658002}
}

@article{Nanni2023ExploringTP,
  title={Exploring the Potential of Ensembles of Deep Learning Networks for Image Segmentation},
  author={Loris Nanni and Alessandra Lumini and Carlo Fantozzi},
  journal={Inf.},
  year={2023},
  volume={14},
  pages={657},
  url={https://api.semanticscholar.org/CorpusID:266216418}
}

@ARTICLE{nun2016ENSEMBLELM,
       author = {{Nun}, Isadora and {Protopapas}, Pavlos and {Sim}, Brandon and {Chen}, Wesley},
        title = "{Ensemble Learning Method for Outlier Detection and its Application to Astronomical Light Curves}",
      journal = {\aj},
         year = 2016,
        month = sep,
       volume = {152},
       number = {3},
          eid = {71},
        pages = {71},
          doi = {10.3847/0004-6256/152/3/71},
       adsurl = {https://ui.adsabs.harvard.edu/abs/2016AJ....152...71N}
}

@Article{Pagliaro2023ApplicationOM,
AUTHOR = {Pagliaro, Antonio and Cusumano, Giancarlo and La Barbera, Antonino and La Parola, Valentina and Lombardi, Saverio},
TITLE = {Application of Machine Learning Ensemble Methods to ASTRI Mini-Array Cherenkov Event Reconstruction},
JOURNAL = {Applied Sciences},
VOLUME = {13},
YEAR = {2023},
NUMBER = {14},
ARTICLE-NUMBER = {8172},
URL = {https://www.mdpi.com/2076-3417/13/14/8172},
ISSN = {2076-3417},
DOI = {10.3390/app13148172}
}

@ARTICLE{Priyadarshini2021ACN,
       author = {{Priyadarshini}, Ishaani and {Puri}, Vikram},
        title = "{A convolutional neural network (CNN) based ensemble model for exoplanet detection}",
      journal = {Earth Science Informatics},
         year = 2021,
        month = jun,
       volume = {14},
       number = {2},
        pages = {735-747},
          doi = {10.1007/s12145-021-00579-5},
       adsurl = {https://ui.adsabs.harvard.edu/abs/2021EScIn..14..735P}
}

@ARTICLE{Zeraatgari2024MachineLP,
       author = {{Zeraatgari}, Fatemeh Zahra and {Hafezianzadeh}, Fatemeh and {Zhang}, Yanxia and {Mei}, Liquan and {Ayubinia}, Ashraf and {Mosallanezhad}, Amin and {Zhang}, Jingyi},
        title = "{Machine learning-based photometric classification of galaxies, quasars, emission-line galaxies, and stars}",
      journal = {\mnras},
         year = 2024,
        month = jan,
       volume = {527},
       number = {3},
        pages = {4677-4689},
          doi = {10.1093/mnras/stad3436},
archivePrefix = {arXiv},
       eprint = {2311.02951},
 primaryClass = {astro-ph.GA},
       adsurl = {https://ui.adsabs.harvard.edu/abs/2024MNRAS.527.4677Z}
}

@article{Breiman1996BaggingP,
  title={Bagging Predictors},
  author={L. Breiman},
  journal={Machine Learning},
  year={1996},
  volume={24},
  pages={123-140},
  url={https://api.semanticscholar.org/CorpusID:47328136}
}

@article{Boosting2000,
author = {Jerome Friedman and Trevor Hastie and Robert Tibshirani},
title = {{Additive logistic regression: a statistical view of boosting (With discussion and a rejoinder by the authors)}},
volume = {28},
journal = {The Annals of Statistics},
number = {2},
publisher = {Institute of Mathematical Statistics},
pages = {337 -- 407},
year = {2000},
doi = {10.1214/aos/1016218223},
URL = {https://doi.org/10.1214/aos/1016218223}
}

@ARTICLE{RandomForest2001,
       author = {{Breiman}, Leo},
        title = "{Random Forests.}",
      journal = {Machine Learning},
         year = 2001,
        month = jan,
       volume = {45},
        pages = {5-32},
          doi = {10.1023/A:1010933404324},
       adsurl = {https://ui.adsabs.harvard.edu/abs/2001MachL..45....5B}
}

@article{Ren2016EnsembleCA,
  title={Ensemble Classification and Regression-Recent Developments, Applications and Future Directions [Review Article]},
  author={Ye Ren and Le Zhang and Ponnuthurai Nagaratnam Suganthan},
  journal={IEEE Computational Intelligence Magazine},
  year={2016},
  volume={11},
  pages={41-53},
  url={https://api.semanticscholar.org/CorpusID:14010214}
}

@ARTICLE{EnsembleLearning2022,
  author={Mienye, Ibomoiye Domor and Sun, Yanxia},
  journal={IEEE Access}, 
  title={A Survey of Ensemble Learning: Concepts, Algorithms, Applications, and Prospects}, 
  year={2022},
  volume={10},
  number={},
  pages={99129-99149},
  doi={10.1109/ACCESS.2022.3207287}
}

@INPROCEEDINGS{Fine-tuning2017,
  author={Yin, Xiangnan and Chen, Weihai and Wu, Xingming and Yue, Haosong},
  booktitle={2017 12th IEEE Conference on Industrial Electronics and Applications (ICIEA)}, 
  title={Fine-tuning and visualization of convolutional neural networks}, 
  year={2017},
  volume={},
  number={},
  pages={1310-1315},
  doi={10.1109/ICIEA.2017.8283041}}

@book{Serra1983ImageAA,
  title={Image Analysis and Mathematical Morphology},
  author={Serra, J.P. and Serra, J.},
  isbn={9780126372410},
  lccn={81066397},
  series={Image Analysis and Mathematical Morphology},
  url={https://books.google.ru/books?id=BpdTAAAAYAAJ},
  year={1982},
  publisher={Academic Press}
}

@ARTICLE{Alina_etal_2022,
       author = {{Alina}, D. and {Shomanov}, A. and {Baimukhametova}, S.},
        title = "{MaLeFiSenta: Machine Learning for FilamentS Identification and orientation in the ISM}",
      journal = {arXiv e-prints},
         year = 2022,
        month = may,
          eid = {arXiv:2205.00683},
        pages = {arXiv:2205.00683},
          doi = {10.48550/arXiv.2205.00683},
archivePrefix = {arXiv},
       eprint = {2205.00683},
 primaryClass = {astro-ph.IM},
       adsurl = {https://ui.adsabs.harvard.edu/abs/2022arXiv220500683A}
}

@ARTICLE{Zavagno_etal_2023,
       author = {{Zavagno}, A. and {Dup{\'e}}, F. -X. and {Bensaid}, S. and {Schisano}, E. and {Li Causi}, G. and {Gray}, M. and {Molinari}, S. and {Elia}, D. and {Lambert}, J. -C. and {Brescia}, M. and {Arzoumanian}, D. and {Russeil}, D. and {Riccio}, G. and {Cavuoti}, S.},
        title = "{Supervised machine learning on Galactic filaments. Revealing the filamentary structure of the Galactic interstellar medium}",
      journal = {\aap},
         year = 2023,
        month = jan,
       volume = {669},
          eid = {A120},
        pages = {A120},
          doi = {10.1051/0004-6361/202244103},
archivePrefix = {arXiv},
       eprint = {2212.00463},
 primaryClass = {astro-ph.GA},
       adsurl = {https://ui.adsabs.harvard.edu/abs/2023A&A...669A.120Z}
}

@ARTICLE{Marchuk_etal_2022_B/PS,
       author = {{Marchuk}, Alexander A. and {Smirnov}, Anton A. and {Sotnikova}, Natalia Y. and {Bunakalya}, Dmitriy A. and {Savchenko}, Sergey S. and {Reshetnikov}, Vladimir P. and {Usachev}, Pavel A. and {Tikhonenko}, Iliya S. and {Zozulia}, Viktor D. and {Zakharova}, Daria A.},
        title = "{B/PS bulges in DESI Legacy edge-on galaxies - I. Sample building}",
      journal = {\mnras},
         year = 2022,
        month = may,
       volume = {512},
       number = {1},
        pages = {1371-1390},
          doi = {10.1093/mnras/stac599},
archivePrefix = {arXiv},
       eprint = {2203.01154},
 primaryClass = {astro-ph.GA},
       adsurl = {https://ui.adsabs.harvard.edu/abs/2022MNRAS.512.1371M}
}

@ARTICLE{Savchenko_etal_2024,
       author = {{Savchenko}, S.~S. and {Makarov}, D.~I. and {Antipova}, A.~V. and {Tikhonenko}, I.~S.},
        title = "{Search for the edge-on galaxies using an artificial neural network}",
      journal = {Astronomy and Computing},
         year = 2024,
        month = jan,
       volume = {46},
          eid = {100771},
        pages = {100771},
          doi = {10.1016/j.ascom.2023.100771},
archivePrefix = {arXiv},
       eprint = {2312.02742},
 primaryClass = {astro-ph.IM},
       adsurl = {https://ui.adsabs.harvard.edu/abs/2024A&C....4600771S}
}

@ARTICLE{Richards_etal_gabor_attention_2024,
       author = {{Richards}, Felix and {Paiement}, Adeline and {Xie}, Xianghua and {Sola}, Elisabeth and {Duc}, Pierre-Alain},
        title = "{Multi-scale gridded Gabor attention for cirrus segmentation}",
      journal = {arXiv e-prints},
         year = 2024,
        month = jul,
          eid = {arXiv:2407.08852},
        pages = {arXiv:2407.08852},
          doi = {10.48550/arXiv.2407.08852},
archivePrefix = {arXiv},
       eprint = {2407.08852},
 primaryClass = {cs.CV},
       adsurl = {https://ui.adsabs.harvard.edu/abs/2024arXiv240708852R}
}

@book{berry2005handbook,
  title={The Handbook of Astronomical Image Processing},
  author={Berry, R. and Burnell, J.},
  isbn={9780943396828},
  lccn={2004063730},
  url={https://books.google.ru/books?id=O0fPPAAACAAJ},
  year={2005},
  publisher={Willmann-Bell}
}

@book{Howell_2006, 
    place={Cambridge}, 
    edition={2}, 
    series={Cambridge Observing Handbooks for Research Astronomers}, 
    title={Handbook of CCD Astronomy}, 
    publisher={Cambridge University Press}, 
    author={Howell, Steve B.}, 
    year={2006}, 
    collection={Cambridge Observing Handbooks for Research Astronomers}
}

@book{Gonzalez_2006_handbook,
author = {Gonzalez, Rafael C. and Woods, Richard E.},
title = {Digital Image Processing (3rd Edition)},
year = {2006},
isbn = {013168728X},
publisher = {Prentice-Hall, Inc.},
address = {USA}
}

@ARTICLE{2014A&A...566A..55P_Planck_Collaboration,
       author = {{Planck Collaboration} and {Abergel}, A. and {Ade}, P.~A.~R. and {Aghanim}, N. and {Alves}, M.~I.~R. and {Aniano}, G. and {Arnaud}, M. and {Ashdown}, M. and {Aumont}, J. and {Baccigalupi}, C. and {Banday}, A.~J. and {Barreiro}, R.~B. and {Bartlett}, J.~G. and {Battaner}, E. and {Benabed}, K. and {Benoit-L{\'e}vy}, A. and {Bernard}, J. -P. and {Bersanelli}, M. and {Bielewicz}, P. and {Bobin}, J. and {Bonaldi}, A. and {Bond}, J.~R. and {Bouchet}, F.~R. and {Boulanger}, F. and {Burigana}, C. and {Cardoso}, J. -F. and {Catalano}, A. and {Chamballu}, A. and {Chiang}, H.~C. and {Christensen}, P.~R. and {Clements}, D.~L. and {Colombi}, S. and {Colombo}, L.~P.~L. and {Couchot}, F. and {Crill}, B.~P. and {Cuttaia}, F. and {Danese}, L. and {Davis}, R.~J. and {de Bernardis}, P. and {de Rosa}, A. and {de Zotti}, G. and {Delabrouille}, J. and {D{\'e}sert}, F. -X. and {Dickinson}, C. and {Diego}, J.~M. and {Dole}, H. and {Donzelli}, S. and {Dor{\'e}}, O. and {Douspis}, M. and {Dupac}, X. and {Efstathiou}, G. and {En{\ss}lin}, T.~A. and {Eriksen}, H.~K. and {Falgarone}, E. and {Finelli}, F. and {Forni}, O. and {Frailis}, M. and {Franceschi}, E. and {Galeotta}, S. and {Ganga}, K. and {Ghosh}, T. and {Giard}, M. and {Giraud-H{\'e}raud}, Y. and {Gonz{\'a}lez-Nuevo}, J. and {G{\'o}rski}, K.~M. and {Gregorio}, A. and {Gruppuso}, A. and {Guillet}, V. and {Hansen}, F.~K. and {Harrison}, D. and {Helou}, G. and {Henrot-Versill{\'e}}, S. and {Hern{\'a}ndez-Monteagudo}, C. and {Herranz}, D. and {Hildebrandt}, S.~R. and {Hivon}, E. and {Hobson}, M. and {Holmes}, W.~A. and {Hornstrup}, A. and {Hovest}, W. and {Huffenberger}, K.~M. and {Jaffe}, A.~H. and {Jaffe}, T.~R. and {Joncas}, G. and {Jones}, A. and {Jones}, W.~C. and {Juvela}, M. and {Kalberla}, P. and {Keih{\"a}nen}, E. and {Kerp}, J. and {Keskitalo}, R. and {Kisner}, T.~S. and {Kneissl}, R. and {Knoche}, J. and {Kunz}, M. and {Kurki-Suonio}, H. and {Lagache}, G. and {L{\"a}hteenm{\"a}ki}, A. and {Lamarre}, J. -M. and {Lasenby}, A. and {Lawrence}, C.~R. and {Leonardi}, R. and {Levrier}, F. and {Liguori}, M. and {Lilje}, P.~B. and {Linden-V{\o}rnle}, M. and {L{\'o}pez-Caniego}, M. and {Lubin}, P.~M. and {Mac{\'\i}as-P{\'e}rez}, J.~F. and {Maffei}, B. and {Maino}, D. and {Mandolesi}, N. and {Maris}, M. and {Marshall}, D.~J. and {Martin}, P.~G. and {Mart{\'\i}nez-Gonz{\'a}lez}, E. and {Masi}, S. and {Massardi}, M. and {Matarrese}, S. and {Mazzotta}, P. and {Melchiorri}, A. and {Mendes}, L. and {Mennella}, A. and {Migliaccio}, M. and {Mitra}, S. and {Miville-Desch{\^e}nes}, M. -A. and {Moneti}, A. and {Montier}, L. and {Morgante}, G. and {Mortlock}, D. and {Munshi}, D. and {Murphy}, J.~A. and {Naselsky}, P. and {Nati}, F. and {Natoli}, P. and {Noviello}, F. and {Novikov}, D. and {Novikov}, I. and {Oxborrow}, C.~A. and {Pagano}, L. and {Pajot}, F. and {Paoletti}, D. and {Pasian}, F. and {Perdereau}, O. and {Perotto}, L. and {Perrotta}, F. and {Piacentini}, F. and {Piat}, M. and {Pierpaoli}, E. and {Pietrobon}, D. and {Plaszczynski}, S. and {Pointecouteau}, E. and {Polenta}, G. and {Ponthieu}, N. and {Popa}, L. and {Pratt}, G.~W. and {Prunet}, S. and {Puget}, J. -L. and {Rachen}, J.~P. and {Reach}, W.~T. and {Rebolo}, R. and {Reinecke}, M. and {Remazeilles}, M. and {Renault}, C. and {Ricciardi}, S. and {Riller}, T. and {Ristorcelli}, I. and {Rocha}, G. and {Rosset}, C. and {Roudier}, G. and {Rusholme}, B. and {Sandri}, M. and {Savini}, G. and {Spencer}, L.~D. and {Starck}, J. -L. and {Sureau}, F. and {Sutton}, D. and {Suur-Uski}, A. -S. and {Sygnet}, J. -F. and {Tauber}, J.~A. and {Terenzi}, L. and {Toffolatti}, L. and {Tomasi}, M. and {Tristram}, M. and {Tucci}, M. and {Umana}, G. and {Valenziano}, L. and {Valiviita}, J. and {Van Tent}, B. and {Verstraete}, L. and {Vielva}, P. and {Villa}, F. and {Wade}, L.~A. and {Wandelt}, B.~D. and {Winkel}, B. and {Yvon}, D. and {Zacchei}, A. and {Zonca}, A.},
        title = "{Planck intermediate results. XVII. Emission of dust in the diffuse interstellar medium from the far-infrared to microwave frequencies}",
      journal = {\aap},
         year = 2014,
        month = jun,
       volume = {566},
          eid = {A55},
        pages = {A55},
          doi = {10.1051/0004-6361/201323270},
archivePrefix = {arXiv},
       eprint = {1312.5446},
 primaryClass = {astro-ph.GA},
       adsurl = {https://ui.adsabs.harvard.edu/abs/2014A&A...566A..55P}
}

@ARTICLE{2017A&A...597A.130B_Bianchi,
       author = {{Bianchi}, S. and {Giovanardi}, C. and {Smith}, M.~W.~L. and {Fritz}, J. and {Davies}, J.~I. and {Haynes}, M.~P. and {Giovanelli}, R. and {Baes}, M. and {Bocchio}, M. and {Boissier}, S. and {Boquien}, M. and {Boselli}, A. and {Casasola}, V. and {Clark}, C.~J.~R. and {De Looze}, I. and {di Serego Alighieri}, S. and {Grossi}, M. and {Jones}, A.~P. and {Hughes}, T.~M. and {Hunt}, L.~K. and {Madden}, S. and {Magrini}, L. and {Pappalardo}, C. and {Ysard}, N. and {Zibetti}, S.},
        title = "{The Herschel Virgo Cluster Survey. XX. Dust and gas in the foreground Galactic cirrus}",
      journal = {\aap},
         year = 2017,
        month = jan,
       volume = {597},
          eid = {A130},
        pages = {A130},
          doi = {10.1051/0004-6361/201629013},
archivePrefix = {arXiv},
       eprint = {1609.05941},
 primaryClass = {astro-ph.GA},
       adsurl = {https://ui.adsabs.harvard.edu/abs/2017A&A...597A.130B}
}

@ARTICLE{1994A&A...291L...5B_Bernard,
       author = {{Bernard}, J.~P. and {Boulanger}, F. and {Desert}, F.~X. and {Giard}, M. and {Helou}, G. and {Puget}, J.~L.},
        title = "{Dust emission of galactic cirrus from DIRBE observations.}",
      journal = {\aap},
         year = 1994,
        month = nov,
       volume = {291},
        pages = {L5-L8},
       adsurl = {https://ui.adsabs.harvard.edu/abs/1994A&A...291L...5B}
}

@ARTICLE{1998_Szomoru,
       author = {{Szomoru}, Arpad and {Guhathakurta}, Puragra},
        title = "{Optical Spectroscopy of Galactic Cirrus Clouds: Extended Red Emission in the Diffuse Interstellar Medium}",
      journal = {\apjl},
         year = 1998,
        month = feb,
       volume = {494},
       number = {1},
        pages = {L93-L97},
          doi = {10.1086/311156},
archivePrefix = {arXiv},
       eprint = {astro-ph/9712056},
 primaryClass = {astro-ph},
       adsurl = {https://ui.adsabs.harvard.edu/abs/1998ApJ...494L..93S}
}

@INPROCEEDINGS{SWarp_2002ASPC..281..228B,
       author = {{Bertin}, Emmanuel and {Mellier}, Yannick and {Radovich}, Mario and {Missonnier}, Gilles and {Didelon}, Pierre and {Morin}, Bertrand},
        title = "{The TERAPIX Pipeline}",
    booktitle = {Astronomical Data Analysis Software and Systems XI},
         year = 2002,
       editor = {{Bohlender}, David A. and {Durand}, Daniel and {Handley}, Thomas H.},
       series = {Astronomical Society of the Pacific Conference Series},
       volume = {281},
        month = jan,
        pages = {228},
       adsurl = {https://ui.adsabs.harvard.edu/abs/2002ASPC..281..228B}
}

@ARTICLE{North_field_2011AJ....142..133G,
       author = {{Geller}, Margaret J. and {Diaferio}, Antonaldo and {Kurtz}, Michael J.},
        title = "{Mapping the Universe: The 2010 Russell Lecture}",
      journal = {\aj},
         year = 2011,
        month = oct,
       volume = {142},
       number = {4},
          eid = {133},
        pages = {133},
          doi = {10.1088/0004-6256/142/4/133},
archivePrefix = {arXiv},
       eprint = {1110.1380},
 primaryClass = {astro-ph.CO},
       adsurl = {https://ui.adsabs.harvard.edu/abs/2011AJ....142..133G}
}

@ARTICLE{2019A&A...621A.133B,
       author = {{Borlaff}, Alejandro and {Trujillo}, Ignacio and {Rom{\'a}n}, Javier and {Beckman}, John E. and {Eliche-Moral}, M. Carmen and {Infante-S{\'a}inz}, Ra{\'u}l and {Lumbreras-Calle}, Alejandro and {de Almagro}, Rodrigo Takuro Sato Mart{\'\i}n and {G{\'o}mez-Guijarro}, Carlos and {Cebri{\'a}n}, Mar{\'\i}a and {Dorta}, Antonio and {Cardiel}, Nicol{\'a}s and {Akhlaghi}, Mohammad and {Mart{\'\i}nez-Lombilla}, Cristina},
        title = "{The missing light of the Hubble Ultra Deep Field}",
      journal = {\aap},
         year = 2019,
        month = jan,
       volume = {621},
          eid = {A133},
        pages = {A133},
          doi = {10.1051/0004-6361/201834312},
archivePrefix = {arXiv},
       eprint = {1810.00002},
 primaryClass = {astro-ph.GA},
       adsurl = {https://ui.adsabs.harvard.edu/abs/2019A&A...621A.133B}
}

@INPROCEEDINGS{DS9_2003ASPC..295..489J,
       author = {{Joye}, W.~A. and {Mandel}, E.},
        title = "{New Features of SAOImage DS9}",
    booktitle = {Astronomical Data Analysis Software and Systems XII},
         year = 2003,
       editor = {{Payne}, H.~E. and {Jedrzejewski}, R.~I. and {Hook}, R.~N.},
       series = {Astronomical Society of the Pacific Conference Series},
       volume = {295},
        month = jan,
        pages = {489},
       adsurl = {https://ui.adsabs.harvard.edu/abs/2003ASPC..295..489J}
}

@article{IoU1901,
    author = {Jaccard, Paul},
    year = {1901},
    month = {01},
    pages = {241-72},
    title = {Distribution de la Flore Alpine dans le Bassin des Dranses et dans quelques régions voisines.},
    volume = {37},
    journal = {Bulletin de la Societe Vaudoise des Sciences Naturelles},
    doi = {10.5169/seals-266440}
}

@ARTICLE{Low_etal1984,
       author = {{Low}, F.~J. and {Beintema}, D.~A. and {Gautier}, T.~N. and {Gillett}, F.~C. and {Beichman}, C.~A. and {Neugebauer}, G. and {Young}, E. and {Aumann}, H.~H. and {Boggess}, N. and {Emerson}, J.~P. and {Habing}, H.~J. and {Hauser}, M.~G. and {Houck}, J.~R. and {Rowan-Robinson}, M. and {Soifer}, B.~T. and {Walker}, R.~G. and {Wesselius}, P.~R.},
        title = "{Infrared cirrus: new components of the extended infrared emission.}",
      journal = {\apjl},
         year = 1984,
        month = mar,
       volume = {278},
        pages = {L19-L22},
          doi = {10.1086/184213},
       adsurl = {https://ui.adsabs.harvard.edu/abs/1984ApJ...278L..19L}
}

@ARTICLE{Kiss_etal2001,
       author = {{Kiss}, Cs. and {{\'A}brah{\'a}m}, P. and {Klaas}, U. and {Juvela}, M. and {Lemke}, D.},
        title = "{Sky confusion noise in the far-infrared: Cirrus, galaxies and the cosmic far-infrared background}",
      journal = {\aap},
         year = 2001,
        month = dec,
       volume = {379},
        pages = {1161-1169},
          doi = {10.1051/0004-6361:20011394},
archivePrefix = {arXiv},
       eprint = {astro-ph/0110143},
 primaryClass = {astro-ph},
       adsurl = {https://ui.adsabs.harvard.edu/abs/2001A&A...379.1161K}
}

@ARTICLE{Kiss_etal2003,
       author = {{Kiss}, Cs. and {{\'A}brah{\'a}m}, P. and {Klaas}, U. and {Lemke}, D. and {H{\'e}raudeau}, Ph. and {del Burgo}, C. and {Herbstmeier}, U.},
        title = "{Small-scale structure of the galactic cirrus emission}",
      journal = {\aap},
         year = 2003,
        month = feb,
       volume = {399},
        pages = {177-185},
          doi = {10.1051/0004-6361:20021787},
archivePrefix = {arXiv},
       eprint = {astro-ph/0212094},
 primaryClass = {astro-ph},
       adsurl = {https://ui.adsabs.harvard.edu/abs/2003A&A...399..177K}
}

@ARTICLE{Martin_etal2010,
       author = {{Martin}, P.~G. and {Miville-Desch{\^e}nes}, M. -A. and {Roy}, A. and {Bernard}, J. -P. and {Molinari}, S. and {Billot}, N. and {Brunt}, C. and {Calzoletti}, L. and {Digiorgio}, A.~M. and {Elia}, D. and {Faustini}, F. and {Joncas}, G. and {Mottram}, J.~C. and {Natoli}, P. and {Noriega-Crespo}, A. and {Paladini}, R. and {Robitaille}, J.~F. and {Strafella}, F. and {Traficante}, A. and {Veneziani}, M.},
        title = "{Direct estimate of cirrus noise in Herschel Hi-GAL images}",
      journal = {\aap},
         year = 2010,
        month = jul,
       volume = {518},
          eid = {L105},
        pages = {L105},
          doi = {10.1051/0004-6361/201014684},
archivePrefix = {arXiv},
       eprint = {1005.3076},
 primaryClass = {astro-ph.GA},
       adsurl = {https://ui.adsabs.harvard.edu/abs/2010A&A...518L.105M}
}

@ARTICLE{Planck_etal2011,
       author = {{Planck Collaboration} and {Ade}, P.~A.~R. and {Aghanim}, N. and {Arnaud}, M. and {Ashdown}, M. and {Aumont}, J. and {Baccigalupi}, C. and {Balbi}, A. and {Banday}, A.~J. and {Barreiro}, R.~B. and {Bartlett}, J.~G. and {Battaner}, E. and {Benabed}, K. and {Beno{\^\i}t}, A. and {Bernard}, J. -P. and {Bersanelli}, M. and {Bhatia}, R. and {Bock}, J.~J. and {Bonaldi}, A. and {Bond}, J.~R. and {Borrill}, J. and {Bouchet}, F.~R. and {Boulanger}, F. and {Bucher}, M. and {Burigana}, C. and {Cabella}, P. and {Cantalupo}, C.~M. and {Cardoso}, J. -F. and {Catalano}, A. and {Cay{\'o}n}, L. and {Challinor}, A. and {Chamballu}, A. and {Chiang}, L. -Y. and {Christensen}, P.~R. and {Clements}, D.~L. and {Colombi}, S. and {Couchot}, F. and {Coulais}, A. and {Crill}, B.~P. and {Cuttaia}, F. and {Danese}, L. and {Davies}, R.~D. and {de Bernardis}, P. and {de Gasperis}, G. and {de Rosa}, A. and {de Zotti}, G. and {Delabrouille}, J. and {Delouis}, J. -M. and {D{\'e}sert}, F. -X. and {Dickinson}, C. and {Doi}, Y. and {Donzelli}, S. and {Dor{\'e}}, O. and {D{\"o}rl}, U. and {Douspis}, M. and {Dupac}, X. and {Efstathiou}, G. and {En{\ss}lin}, T.~A. and {Falgarone}, E. and {Finelli}, F. and {Forni}, O. and {Frailis}, M. and {Franceschi}, E. and {Galeotta}, S. and {Ganga}, K. and {Giard}, M. and {Giardino}, G. and {Giraud-H{\'e}raud}, Y. and {Gonz{\'a}lez-Nuevo}, J. and {G{\'o}rski}, K.~M. and {Gratton}, S. and {Gregorio}, A. and {Gruppuso}, A. and {Hansen}, F.~K. and {Harrison}, D. and {Helou}, G. and {Henrot-Versill{\'e}}, S. and {Herranz}, D. and {Hildebrandt}, S.~R. and {Hivon}, E. and {Hobson}, M. and {Holmes}, W.~A. and {Hovest}, W. and {Hoyland}, R.~J. and {Huffenberger}, K.~M. and {Ikeda}, N. and {Jaffe}, A.~H. and {Jones}, W.~C. and {Juvela}, M. and {Keih{\"a}nen}, E. and {Keskitalo}, R. and {Kisner}, T.~S. and {Kitamura}, Y. and {Kneissl}, R. and {Knox}, L. and {Kurki-Suonio}, H. and {Lagache}, G. and {Lamarre}, J. -M. and {Lasenby}, A. and {Laureijs}, R.~J. and {Lawrence}, C.~R. and {Leach}, S. and {Leonardi}, R. and {Leroy}, C. and {Linden-V{\o}rnle}, M. and {L{\'o}pez-Caniego}, M. and {Lubin}, P.~M. and {Mac{\'\i}as-P{\'e}rez}, J.~F. and {MacTavish}, C.~J. and {Maffei}, B. and {Malinen}, J. and {Mandolesi}, N. and {Mann}, R. and {Maris}, M. and {Marshall}, D.~J. and {Martin}, P. and {Mart{\'\i}nez-Gonz{\'a}lez}, E. and {Masi}, S. and {Matarrese}, S. and {Matthai}, F. and {Mazzotta}, P. and {McGehee}, P. and {Melchiorri}, A. and {Mendes}, L. and {Mennella}, A. and {Meny}, C. and {Mitra}, S. and {Miville-Desch{\^e}nes}, M. -A. and {Moneti}, A. and {Montier}, L. and {Morgante}, G. and {Mortlock}, D. and {Munshi}, D. and {Murphy}, A. and {Naselsky}, P. and {Nati}, F. and {Natoli}, P. and {Netterfield}, C.~B. and {N{\o}rgaard-Nielsen}, H.~U. and {Noviello}, F. and {Novikov}, D. and {Novikov}, I. and {Osborne}, S. and {Pagani}, L. and {Pajot}, F. and {Paladini}, R. and {Pasian}, F. and {Patanchon}, G. and {Pelkonen}, V. -M. and {Perdereau}, O. and {Perotto}, L. and {Perrotta}, F. and {Piacentini}, F. and {Piat}, M. and {Plaszczynski}, S. and {Pointecouteau}, E. and {Polenta}, G. and {Ponthieu}, N. and {Poutanen}, T. and {Pr{\'e}zeau}, G. and {Prunet}, S. and {Puget}, J. -L. and {Reach}, W.~T. and {Rebolo}, R. and {Reinecke}, M. and {Renault}, C. and {Ricciardi}, S. and {Riller}, T. and {Ristorcelli}, I. and {Rocha}, G. and {Rosset}, C. and {Rowan-Robinson}, M. and {Rubi{\~n}o-Mart{\'\i}n}, J.~A. and {Rusholme}, B. and {Sandri}, M. and {Santos}, D. and {Savini}, G. and {Scott}, D. and {Seiffert}, M.~D. and {Smoot}, G.~F. and {Starck}, J. -L. and {Stivoli}, F. and {Stolyarov}, V. and {Sudiwala}, R. and {Sygnet}, J. -F. and {Tauber}, J.~A. and {Terenzi}, L. and {Toffolatti}, L. and {Tomasi}, M. and {Torre}, J. -P. and {Toth}, V. and {Tristram}, M. and {Tuovinen}, J. and {Umana}, G. and {Valenziano}, L. and {Vielva}, P. and {Villa}, F. and {Vittorio}, N. and {Wade}, L.~A. and {Wandelt}, B.~D. and {Ysard}, N. and {Yvon}, D. and {Zacchei}, A. and {Zonca}, A.},
        title = "{Planck early results. XXII. The submillimetre properties of a sample of Galactic cold clumps}",
      journal = {\aap},
         year = 2011,
        month = dec,
       volume = {536},
          eid = {A22},
        pages = {A22},
          doi = {10.1051/0004-6361/201116481},
archivePrefix = {arXiv},
       eprint = {1101.2034},
 primaryClass = {astro-ph.GA},
       adsurl = {https://ui.adsabs.harvard.edu/abs/2011A&A...536A..22P}
}

@ARTICLE{Penin_etal2012,
       author = {{P{\'e}nin}, A. and {Lagache}, G. and {Noriega-Crespo}, A. and {Grain}, J. and {Miville-Desch{\^e}nes}, M. -A. and {Ponthieu}, N. and {Martin}, P. and {Blagrave}, K. and {Lockman}, F.~J.},
        title = "{An accurate measurement of the anisotropies and mean level of the cosmic infrared background at 100 {\ensuremath{\mu}}m and 160 {\ensuremath{\mu}}m}",
      journal = {\aap},
         year = 2012,
        month = jul,
       volume = {543},
          eid = {A123},
        pages = {A123},
          doi = {10.1051/0004-6361/201015929},
archivePrefix = {arXiv},
       eprint = {1105.1463},
 primaryClass = {astro-ph.CO},
       adsurl = {https://ui.adsabs.harvard.edu/abs/2012A&A...543A.123P}
}

@ARTICLE{Schisano_etal2020,
       author = {{Schisano}, Eugenio and {Molinari}, S. and {Elia}, D. and {Benedettini}, M. and {Olmi}, L. and {Pezzuto}, S. and {Traficante}, A. and {Brescia}, M. and {Cavuoti}, S. and {di Giorgio}, A.~M. and {Liu}, S.~J. and {Moore}, T.~J.~T. and {Noriega-Crespo}, A. and {Riccio}, G. and {Baldeschi}, A. and {Becciani}, U. and {Peretto}, N. and {Merello}, M. and {Vitello}, F. and {Zavagno}, A. and {Beltr{\'a}n}, M.~T. and {Cambr{\'e}sy}, L. and {Eden}, D.~J. and {Li Causi}, G. and {Molinaro}, M. and {Palmeirim}, P. and {Sciacca}, E. and {Testi}, L. and {Umana}, G. and {Whitworth}, A.~P.},
        title = "{The Hi-GAL catalogue of dusty filamentary structures in the Galactic plane}",
      journal = {\mnras},
         year = 2020,
        month = mar,
       volume = {492},
       number = {4},
        pages = {5420-5456},
          doi = {10.1093/mnras/stz3466},
archivePrefix = {arXiv},
       eprint = {1912.04020},
 primaryClass = {astro-ph.GA},
       adsurl = {https://ui.adsabs.harvard.edu/abs/2020MNRAS.492.5420S}
}

@ARTICLE{Vacouleurs_1955,
       author = {{\noopsort{Vaucouleurs}}{de Vaucouleurs}, G.},
        title = "{Emission nebulosities near the south pole}",
      journal = {The Observatory},
         year = 1955,
        month = jun,
       volume = {75},
        pages = {129-130},
       adsurl = {https://ui.adsabs.harvard.edu/abs/1955Obs....75..129D}
}

@ARTICLE{Vacouleurs_1960,
       author = {{\noopsort{Vaucouleurs}}{de Vaucouleurs}, G.},
        title = "{Emission nebulosities near the south pole - II}",
      journal = {The Observatory},
         year = 1960,
        month = jun,
       volume = {80},
        pages = {106-109},
       adsurl = {https://ui.adsabs.harvard.edu/abs/1960Obs....80..106D}
}

@ARTICLE{Vaucouleurs_Freeman_1972,
       author = {{\noopsort{Vaucouleurs}}{de Vaucouleurs}, G. and {Freeman}, K.~C.},
        title = "{Structure and dynamics of barred spiral galaxies, in particular of the Magellanic type}",
      journal = {Vistas in Astronomy},
         year = 1972,
        month = jan,
       volume = {14},
       number = {1},
        pages = {163-294},
          doi = {10.1016/0083-6656(72)90026-8},
       adsurl = {https://ui.adsabs.harvard.edu/abs/1972VA.....14..163D}
}

@ARTICLE{Sandage_1976,
       author = {{Sandage}, Allen},
        title = "{High-latitude reflection nebulosities illuminated by the galactic plane}",
      journal = {\aj},
         year = 1976,
        month = nov,
       volume = {81},
        pages = {954},
          doi = {10.1086/111975},
       adsurl = {https://ui.adsabs.harvard.edu/abs/1976AJ.....81..954S}
}

@ARTICLE{Mattila_1979,
       author = {{Mattila}, K.},
        title = "{Optical extinction and surface brightness observations of the dark nebulae Lynds 134 and Lynds 1778/1780.}",
      journal = {\aap},
         year = 1979,
        month = oct,
       volume = {78},
        pages = {253-263},
       adsurl = {https://ui.adsabs.harvard.edu/abs/1979A&A....78..253M}
}

@ARTICLE{Vries_1985,
       author = {{\noopsort{Vries}}{de Vries}, C.~P. and {Le Poole}, R.~S.},
        title = "{Comparison of optical appearance and infrared emission of some high latitude extended dust clouds}",
      journal = {\aap},
         year = 1985,
        month = apr,
       volume = {145},
       number = {2},
        pages = {L7-L9},
       adsurl = {https://ui.adsabs.harvard.edu/abs/1985A&A...145L...7D}
}

@ARTICLE{Ienaka_etal2013,
       author = {{Ienaka}, N. and {Kawara}, K. and {Matsuoka}, Y. and {Sameshima}, H. and {Oyabu}, S. and {Tsujimoto}, T. and {Peterson}, B.~A.},
        title = "{Diffuse Galactic Light in the Field of the Translucent High Galactic Latitude Cloud MBM32}",
      journal = {\apj},
         year = 2013,
        month = apr,
       volume = {767},
       number = {1},
          eid = {80},
        pages = {80},
          doi = {10.1088/0004-637X/767/1/80},
archivePrefix = {arXiv},
       eprint = {1303.0938},
 primaryClass = {astro-ph.GA},
       adsurl = {https://ui.adsabs.harvard.edu/abs/2013ApJ...767...80I}
}

@ARTICLE{Miville-Deschenes_etal2016,
       author = {{Miville-Desch{\^e}nes}, M. -A. and {Duc}, P. -A. and {Marleau}, F. and {Cuillandre}, J. -C. and {Didelon}, P. and {Gwyn}, S. and {Karabal}, E.},
        title = "{Probing interstellar turbulence in cirrus with deep optical imaging: no sign of energy dissipation at 0.01 pc scale}",
      journal = {\aap},
         year = 2016,
        month = aug,
       volume = {593},
          eid = {A4},
        pages = {A4},
          doi = {10.1051/0004-6361/201628503},
archivePrefix = {arXiv},
       eprint = {1605.08360},
 primaryClass = {astro-ph.GA},
       adsurl = {https://ui.adsabs.harvard.edu/abs/2016A&A...593A...4M}
}

@ARTICLE{Haikala_etal_1995,
       author = {{Haikala}, Lauri K. and {Mattila}, Kalevi and {Bowyer}, Stuart and {Sasseen}, Timothy P. and {Lampton}, Michael and {Knude}, Jens},
        title = "{Discovery and Imaging of a Galactic Cirrus Cloud with the Far Ultraviolet Space Telescope}",
      journal = {\apjl},
         year = 1995,
        month = apr,
       volume = {443},
        pages = {L33},
          doi = {10.1086/187829},
       adsurl = {https://ui.adsabs.harvard.edu/abs/1995ApJ...443L..33H}
}

@ARTICLE{Gillmon_Shull2006,
       author = {{Gillmon}, Kristen and {Shull}, J. Michael},
        title = "{Molecular Hydrogen in Infrared Cirrus}",
      journal = {\apj},
         year = 2006,
        month = jan,
       volume = {636},
       number = {2},
        pages = {908-915},
          doi = {10.1086/498055},
archivePrefix = {arXiv},
       eprint = {astro-ph/0507587},
 primaryClass = {astro-ph},
       adsurl = {https://ui.adsabs.harvard.edu/abs/2006ApJ...636..908G}
}

@ARTICLE{Boissier_2015,
       author = {{Boissier}, S. and {Boselli}, A. and {Voyer}, E. and {Bianchi}, S. and {Pappalardo}, C. and {Guhathakurta}, P. and {Heinis}, S. and {Cortese}, L. and {Duc}, P. -A. and {Cuillandre}, J. -C. and {Davies}, J.~I. and {Smith}, M.~W.~L.},
        title = "{The GALEX Ultraviolet Virgo Cluster Survey (GUViCS). V. Ultraviolet diffuse emission and cirrus properties in the Virgo cluster direction}",
      journal = {\aap},
         year = 2015,
        month = jul,
       volume = {579},
          eid = {A29},
        pages = {A29},
          doi = {10.1051/0004-6361/201526089},
archivePrefix = {arXiv},
       eprint = {1504.06111},
 primaryClass = {astro-ph.GA},
       adsurl = {https://ui.adsabs.harvard.edu/abs/2015A&A...579A..29B}
}

@ARTICLE{Akshaya_etal2019,
       author = {{Akshaya}, M.~S. and {Murthy}, Jayant and {Ravichandran}, S. and {Henry}, R.~C. and {Overduin}, James},
        title = "{Components of the diffuse ultraviolet radiation at high latitudes}",
      journal = {\mnras},
         year = 2019,
        month = oct,
       volume = {489},
       number = {1},
        pages = {1120-1126},
          doi = {10.1093/mnras/stz2186},
archivePrefix = {arXiv},
       eprint = {1908.02260},
 primaryClass = {astro-ph.GA},
       adsurl = {https://ui.adsabs.harvard.edu/abs/2019MNRAS.489.1120A}
}

@ARTICLE{Weiland_etal1986,
       author = {{Weiland}, Janet L. and {Blitz}, Leo and {Dwek}, Eli and {Hauser}, M.~G. and {Magnani}, Loris and {Rickard}, Lee J.},
        title = "{Infrared Cirrus and High-Latitude Molecular Clouds}",
      journal = {\apjl},
         year = 1986,
        month = jul,
       volume = {306},
        pages = {L101},
          doi = {10.1086/184714},
       adsurl = {https://ui.adsabs.harvard.edu/abs/1986ApJ...306L.101W}
}

@ARTICLE{deVries_etal1987,
       author = {{\noopsort{Vries}}{de Vries}, Hendrik Willem and {Heithausen}, Andreas and {Thaddeus}, Patrick},
        title = "{Molecular and Atomic Clouds Associated with Infrared Cirrus in Ursa Major}",
      journal = {\apj},
         year = 1987,
        month = aug,
       volume = {319},
        pages = {723},
          doi = {10.1086/165492},
       adsurl = {https://ui.adsabs.harvard.edu/abs/1987ApJ...319..723D}
}

@ARTICLE{Brandt_Draine_2012,
       author = {{Brandt}, Timothy D. and {Draine}, B.~T.},
        title = "{The Spectrum of the Diffuse Galactic Light: The Milky Way in Scattered Light}",
      journal = {\apj},
         year = 2012,
        month = jan,
       volume = {744},
       number = {2},
          eid = {129},
        pages = {129},
          doi = {10.1088/0004-637X/744/2/129},
archivePrefix = {arXiv},
       eprint = {1109.4175},
 primaryClass = {astro-ph.GA},
       adsurl = {https://ui.adsabs.harvard.edu/abs/2012ApJ...744..129B}
}

@ARTICLE{Chellew_etal_2022,
       author = {{Chellew}, Blake and {Brandt}, Timothy D. and {Hensley}, Brandon S. and {Draine}, Bruce T. and {Matthaey}, Eve},
        title = "{An Optical Spectrum of the Diffuse Galactic Light from BOSS and IRIS}",
      journal = {arXiv e-prints},
         year = 2022,
        month = jan,
          eid = {arXiv:2201.01378},
        pages = {arXiv:2201.01378},
archivePrefix = {arXiv},
       eprint = {2201.01378},
 primaryClass = {astro-ph.GA},
       adsurl = {https://ui.adsabs.harvard.edu/abs/2022arXiv220101378C}
}

@article{Zubko_2004,
	doi = {10.1086/382351},
	url = {https://doi.org/10.1086/382351},
	year = 2004,
	month = {jun},
	publisher = {American Astronomical Society},
	volume = {152},
	number = {2},
	pages = {211--249},
	author = {Viktor Zubko and Eli Dwek and Richard G. Arendt},
	title = {Interstellar Dust Models Consistent with Extinction, Emission, and Abundance Constraints},
	journal = {The Astrophysical Journal Supplement Series}
}

@article{Weingartner2001DustGD,
  title={Dust Grain-Size Distributions and Extinction in the Milky Way, Large Magellanic Cloud, and Small Magellanic Cloud},
  author={Joseph C. Weingartner and Bruce T. Draine},
  journal={The Astrophysical Journal},
  year={2001},
  volume={548},
  pages={296 - 309},
  url={https://api.semanticscholar.org/CorpusID:121359249}
}

@ARTICLE{Bazell_Desert1988,
       author = {{Bazell}, D. and {Desert}, F.~X.},
        title = "{Fractal Structure of Interstellar Cirrus}",
      journal = {\apj},
         year = 1988,
        month = oct,
       volume = {333},
        pages = {353},
          doi = {10.1086/166751},
       adsurl = {https://ui.adsabs.harvard.edu/abs/1988ApJ...333..353B}
}

@ARTICLE{Falgarone_etal1991,
       author = {{Falgarone}, E. and {Phillips}, T.~G. and {Walker}, C.~K.},
        title = "{The Edges of Molecular Clouds: Fractal Boundaries and Density Structure}",
      journal = {\apj},
         year = 1991,
        month = sep,
       volume = {378},
        pages = {186},
          doi = {10.1086/170419},
       adsurl = {https://ui.adsabs.harvard.edu/abs/1991ApJ...378..186F}
}

@ARTICLE{Hetem_Lepine1993,
       author = {{Hetem}, A., Jr. and {Lepine}, J.~R.~D.},
        title = "{Fractal 3-D simulations of molecular clouds.}",
      journal = {\aap},
         year = 1993,
        month = mar,
       volume = {270},
        pages = {451-461},
       adsurl = {https://ui.adsabs.harvard.edu/abs/1993A&A...270..451H}
}

@ARTICLE{Vogelaar_Wakker1994,
       author = {{Vogelaar}, M.~G.~R. and {Wakker}, B.~P.},
        title = "{Measuring the fractal structure of interstellar clouds.}",
      journal = {\aap},
         year = 1994,
        month = nov,
       volume = {291},
        pages = {557-568},
       adsurl = {https://ui.adsabs.harvard.edu/abs/1994A&A...291..557V}
}

@ARTICLE{Elmegreen_Falgarone1996,
       author = {{Elmegreen}, Bruce G. and {Falgarone}, Edith},
        title = "{A Fractal Origin for the Mass Spectrum of Interstellar Clouds}",
      journal = {\apj},
         year = 1996,
        month = nov,
       volume = {471},
        pages = {816},
          doi = {10.1086/178009},
       adsurl = {https://ui.adsabs.harvard.edu/abs/1996ApJ...471..816E}
}

@ARTICLE{Sanchez_etal2005,
       author = {{S{\'a}nchez}, N{\'e}stor and {Alfaro}, Emilio J. and {P{\'e}rez}, Enrique},
        title = "{The Fractal Dimension of Projected Clouds}",
      journal = {\apj},
         year = 2005,
        month = jun,
       volume = {625},
       number = {2},
        pages = {849-856},
          doi = {10.1086/429553},
archivePrefix = {arXiv},
       eprint = {astro-ph/0501573},
 primaryClass = {astro-ph},
       adsurl = {https://ui.adsabs.harvard.edu/abs/2005ApJ...625..849S}
}

@ARTICLE{Juvela_etal2018,
       author = {{Juvela}, M. and {Malinen}, J. and {Montillaud}, J. and {Pelkonen}, V. -M. and {Ristorcelli}, I. and {T{\'o}th}, L.~V.},
        title = "{Galactic cold cores. IX. Column density structures and radiative-transfer modelling}",
      journal = {\aap},
         year = 2018,
        month = jun,
       volume = {614},
          eid = {A83},
        pages = {A83},
          doi = {10.1051/0004-6361/201630304},
archivePrefix = {arXiv},
       eprint = {1801.02419},
 primaryClass = {astro-ph.GA},
       adsurl = {https://ui.adsabs.harvard.edu/abs/2018A&A...614A..83J}
}

@ARTICLE{Padoan_etal_2001,
       author = {{Padoan}, Paolo and {Juvela}, Mika and {Goodman}, Alyssa A. and {Nordlund}, {\r{A}}ke},
        title = "{The Turbulent Shock Origin of Proto-Stellar Cores}",
      journal = {\apj},
         year = 2001,
        month = may,
       volume = {553},
       number = {1},
        pages = {227-234},
          doi = {10.1086/320636},
archivePrefix = {arXiv},
       eprint = {astro-ph/0011122},
 primaryClass = {astro-ph},
       adsurl = {https://ui.adsabs.harvard.edu/abs/2001ApJ...553..227P}
}

@ARTICLE{Kowal_Lazarian_2007,
       author = {{Kowal}, Grzegorz and {Lazarian}, A.},
        title = "{Scaling Relations of Compressible MHD Turbulence}",
      journal = {\apjl},
         year = 2007,
        month = sep,
       volume = {666},
       number = {2},
        pages = {L69-L72},
          doi = {10.1086/521788},
archivePrefix = {arXiv},
       eprint = {0705.2464},
 primaryClass = {astro-ph},
       adsurl = {https://ui.adsabs.harvard.edu/abs/2007ApJ...666L..69K}
}

@ARTICLE{2014arXiv1406.2661G,
       author = {{Goodfellow}, Ian J. and {Pouget-Abadie}, Jean and {Mirza}, Mehdi and {Xu}, Bing and {Warde-Farley}, David and {Ozair}, Sherjil and {Courville}, Aaron and {Bengio}, Yoshua},
        title = "{Generative Adversarial Networks}",
      journal = {arXiv e-prints},
         year = 2014,
        month = jun,
          eid = {arXiv:1406.2661},
        pages = {arXiv:1406.2661},
          doi = {10.48550/arXiv.1406.2661},
archivePrefix = {arXiv},
       eprint = {1406.2661},
 primaryClass = {stat.ML},
       adsurl = {https://ui.adsabs.harvard.edu/abs/2014arXiv1406.2661G}
}

@ARTICLE{Federrath_etal2009,
       author = {{Federrath}, Christoph and {Klessen}, Ralf S. and {Schmidt}, Wolfram},
        title = "{The Fractal Density Structure in Supersonic Isothermal Turbulence: Solenoidal Versus Compressive Energy Injection}",
      journal = {\apj},
         year = 2009,
        month = feb,
       volume = {692},
       number = {1},
        pages = {364-374},
          doi = {10.1088/0004-637X/692/1/364},
archivePrefix = {arXiv},
       eprint = {0710.1359},
 primaryClass = {astro-ph},
       adsurl = {https://ui.adsabs.harvard.edu/abs/2009ApJ...692..364F}
}

@ARTICLE{Konstandin_etal2016,
       author = {{Konstandin}, L. and {Schmidt}, W. and {Girichidis}, P. and {Peters}, T. and {Shetty}, R. and {Klessen}, R.~S.},
        title = "{Mach number study of supersonic turbulence: the properties of the density field}",
      journal = {\mnras},
         year = 2016,
        month = aug,
       volume = {460},
       number = {4},
        pages = {4483-4491},
          doi = {10.1093/mnras/stw1313},
archivePrefix = {arXiv},
       eprint = {1506.03834},
 primaryClass = {astro-ph.SR},
       adsurl = {https://ui.adsabs.harvard.edu/abs/2016MNRAS.460.4483K}
}

@ARTICLE{Beattie_etal2019a,
       author = {{Beattie}, James R. and {Federrath}, Christoph and {Klessen}, Ralf S.},
        title = "{The relation between the true and observed fractal dimensions of turbulent clouds}",
      journal = {\mnras},
         year = 2019,
        month = aug,
       volume = {487},
       number = {2},
        pages = {2070-2081},
          doi = {10.1093/mnras/stz1416},
archivePrefix = {arXiv},
       eprint = {1905.04979},
 primaryClass = {astro-ph.GA},
       adsurl = {https://ui.adsabs.harvard.edu/abs/2019MNRAS.487.2070B}
}

@ARTICLE{Beattie_etal2019b,
       author = {{Beattie}, James R. and {Federrath}, Christoph and {Klessen}, Ralf S. and {Schneider}, Nicola},
        title = "{The relation between the turbulent Mach number and observed fractal dimensions of turbulent clouds}",
      journal = {\mnras},
         year = 2019,
        month = sep,
       volume = {488},
       number = {2},
        pages = {2493-2502},
          doi = {10.1093/mnras/stz1853},
archivePrefix = {arXiv},
       eprint = {1907.01689},
 primaryClass = {astro-ph.GA},
       adsurl = {https://ui.adsabs.harvard.edu/abs/2019MNRAS.488.2493B}
}

@article{Koyama_Inutsuka_2000,
	doi = {10.1086/308594},
	url = {https://doi.org/10.1086/308594},
	year = 2000,
	month = {apr},
	publisher = {American Astronomical Society},
	volume = {532},
	number = {2},
	pages = {980--993},
	author = {Hiroshi Koyama and Shu-Ichiro Inutsuka},
	title = {Molecular Cloud Formation in Shock-compressed Layers},
	journal = {The Astrophysical Journal}
}

@article{Vazquez_Semadeni_2007,
	doi = {10.1086/510771},
	url = {https://doi.org/10.1086/510771},
	year = 2007,
	month = {mar},
	publisher = {American Astronomical Society},
	volume = {657},
	number = {2},
	pages = {870--883},
	author = {Enrique Vazquez-Semadeni and Gilberto C. Gomez and A. Katharina Jappsen and Javier Ballesteros-Paredes and Ricardo F. Gonzalez and Ralf S. Klessen},
	title = {Molecular Cloud Evolution. {II}. From Cloud Formation to the Early Stages of Star Formation in Decaying Conditions},
	journal = {The Astrophysical Journal}
}

@article{Nagai_1998,
	doi = {10.1086/306249},
	url = {https://doi.org/10.1086/306249},
	year = 1998,
	month = {oct},
	publisher = {American Astronomical Society},
	volume = {506},
	number = {1},
	pages = {306--322},
	author = {Tomoya Nagai and Shu-ichiro Inutsuka and Shoken M. Miyama},
	title = {An Origin of Filamentary Structure in Molecular Clouds},
	journal = {The Astrophysical Journal}
}

@ARTICLE{Hennebelle_2013,
       author = {{Hennebelle}, Patrick},
        title = "{On the origin of non-self-gravitating filaments in the ISM}",
      journal = {\aap},
         year = 2013,
        month = aug,
       volume = {556},
          eid = {A153},
        pages = {A153},
          doi = {10.1051/0004-6361/201321292},
archivePrefix = {arXiv},
       eprint = {1306.5452},
 primaryClass = {astro-ph.GA},
       adsurl = {https://ui.adsabs.harvard.edu/abs/2013A&A...556A.153H}
}

@ARTICLE{Cortese_etal_2010,
       author = {{Cortese}, L. and {Bendo}, G.~J. and {Isaak}, K.~G. and {Davies}, J.~I. and {Kent}, B.~R.},
        title = "{Diffuse far-infrared and ultraviolet emission in the NGC 4435/4438 system: tidal stream or Galactic cirrus?}",
      journal = {\mnras},
         year = 2010,
        month = mar,
       volume = {403},
       number = {1},
        pages = {L26-L30},
          doi = {10.1111/j.1745-3933.2009.00808.x},
archivePrefix = {arXiv},
       eprint = {1001.0980},
 primaryClass = {astro-ph.CO},
       adsurl = {https://ui.adsabs.harvard.edu/abs/2010MNRAS.403L..26C}
}

@ARTICLE{Sollima_etal_2010,
       author = {{Sollima}, A. and {Gil de Paz}, A. and {Martinez-Delgado}, D. and {Gabany}, R.~J. and {Gallego-Laborda}, J.~J. and {Hallas}, T.},
        title = "{A multi-wavelength analysis of M 81: insight on the nature of Arp's loop}",
      journal = {\aap},
         year = 2010,
        month = jun,
       volume = {516},
          eid = {A83},
        pages = {A83},
          doi = {10.1051/0004-6361/201014085},
archivePrefix = {arXiv},
       eprint = {1004.1610},
 primaryClass = {astro-ph.CO},
       adsurl = {https://ui.adsabs.harvard.edu/abs/2010A&A...516A..83S}
}

@ARTICLE{Rudick_etal_2010,
       author = {{Rudick}, Craig S. and {Mihos}, J. Christopher and {Harding}, Paul and {Feldmeier}, John J. and {Janowiecki}, Steven and {Morrison}, Heather L.},
        title = "{Optical Colors of Intracluster Light in the Virgo Cluster Core}",
      journal = {\apj},
         year = 2010,
        month = sep,
       volume = {720},
       number = {1},
        pages = {569-580},
          doi = {10.1088/0004-637X/720/1/569},
archivePrefix = {arXiv},
       eprint = {1003.4500},
 primaryClass = {astro-ph.CO},
       adsurl = {https://ui.adsabs.harvard.edu/abs/2010ApJ...720..569R}
}

@ARTICLE{Davies_etal_2010,
       author = {{Davies}, J.~I. and {Wilson}, C.~D. and {Auld}, R. and {Baes}, M. and {Barlow}, M.~J. and {Bendo}, G.~J. and {Bock}, J.~J. and {Boselli}, A. and {Bradford}, M. and {Buat}, V. and {Castro-Rodriguez}, N. and {Chanial}, P. and {Charlot}, S. and {Ciesla}, L. and {Clements}, D.~L. and {Cooray}, A. and {Cormier}, D. and {Cortese}, L. and {Dwek}, E. and {Eales}, S.~A. and {Elbaz}, D. and {Galametz}, M. and {Galliano}, F. and {Gear}, W.~K. and {Glenn}, J. and {Gomez}, H.~L. and {Griffin}, M. and {Hony}, S. and {Isaak}, K.~G. and {Levenson}, L.~R. and {Lu}, N. and {Madden}, S. and {O'Halloran}, B. and {Okumura}, K. and {Oliver}, S. and {Page}, M.~J. and {Panuzzo}, P. and {Papageorgiou}, A. and {Parkin}, T.~J. and {Perez-Fournon}, I. and {Pohlen}, M. and {Rangwala}, N. and {Rigby}, E.~E. and {Roussel}, H. and {Rykala}, A. and {Sacchi}, N. and {Sauvage}, M. and {Schulz}, B. and {Schirm}, M.~R.~P. and {Smith}, M.~W.~L. and {Spinoglio}, L. and {Stevens}, J.~A. and {Srinivasan}, S. and {Symeonidis}, M. and {Trichas}, M. and {Vaccari}, M. and {Vigroux}, L. and {Wozniak}, H. and {Wright}, G.~S. and {Zeilinger}, W.~W.},
        title = "{On the origin of M81 group extended dust emission}",
      journal = {\mnras},
         year = 2010,
        month = nov,
       volume = {409},
       number = {1},
        pages = {102-108},
          doi = {10.1111/j.1365-2966.2010.17774.x},
archivePrefix = {arXiv},
       eprint = {1010.4770},
 primaryClass = {astro-ph.CO},
       adsurl = {https://ui.adsabs.harvard.edu/abs/2010MNRAS.409..102D}
}

@ARTICLE{Duc_etal_2018,
       author = {{Duc}, Pierre-Alain and {Cuillandre}, Jean-Charles and {Renaud}, Florent},
        title = "{Revisiting Stephan's Quintet with deep optical images}",
      journal = {\mnras},
         year = 2018,
        month = mar,
       volume = {475},
       number = {1},
        pages = {L40-L44},
          doi = {10.1093/mnrasl/sly004},
archivePrefix = {arXiv},
       eprint = {1712.07145},
 primaryClass = {astro-ph.GA},
       adsurl = {https://ui.adsabs.harvard.edu/abs/2018MNRAS.475L..40D}
}

@ARTICLE{Barrena_etal_2018,
       author = {{Barrena}, R. and {Streblyanska}, A. and {Ferragamo}, A. and {Rubi{\~n}o-Mart{\'\i}n}, J.~A. and {Aguado-Barahona}, A. and {Tramonte}, D. and {G{\'e}nova-Santos}, R.~T. and {Hempel}, A. and {Lietzen}, H. and {Aghanim}, N. and {Arnaud}, M. and {B{\"o}hringer}, H. and {Chon}, G. and {Democles}, J. and {Dahle}, H. and {Douspis}, M. and {Lasenby}, A.~N. and {Mazzotta}, P. and {Melin}, J.~B. and {Pointecouteau}, E. and {Pratt}, G.~W. and {Rossetti}, M. and {van der Burg}, R.~F.~J.},
        title = "{Optical validation and characterization of Planck PSZ1 sources at the Canary Islands observatories. I. First year of ITP13 observations}",
      journal = {\aap},
         year = 2018,
        month = aug,
       volume = {616},
          eid = {A42},
        pages = {A42},
          doi = {10.1051/0004-6361/201732315},
archivePrefix = {arXiv},
       eprint = {1803.05764},
 primaryClass = {astro-ph.CO},
       adsurl = {https://ui.adsabs.harvard.edu/abs/2018A&A...616A..42B}
}

@article{Molinari_etal_2010,
	doi = {10.1086/651314},
	url = {https://doi.org/10.1086/651314},
	year = 2010,
	month = {mar},
	publisher = {{IOP} Publishing},
	volume = {122},
	number = {889},
	pages = {314--325},
	author = {S. Molinari and B. Swinyard and J. Bally and M. Barlow and J.-P. Bernard and P. Martin and T. Moore and A. Noriega-Crespo and R. Plume and L. Testi and A. Zavagno and A. Abergel and B. Ali and P. Andr{\'{e}} and J.-P. Baluteau and M. Benedettini and O. Bern{\'{e}} and N. P. Billot and J. Blommaert and S. Bontemps and F. Boulanger and J. Brand and C. Brunt and M. Burton and L. Campeggio and S. Carey and P. Caselli and R. Cesaroni and J. Cernicharo and S. Chakrabarti and A. Chrysostomou and C. Codella and M. Cohen and M. Compiegne and C. J. Davis and P. de Bernardis and G. de Gasperis and J. Di Francesco and A. M. di Giorgio and D. Elia and F. Faustini and J. F. Fischera and Y. Fukui and G. A. Fuller and K. Ganga and P. Garcia-Lario and M. Giard and G. Giardino and J: Glenn and P. Goldsmith and M. Griffin and M. Hoare and M. Huang and B. Jiang and C. Joblin and G. Joncas and M. Juvela and J. Kirk and G. Lagache and J. Z. Li and T. L. Lim and S. D. Lord and P. W. Lucas and B. Maiolo and M. Marengo and D. Marshall and S. Masi and F. Massi and M. Matsuura and C. Meny and V. Minier and M.-A. Miville-Desch{\^{e}}nes and L. Montier and F. Motte and T. G. Müller and P. Natoli and J. Neves and L. Olmi and R. Paladini and D. Paradis and M. Pestalozzi and S. Pezzuto and F. Piacentini and M. Pomar{\`{e}}s and C. C. Popescu and W. T. Reach and J. Richer and I. Ristorcelli and A. Roy and P. Royer and D. Russeil and P. Saraceno and M. Sauvage and P. Schilke and N. Schneider-Bontemps and F. Schuller and B. Schultz and D. S. Shepherd and B. Sibthorpe and H. A. Smith and M. D. Smith and L. Spinoglio and D. Stamatellos and F. Strafella and G. Stringfellow and E. Sturm and R. Taylor and M. A. Thompson and R. J. Tuffs and G. Umana and L. Valenziano and R. Vavrek and S. Viti and C. Waelkens and D. Ward-Thompson and G. White and F. Wyrowski and H. W. Yorke and Q. Zhang},
	title = {Hi-{GAL}: The Herschel Infrared Galactic Plane Survey},
	journal = {Publications of the Astronomical Society of the Pacific}
}

@ARTICLE{Planck_colab_2016,
       author = {{Planck Collaboration} and {Adam}, R. and {Ade}, P.~A.~R. and {Aghanim}, N. and {Alves}, M.~I.~R. and {Arnaud}, M. and {Arzoumanian}, D. and {Ashdown}, M. and {Aumont}, J. and {Baccigalupi}, C. and {Banday}, A.~J. and {Barreiro}, R.~B. and {Bartolo}, N. and {Battaner}, E. and {Benabed}, K. and {Benoit-L{\'e}vy}, A. and {Bernard}, J. -P. and {Bersanelli}, M. and {Bielewicz}, P. and {Bonaldi}, A. and {Bonavera}, L. and {Bond}, J.~R. and {Borrill}, J. and {Bouchet}, F.~R. and {Boulanger}, F. and {Bracco}, A. and {Burigana}, C. and {Butler}, R.~C. and {Calabrese}, E. and {Cardoso}, J. -F. and {Catalano}, A. and {Chamballu}, A. and {Chiang}, H.~C. and {Christensen}, P.~R. and {Colombi}, S. and {Colombo}, L.~P.~L. and {Combet}, C. and {Couchot}, F. and {Crill}, B.~P. and {Curto}, A. and {Cuttaia}, F. and {Danese}, L. and {Davies}, R.~D. and {Davis}, R.~J. and {de Bernardis}, P. and {de Rosa}, A. and {de Zotti}, G. and {Delabrouille}, J. and {Dickinson}, C. and {Diego}, J.~M. and {Dole}, H. and {Donzelli}, S. and {Dor{\'e}}, O. and {Douspis}, M. and {Ducout}, A. and {Dupac}, X. and {Efstathiou}, G. and {Elsner}, F. and {En{\ss}lin}, T.~A. and {Eriksen}, H.~K. and {Falgarone}, E. and {Ferri{\`e}re}, K. and {Finelli}, F. and {Forni}, O. and {Frailis}, M. and {Fraisse}, A.~A. and {Franceschi}, E. and {Frejsel}, A. and {Galeotta}, S. and {Galli}, S. and {Ganga}, K. and {Ghosh}, T. and {Giard}, M. and {Gjerl{\o}w}, E. and {Gonz{\'a}lez-Nuevo}, J. and {G{\'o}rski}, K.~M. and {Gregorio}, A. and {Gruppuso}, A. and {Guillet}, V. and {Hansen}, F.~K. and {Hanson}, D. and {Harrison}, D.~L. and {Henrot-Versill{\'e}}, S. and {Hern{\'a}ndez-Monteagudo}, C. and {Herranz}, D. and {Hildebrandt}, S.~R. and {Hivon}, E. and {Hobson}, M. and {Holmes}, W.~A. and {Hovest}, W. and {Huffenberger}, K.~M. and {Hurier}, G. and {Jaffe}, A.~H. and {Jaffe}, T.~R. and {Jones}, W.~C. and {Juvela}, M. and {Keih{\"a}nen}, E. and {Keskitalo}, R. and {Kisner}, T.~S. and {Kneissl}, R. and {Knoche}, J. and {Kunz}, M. and {Kurki-Suonio}, H. and {Lagache}, G. and {Lamarre}, J. -M. and {Lasenby}, A. and {Lattanzi}, M. and {Lawrence}, C.~R. and {Leonardi}, R. and {Levrier}, F. and {Liguori}, M. and {Lilje}, P.~B. and {Linden-V{\o}rnle}, M. and {L{\'o}pez-Caniego}, M. and {Lubin}, P.~M. and {Mac{\'\i}as-P{\'e}rez}, J.~F. and {Maffei}, B. and {Maino}, D. and {Mandolesi}, N. and {Maris}, M. and {Marshall}, D.~J. and {Martin}, P.~G. and {Mart{\'\i}nez-Gonz{\'a}lez}, E. and {Masi}, S. and {Matarrese}, S. and {Mazzotta}, P. and {Melchiorri}, A. and {Mendes}, L. and {Mennella}, A. and {Migliaccio}, M. and {Miville-Desch{\^e}nes}, M. -A. and {Moneti}, A. and {Montier}, L. and {Morgante}, G. and {Mortlock}, D. and {Munshi}, D. and {Murphy}, J.~A. and {Naselsky}, P. and {Natoli}, P. and {N{\o}rgaard-Nielsen}, H.~U. and {Noviello}, F. and {Novikov}, D. and {Novikov}, I. and {Oppermann}, N. and {Oxborrow}, C.~A. and {Pagano}, L. and {Pajot}, F. and {Paoletti}, D. and {Pasian}, F. and {Perdereau}, O. and {Perotto}, L. and {Perrotta}, F. and {Pettorino}, V. and {Piacentini}, F. and {Piat}, M. and {Plaszczynski}, S. and {Pointecouteau}, E. and {Polenta}, G. and {Ponthieu}, N. and {Popa}, L. and {Pratt}, G.~W. and {Prunet}, S. and {Puget}, J. -L. and {Rachen}, J.~P. and {Reach}, W.~T. and {Reinecke}, M. and {Remazeilles}, M. and {Renault}, C. and {Ristorcelli}, I. and {Rocha}, G. and {Roudier}, G. and {Rubi{\~n}o-Mart{\'\i}n}, J.~A. and {Rusholme}, B. and {Sandri}, M. and {Santos}, D. and {Savini}, G. and {Scott}, D. and {Soler}, J.~D. and {Spencer}, L.~D. and {Stolyarov}, V. and {Sudiwala}, R. and {Sunyaev}, R. and {Sutton}, D. and {Suur-Uski}, A. -S. and {Sygnet}, J. -F. and {Tauber}, J.~A. and {Terenzi}, L. and {Toffolatti}, L. and {Tomasi}, M. and {Tristram}, M. and {Tucci}, M. and {Umana}, G. and {Valenziano}, L. and {Valiviita}, J. and {Van Tent}, B. and {Vielva}, P. and {Villa}, F. and {Wade}, L.~A. and {Wandelt}, B.~D. and {Wehus}, I.~K. and {Wiesemeyer}, H. and {Yvon}, D. and {Zacchei}, A. and {Zonca}, A.},
        title = "{Planck intermediate results. XXXII. The relative orientation between the magnetic field and structures traced by interstellar dust}",
      journal = {\aap},
         year = 2016,
        month = feb,
       volume = {586},
          eid = {A135},
        pages = {A135},
          doi = {10.1051/0004-6361/201425044},
archivePrefix = {arXiv},
       eprint = {1409.6728},
 primaryClass = {astro-ph.GA},
       adsurl = {https://ui.adsabs.harvard.edu/abs/2016A&A...586A.135P}
}

@ARTICLE{Soler_etal_2022,
       author = {{Soler}, J.~D. and {Miville-Desch{\^e}nes}, M. -A. and {Molinari}, S. and {Klessen}, R.~S. and {Hennebelle}, P. and {Testi}, L. and {McClure-Griffiths}, N.~M. and {Beuther}, H. and {Elia}, D. and {Schisano}, E. and {Traficante}, A. and {Girichidis}, P. and {Glover}, S.~C.~O. and {Smith}, R.~J. and {Sormani}, M. and {Tre{\ss}}, R.},
        title = "{The Galactic dynamics revealed by the filamentary structure in atomic hydrogen emission}",
      journal = {\aap},
         year = 2022,
        month = jun,
       volume = {662},
          eid = {A96},
        pages = {A96},
          doi = {10.1051/0004-6361/202243334},
archivePrefix = {arXiv},
       eprint = {2205.10426},
 primaryClass = {astro-ph.GA},
       adsurl = {https://ui.adsabs.harvard.edu/abs/2022A&A...662A..96S}
}

@ARTICLE{Menshchikov_2013,
       author = {{Men'shchikov}, A.},
        title = "{A multi-scale filament extraction method: getfilaments}",
      journal = {\aap},
         year = 2013,
        month = dec,
       volume = {560},
          eid = {A63},
        pages = {A63},
          doi = {10.1051/0004-6361/201321885},
archivePrefix = {arXiv},
       eprint = {1309.2170},
 primaryClass = {astro-ph.GA},
       adsurl = {https://ui.adsabs.harvard.edu/abs/2013A&A...560A..63M}
}

@ARTICLE{Salji_etal_2015,
       author = {{Salji}, C.~J. and {Richer}, J.~S. and {Buckle}, J.~V. and {di Francesco}, J. and {Hatchell}, J. and {Hogerheijde}, M. and {Johnstone}, D. and {Kirk}, H. and {Ward-Thompson}, D. and {JCMT GBS Consortium}},
        title = "{The JCMT Gould Belt Survey: properties of star-forming filaments in Orion A North}",
      journal = {\mnras},
         year = 2015,
        month = may,
       volume = {449},
       number = {2},
        pages = {1782-1796},
          doi = {10.1093/mnras/stv369},
       adsurl = {https://ui.adsabs.harvard.edu/abs/2015MNRAS.449.1782S}
}

@ARTICLE{Koch_Rosolowsky_2015,
       author = {{Koch}, Eric W. and {Rosolowsky}, Erik W.},
        title = "{Filament identification through mathematical morphology}",
      journal = {\mnras},
         year = 2015,
        month = oct,
       volume = {452},
       number = {4},
        pages = {3435-3450},
          doi = {10.1093/mnras/stv1521},
archivePrefix = {arXiv},
       eprint = {1507.02289},
 primaryClass = {astro-ph.GA},
       adsurl = {https://ui.adsabs.harvard.edu/abs/2015MNRAS.452.3435K}
}

@ARTICLE{Andre_etal_2010,
       author = {{Andr{\'e}}, Ph. and {Men'shchikov}, A. and {Bontemps}, S. and {K{\"o}nyves}, V. and {Motte}, F. and {Schneider}, N. and {Didelon}, P. and {Minier}, V. and {Saraceno}, P. and {Ward-Thompson}, D. and {di Francesco}, J. and {White}, G. and {Molinari}, S. and {Testi}, L. and {Abergel}, A. and {Griffin}, M. and {Henning}, Th. and {Royer}, P. and {Mer{\'\i}n}, B. and {Vavrek}, R. and {Attard}, M. and {Arzoumanian}, D. and {Wilson}, C.~D. and {Ade}, P. and {Aussel}, H. and {Baluteau}, J. -P. and {Benedettini}, M. and {Bernard}, J. -Ph. and {Blommaert}, J.~A.~D.~L. and {Cambr{\'e}sy}, L. and {Cox}, P. and {di Giorgio}, A. and {Hargrave}, P. and {Hennemann}, M. and {Huang}, M. and {Kirk}, J. and {Krause}, O. and {Launhardt}, R. and {Leeks}, S. and {Le Pennec}, J. and {Li}, J.~Z. and {Martin}, P.~G. and {Maury}, A. and {Olofsson}, G. and {Omont}, A. and {Peretto}, N. and {Pezzuto}, S. and {Prusti}, T. and {Roussel}, H. and {Russeil}, D. and {Sauvage}, M. and {Sibthorpe}, B. and {Sicilia-Aguilar}, A. and {Spinoglio}, L. and {Waelkens}, C. and {Woodcraft}, A. and {Zavagno}, A.},
        title = "{From filamentary clouds to prestellar cores to the stellar IMF: Initial highlights from the Herschel Gould Belt Survey}",
      journal = {\aap},
         year = 2010,
        month = jul,
       volume = {518},
          eid = {L102},
        pages = {L102},
          doi = {10.1051/0004-6361/201014666},
archivePrefix = {arXiv},
       eprint = {1005.2618},
 primaryClass = {astro-ph.GA},
       adsurl = {https://ui.adsabs.harvard.edu/abs/2010A&A...518L.102A}
}

@ARTICLE{Danieli_etal_2020,
       author = {{Danieli}, Shany and {Lokhorst}, Deborah and {Zhang}, Jielai and {Merritt}, Allison and {van Dokkum}, Pieter and {Abraham}, Roberto and {Conroy}, Charlie and {Gilhuly}, Colleen and {Greco}, Johnny and {Janssens}, Steven and {Li}, Jiaxuan and {Liu}, Qing and {Miller}, Tim B. and {Mowla}, Lamiya},
        title = "{The Dragonfly Wide Field Survey. I. Telescope, Survey Design, and Data Characterization}",
      journal = {\apj},
         year = 2020,
        month = may,
       volume = {894},
       number = {2},
          eid = {119},
        pages = {119},
          doi = {10.3847/1538-4357/ab88a8},
archivePrefix = {arXiv},
       eprint = {1910.14045},
 primaryClass = {astro-ph.GA},
       adsurl = {https://ui.adsabs.harvard.edu/abs/2020ApJ...894..119D}
}

@ARTICLE{Clark_etal_2014,
       author = {{Clark}, S.~E. and {Peek}, J.~E.~G. and {Putman}, M.~E.},
        title = "{Magnetically Aligned H I Fibers and the Rolling Hough Transform}",
      journal = {\apj},
         year = 2014,
        month = jul,
       volume = {789},
       number = {1},
          eid = {82},
        pages = {82},
          doi = {10.1088/0004-637X/789/1/82},
archivePrefix = {arXiv},
       eprint = {1312.1338},
 primaryClass = {astro-ph.GA},
       adsurl = {https://ui.adsabs.harvard.edu/abs/2014ApJ...789...82C}
}

@ARTICLE{Euclid_2011,
       author = {{Laureijs}, R. and {Amiaux}, J. and {Arduini}, S. and {Augu{\`e}res}, J. -L. and {Brinchmann}, J. and {Cole}, R. and {Cropper}, M. and {Dabin}, C. and {Duvet}, L. and {Ealet}, A. and {Garilli}, B. and {Gondoin}, P. and {Guzzo}, L. and {Hoar}, J. and {Hoekstra}, H. and {Holmes}, R. and {Kitching}, T. and {Maciaszek}, T. and {Mellier}, Y. and {Pasian}, F. and {Percival}, W. and {Rhodes}, J. and {Saavedra Criado}, G. and {Sauvage}, M. and {Scaramella}, R. and {Valenziano}, L. and {Warren}, S. and {Bender}, R. and {Castander}, F. and {Cimatti}, A. and {Le F{\`e}vre}, O. and {Kurki-Suonio}, H. and {Levi}, M. and {Lilje}, P. and {Meylan}, G. and {Nichol}, R. and {Pedersen}, K. and {Popa}, V. and {Rebolo Lopez}, R. and {Rix}, H. -W. and {Rottgering}, H. and {Zeilinger}, W. and {Grupp}, F. and {Hudelot}, P. and {Massey}, R. and {Meneghetti}, M. and {Miller}, L. and {Paltani}, S. and {Paulin-Henriksson}, S. and {Pires}, S. and {Saxton}, C. and {Schrabback}, T. and {Seidel}, G. and {Walsh}, J. and {Aghanim}, N. and {Amendola}, L. and {Bartlett}, J. and {Baccigalupi}, C. and {Beaulieu}, J. -P. and {Benabed}, K. and {Cuby}, J. -G. and {Elbaz}, D. and {Fosalba}, P. and {Gavazzi}, G. and {Helmi}, A. and {Hook}, I. and {Irwin}, M. and {Kneib}, J. -P. and {Kunz}, M. and {Mannucci}, F. and {Moscardini}, L. and {Tao}, C. and {Teyssier}, R. and {Weller}, J. and {Zamorani}, G. and {Zapatero Osorio}, M.~R. and {Boulade}, O. and {Foumond}, J.~J. and {Di Giorgio}, A. and {Guttridge}, P. and {James}, A. and {Kemp}, M. and {Martignac}, J. and {Spencer}, A. and {Walton}, D. and {Bl{\"u}mchen}, T. and {Bonoli}, C. and {Bortoletto}, F. and {Cerna}, C. and {Corcione}, L. and {Fabron}, C. and {Jahnke}, K. and {Ligori}, S. and {Madrid}, F. and {Martin}, L. and {Morgante}, G. and {Pamplona}, T. and {Prieto}, E. and {Riva}, M. and {Toledo}, R. and {Trifoglio}, M. and {Zerbi}, F. and {Abdalla}, F. and {Douspis}, M. and {Grenet}, C. and {Borgani}, S. and {Bouwens}, R. and {Courbin}, F. and {Delouis}, J. -M. and {Dubath}, P. and {Fontana}, A. and {Frailis}, M. and {Grazian}, A. and {Koppenh{\"o}fer}, J. and {Mansutti}, O. and {Melchior}, M. and {Mignoli}, M. and {Mohr}, J. and {Neissner}, C. and {Noddle}, K. and {Poncet}, M. and {Scodeggio}, M. and {Serrano}, S. and {Shane}, N. and {Starck}, J. -L. and {Surace}, C. and {Taylor}, A. and {Verdoes-Kleijn}, G. and {Vuerli}, C. and {Williams}, O.~R. and {Zacchei}, A. and {Altieri}, B. and {Escudero Sanz}, I. and {Kohley}, R. and {Oosterbroek}, T. and {Astier}, P. and {Bacon}, D. and {Bardelli}, S. and {Baugh}, C. and {Bellagamba}, F. and {Benoist}, C. and {Bianchi}, D. and {Biviano}, A. and {Branchini}, E. and {Carbone}, C. and {Cardone}, V. and {Clements}, D. and {Colombi}, S. and {Conselice}, C. and {Cresci}, G. and {Deacon}, N. and {Dunlop}, J. and {Fedeli}, C. and {Fontanot}, F. and {Franzetti}, P. and {Giocoli}, C. and {Garcia-Bellido}, J. and {Gow}, J. and {Heavens}, A. and {Hewett}, P. and {Heymans}, C. and {Holland}, A. and {Huang}, Z. and {Ilbert}, O. and {Joachimi}, B. and {Jennins}, E. and {Kerins}, E. and {Kiessling}, A. and {Kirk}, D. and {Kotak}, R. and {Krause}, O. and {Lahav}, O. and {van Leeuwen}, F. and {Lesgourgues}, J. and {Lombardi}, M. and {Magliocchetti}, M. and {Maguire}, K. and {Majerotto}, E. and {Maoli}, R. and {Marulli}, F. and {Maurogordato}, S. and {McCracken}, H. and {McLure}, R. and {Melchiorri}, A. and {Merson}, A. and {Moresco}, M. and {Nonino}, M. and {Norberg}, P. and {Peacock}, J. and {Pello}, R. and {Penny}, M. and {Pettorino}, V. and {Di Porto}, C. and {Pozzetti}, L. and {Quercellini}, C. and {Radovich}, M. and {Rassat}, A. and {Roche}, N. and {Ronayette}, S. and {Rossetti}, E. and {Sartoris}, B. and {Schneider}, P. and {Semboloni}, E. and {Serjeant}, S. and {Simpson}, F. and {Skordis}, C. and {Smadja}, G. and {Smartt}, S. and {Spano}, P. and {Spiro}, S. and {Sullivan}, M. and {Tilquin}, A. and {Trotta}, R. and {Verde}, L. and {Wang}, Y. and {Williger}, G. and {Zhao}, G. and {Zoubian}, J. and {Zucca}, E.},
        title = "{Euclid Definition Study Report}",
      journal = {arXiv e-prints},
         year = 2011,
        month = oct,
          eid = {arXiv:1110.3193},
        pages = {arXiv:1110.3193},
archivePrefix = {arXiv},
       eprint = {1110.3193},
 primaryClass = {astro-ph.CO},
       adsurl = {https://ui.adsabs.harvard.edu/abs/2011arXiv1110.3193L}
}

@ARTICLE{LSST_2009,
       author = {{LSST Science Collaboration} and {Abell}, Paul A. and {Allison}, Julius and {Anderson}, Scott F. and {Andrew}, John R. and {Angel}, J. Roger P. and {Armus}, Lee and {Arnett}, David and {Asztalos}, S.~J. and {Axelrod}, Tim S. and {Bailey}, Stephen and {Ballantyne}, D.~R. and {Bankert}, Justin R. and {Barkhouse}, Wayne A. and {Barr}, Jeffrey D. and {Barrientos}, L. Felipe and {Barth}, Aaron J. and {Bartlett}, James G. and {Becker}, Andrew C. and {Becla}, Jacek and {Beers}, Timothy C. and {Bernstein}, Joseph P. and {Biswas}, Rahul and {Blanton}, Michael R. and {Bloom}, Joshua S. and {Bochanski}, John J. and {Boeshaar}, Pat and {Borne}, Kirk D. and {Bradac}, Marusa and {Brandt}, W.~N. and {Bridge}, Carrie R. and {Brown}, Michael E. and {Brunner}, Robert J. and {Bullock}, James S. and {Burgasser}, Adam J. and {Burge}, James H. and {Burke}, David L. and {Cargile}, Phillip A. and {Chandrasekharan}, Srinivasan and {Chartas}, George and {Chesley}, Steven R. and {Chu}, You-Hua and {Cinabro}, David and {Claire}, Mark W. and {Claver}, Charles F. and {Clowe}, Douglas and {Connolly}, A.~J. and {Cook}, Kem H. and {Cooke}, Jeff and {Cooray}, Asantha and {Covey}, Kevin R. and {Culliton}, Christopher S. and {de Jong}, Roelof and {de Vries}, Willem H. and {Debattista}, Victor P. and {Delgado}, Francisco and {Dell'Antonio}, Ian P. and {Dhital}, Saurav and {Di Stefano}, Rosanne and {Dickinson}, Mark and {Dilday}, Benjamin and {Djorgovski}, S.~G. and {Dobler}, Gregory and {Donalek}, Ciro and {Dubois-Felsmann}, Gregory and {Durech}, Josef and {Eliasdottir}, Ardis and {Eracleous}, Michael and {Eyer}, Laurent and {Falco}, Emilio E. and {Fan}, Xiaohui and {Fassnacht}, Christopher D. and {Ferguson}, Harry C. and {Fernandez}, Yanga R. and {Fields}, Brian D. and {Finkbeiner}, Douglas and {Figueroa}, Eduardo E. and {Fox}, Derek B. and {Francke}, Harold and {Frank}, James S. and {Frieman}, Josh and {Fromenteau}, Sebastien and {Furqan}, Muhammad and {Galaz}, Gaspar and {Gal-Yam}, A. and {Garnavich}, Peter and {Gawiser}, Eric and {Geary}, John and {Gee}, Perry and {Gibson}, Robert R. and {Gilmore}, Kirk and {Grace}, Emily A. and {Green}, Richard F. and {Gressler}, William J. and {Grillmair}, Carl J. and {Habib}, Salman and {Haggerty}, J.~S. and {Hamuy}, Mario and {Harris}, Alan W. and {Hawley}, Suzanne L. and {Heavens}, Alan F. and {Hebb}, Leslie and {Henry}, Todd J. and {Hileman}, Edward and {Hilton}, Eric J. and {Hoadley}, Keri and {Holberg}, J.~B. and {Holman}, Matt J. and {Howell}, Steve B. and {Infante}, Leopoldo and {Ivezic}, Zeljko and {Jacoby}, Suzanne H. and {Jain}, Bhuvnesh and {R} and {Jedicke} and {Jee}, M. James and {Garrett Jernigan}, J. and {Jha}, Saurabh W. and {Johnston}, Kathryn V. and {Jones}, R. Lynne and {Juric}, Mario and {Kaasalainen}, Mikko and {Styliani} and {Kafka} and {Kahn}, Steven M. and {Kaib}, Nathan A. and {Kalirai}, Jason and {Kantor}, Jeff and {Kasliwal}, Mansi M. and {Keeton}, Charles R. and {Kessler}, Richard and {Knezevic}, Zoran and {Kowalski}, Adam and {Krabbendam}, Victor L. and {Krughoff}, K. Simon and {Kulkarni}, Shrinivas and {Kuhlman}, Stephen and {Lacy}, Mark and {Lepine}, Sebastien and {Liang}, Ming and {Lien}, Amy and {Lira}, Paulina and {Long}, Knox S. and {Lorenz}, Suzanne and {Lotz}, Jennifer M. and {Lupton}, R.~H. and {Lutz}, Julie and {Macri}, Lucas M. and {Mahabal}, Ashish A. and {Mandelbaum}, Rachel and {Marshall}, Phil and {May}, Morgan and {McGehee}, Peregrine M. and {Meadows}, Brian T. and {Meert}, Alan and {Milani}, Andrea and {Miller}, Christopher J. and {Miller}, Michelle and {Mills}, David and {Minniti}, Dante and {Monet}, David and {Mukadam}, Anjum S. and {Nakar}, Ehud and {Neill}, Douglas R. and {Newman}, Jeffrey A. and {Nikolaev}, Sergei and {Nordby}, Martin and {O'Connor}, Paul and {Oguri}, Masamune and {Oliver}, John and {Olivier}, Scot S. and {Olsen}, Julia K. and {Olsen}, Knut and {Olszewski}, Edward W. and {Oluseyi}, Hakeem and {Padilla}, Nelson D. and {Parker}, Alex and {Pepper}, Joshua and {Peterson}, John R. and {Petry}, Catherine and {Pinto}, Philip A. and {Pizagno}, James L. and {Popescu}, Bogdan and {Prsa}, Andrej and {Radcka}, Veljko and {Raddick}, M. Jordan and {Rasmussen}, Andrew and {Rau}, Arne and {Rho}, Jeonghee and {Rhoads}, James E. and {Richards}, Gordon T. and {Ridgway}, Stephen T. and {Robertson}, Brant E. and {Roskar}, Rok and {Saha}, Abhijit and {Sarajedini}, Ata and {Scannapieco}, Evan and {Schalk}, Terry and {Schindler}, Rafe and {Schmidt}, Samuel and {Schmidt}, Sarah and {Schneider}, Donald P. and {Schumacher}, German and {Scranton}, Ryan and {Sebag}, Jacques and {Seppala}, Lynn G. and {Shemmer}, Ohad and {Simon}, Joshua D. and {Sivertz}, M. and {Smith}, Howard A. and {Allyn Smith}, J. and {Smith}, Nathan and {Spitz}, Anna H. and {Stanford}, Adam and {Stassun}, Keivan G. and {Strader}, Jay and {Strauss}, Michael A. and {Stubbs}, Christopher W. and {Sweeney}, Donald W. and {Szalay}, Alex and {Szkody}, Paula and {Takada}, Masahiro and {Thorman}, Paul and {Trilling}, David E. and {Trimble}, Virginia and {Tyson}, Anthony and {Van Berg}, Richard and {Vanden Berk}, Daniel and {VanderPlas}, Jake and {Verde}, Licia and {Vrsnak}, Bojan and {Walkowicz}, Lucianne M. and {Wandelt}, Benjamin D. and {Wang}, Sheng and {Wang}, Yun and {Warner}, Michael and {Wechsler}, Risa H. and {West}, Andrew A. and {Wiecha}, Oliver and {Williams}, Benjamin F. and {Willman}, Beth and {Wittman}, David and {Wolff}, Sidney C. and {Wood-Vasey}, W. Michael and {Wozniak}, Przemek and {Young}, Patrick and {Zentner}, Andrew and {Zhan}, Hu},
        title = "{LSST Science Book, Version 2.0}",
      journal = {arXiv e-prints},
         year = 2009,
        month = dec,
          eid = {arXiv:0912.0201},
        pages = {arXiv:0912.0201},
archivePrefix = {arXiv},
       eprint = {0912.0201},
 primaryClass = {astro-ph.IM},
       adsurl = {https://ui.adsabs.harvard.edu/abs/2009arXiv0912.0201L}
}

@article{opencv_library,
    author = {Bradski, G.},
    journal = {Dr. Dobb's Journal of Software Tools},
    title = {{The OpenCV Library}},
    year = {2000}
}

@article{Kingma2014AdamAM,
  title={Adam: A Method for Stochastic Optimization},
  author={Diederik P. Kingma and Jimmy Ba},
  journal={CoRR},
  year={2014},
  volume={abs/1412.6980},
  url={https://api.semanticscholar.org/CorpusID:6628106}
}

@ARTICLE{2024AJ....168...88Z,
       author = {{Zhao}, Yunning and {Zhang}, Wei and {Ma}, Lin and {Wen}, Shiming and {Wu}, Hong},
        title = "{Galactic Cirri at High Galactic Latitudes. I. Investigating Scatter in Slopes between Optical and Far-infrared Intensities}",
      journal = {\aj},
         year = 2024,
        month = aug,
       volume = {168},
       number = {2},
          eid = {88},
        pages = {88},
          doi = {10.3847/1538-3881/ad58d5},
archivePrefix = {arXiv},
       eprint = {2406.03031},
 primaryClass = {astro-ph.GA},
       adsurl = {https://ui.adsabs.harvard.edu/abs/2024AJ....168...88Z}
}

@ARTICLE{2022ApJ...925..219L,
       author = {{Liu}, Qing and {Abraham}, Roberto and {Gilhuly}, Colleen and {van Dokkum}, Pieter and {Martin}, Peter G. and {Li}, Jiaxuan and {Greco}, Johnny P. and {Lokhorst}, Deborah and {Chen}, Seery and {Danieli}, Shany and {Keim}, Michael A. and {Merritt}, Allison and {Miller}, Tim B. and {Pasha}, Imad and {Polzin}, Ava and {Shen}, Zili and {Zhang}, Jielai},
        title = "{A Method to Characterize the Wide-angle Point-Spread Function of Astronomical Images}",
      journal = {\apj},
         year = 2022,
        month = feb,
       volume = {925},
       number = {2},
          eid = {219},
        pages = {219},
          doi = {10.3847/1538-4357/ac32c6},
archivePrefix = {arXiv},
       eprint = {2110.11598},
 primaryClass = {astro-ph.IM},
       adsurl = {https://ui.adsabs.harvard.edu/abs/2022ApJ...925..219L}
}

@ARTICLE{2025ApJ...979..175L,
       author = {{Liu}, Qing and {Abraham}, Roberto and {Martin}, Peter G. and {Bowman}, William P. and {Dokkum}, Pieter van and {Danieli}, Shany and {Patel}, Ekta and {Janssens}, Steven R. and {Shen}, Zili and {Chen}, Seery and {Karunakaran}, Ananthan and {Keim}, Michael A. and {Lokhorst}, Deborah and {Pasha}, Imad and {Welch}, Douglas L.},
        title = "{Fuzzy Galaxies or Cirrus? Decomposition of Galactic Cirrus in Deep Wide-field Images}",
      journal = {\apj},
         year = 2025,
        month = feb,
       volume = {979},
       number = {2},
          eid = {175},
        pages = {175},
          doi = {10.3847/1538-4357/ad9b25},
       adsurl = {https://ui.adsabs.harvard.edu/abs/2025ApJ...979..175L}
}

@article{Szomoru1999ExtinctionCD,
  title={Extinction Curves, Distances, and Clumpiness of Diffuse Interstellar Dust Clouds,},
  author={Arpad Szomoru and Puragra Guhathakurta},
  journal={The Astronomical Journal},
  year={1999},
  volume={117},
  pages={2226 - 2243},
  url={https://api.semanticscholar.org/CorpusID:15392082}
}

@ARTICLE{Wolf_1923,
       author = {{Wolf}, M.},
        title = "{{\"U}ber den dunklen Nebel NGC 6960}",
      journal = {Astronomische Nachrichten},
         year = 1923,
        month = jul,
       volume = {219},
       number = {7},
        pages = {109},
          doi = {10.1002/asna.19232190702},
       adsurl = {https://ui.adsabs.harvard.edu/abs/1923AN....219..109W}
}

@ARTICLE{Onishi_etal_2018,
       author = {{Onishi}, Yosuke and {Sano}, Kei and {Matsuura}, Shuji and {Jeong}, Woong-Seob and {Pyo}, Jeonghyun and {Kim}, Il-Jong and {Seo}, Hyun Jong and {Han}, Wonyong and {Lee}, DaeHee and {Moon}, Bongkon and {Park}, Wonkee and {Park}, Younsik and {Kim}, MinGyu and {Matsumoto}, Toshio and {Matsuhara}, Hideo and {Nakagawa}, Takao and {Tsumura}, Kohji and {Shirahata}, Mai and {Arai}, Toshiaki and {Ienaka}, Nobuyuki},
        title = "{MIRIS observation of near-infrared diffuse Galactic light}",
      journal = {\pasj},
         year = 2018,
        month = aug,
       volume = {70},
       number = {4},
          eid = {76},
        pages = {76},
          doi = {10.1093/pasj/psy070},
archivePrefix = {arXiv},
       eprint = {1806.08891},
 primaryClass = {astro-ph.GA},
       adsurl = {https://ui.adsabs.harvard.edu/abs/2018PASJ...70...76O}
}

@ARTICLE{WLJ_2015ApJ...811...38W,
       author = {{Wang}, Shu and {Li}, Aigen and {Jiang}, B.~W.},
        title = "{Very Large Interstellar Grains as Evidenced by the Mid-infrared Extinction}",
      journal = {\apj},
         year = 2015,
        month = sep,
       volume = {811},
       number = {1},
          eid = {38},
        pages = {38},
          doi = {10.1088/0004-637X/811/1/38},
archivePrefix = {arXiv},
       eprint = {1508.03403},
 primaryClass = {astro-ph.GA},
       adsurl = {https://ui.adsabs.harvard.edu/abs/2015ApJ...811...38W}
}

@ARTICLE{Bowes_Martin_2023ApJ...959...40B,
       author = {{Bowes}, Shannon K. and {Martin}, Peter G.},
        title = "{Diagnostics from Polarization of Scattered Optical Light from Galactic Infrared Cirrus}",
      journal = {\apj},
         year = 2023,
        month = dec,
       volume = {959},
       number = {1},
          eid = {40},
        pages = {40},
          doi = {10.3847/1538-4357/ad0971},
archivePrefix = {arXiv},
       eprint = {2311.01376},
 primaryClass = {astro-ph.GA},
       adsurl = {https://ui.adsabs.harvard.edu/abs/2023ApJ...959...40B}
}






\end{document}